\documentclass[iop,revtex4,12pt,preprint]{emulateapj_recent}

\usepackage[utf8]{inputenc}

\usepackage{lipsum}
\newsavebox{\myimage}
\usepackage{graphics,graphicx}
\usepackage{helvet}
\usepackage{subfigure}
\usepackage[utf8]{inputenc}
\usepackage[T1]{fontenc}
\usepackage{soul} 
 \usepackage[colorlinks,urlcolor=blue,citecolor=blue,linkcolor=blue,pdfusetitle]{hyperref}
\usepackage{appendix}
\usepackage{amsmath, amssymb, gensymb}
\usepackage{morefloats}
\usepackage{float}
\usepackage{ulem}
\usepackage{url}
\usepackage{natbib}
\usepackage{microtype}
\usepackage{multirow}
\usepackage{morefloats}
\usepackage{here}
\usepackage{xspace}
\usepackage{color}
\usepackage{lipsum}
\usepackage{xspace}
\usepackage[figuresright]{rotating}

\usepackage{float}
\usepackage{stmaryrd}

\usepackage{longtable}

\def\S{\mathcal{S}}

\def\eg {e.g.,\xspace} 
\def\ie {i.e.,\xspace} 

\def\etal {\textit{et al.}\xspace}

\def\HttwotooneSfour {$\rm  H_{2}\,2-1\,S(4) $\xspace}
\def\HtonetoOStwo {$\rm  H_{2}\,1-0\,S2 $\xspace}
\def\HtthreetotwoSfive {$\rm  H_{2}\,3-2\,S(5) $\xspace}
\def\HttwotooneSthree {$\rm  H_{2}\,2-1\,S(3) $\xspace}
\def\HtonetoOSone {$\rm  H_{2}\,1-0\,S1 $\xspace}
\def\HtthreetotwoSfour {$\rm  H_{2}\,3-2\,S(4) $\xspace}
\def\HttwotooneStwo {$\rm  H_{2}\,2-1\,S2 $\xspace}
\def\HtthreetotwoSthree {$\rm  H_{2}\,3-2\,S(3) $\xspace}
\def\HtonetoOSO {$\rm  H_{2}\,1-0\,S0 $\xspace}
\def\HttwotooneSone {$\rm  H_{2}\,2-1\,S1 $\xspace}
\def\HtthreetotwoStwo {$\rm  H_{2}\,3-2\,S2 $\xspace}
\def\HtfourtothreeSthree {$\rm  H_{2}\,4-3\,S(3) $\xspace}
\def\HttwotooneSO {$\rm  H_{2}\,2-1\,S0 $\xspace}
\def\HtthreetotwoSone {$\rm  H_{2}\,3-2\,S1 $\xspace}

\def\Brgamma {Br$\gamma$\xspace}
\def\COfirstband {CO (2$\,\rightarrow\,$0) \xspace}

\begin{document}
~

\title{New insights on the accretion properties of Class 0 protostars \\
from 2 micron spectroscopy}

\shorttitle{Near Infrared spectroscopy of Class 0 protostars}

%
\author{Valentin J. M. Le Gouellec}
\affiliation{NASA Ames Research Center, Space Science and Astrobiology Division M.S. 245-6 Moffett Field, CA 94035, USA}
\affiliation{NASA Postdoctoral Program Fellow}

\author{Thomas P. Greene}
\affiliation{NASA Ames Research Center, Space Science and Astrobiology Division M.S. 245-6 Moffett Field, CA 94035, USA}

\author{Lynne A. Hillenbrand}
\affiliation{Department of Astronomy, MC 249-17, California Institute of Technology, Pasadena, CA 91125, USA}

\author{ Zoe Yates}
\affiliation{NASA Ames Research Center, Space Science and Astrobiology Division M.S. 245-6 Moffett Field, CA 94035, USA}

\shortauthors{Le Gouellec \etal}
\email{valentin.j.legouellec@nasa.gov}

\slugcomment{Accepted for publication in ApJ on 01/29/2024}

\begin{abstract}
Sun-like stars are thought to accrete most of their final mass during the protostellar phase, during which the stellar embryo is surrounded by an infalling dense envelope. We present an analysis of 26 $K$-band spectra of Class 0 protostars, which are the youngest protostars. 18 of these are new observations made with the Keck MOSFIRE instrument. \ion{H}{1} \Brgamma, several H$_2$, and CO $\Delta\,v\;=\;2$ features are detected and analyzed. We detect \Brgamma emission in 62\%, CO overtone emission in 50\%, and H$_2$ emission in 90\% of sources. The \ion{H}{1} and CO emission is associated with accretion while the H$_2$ lines are consistent with shock excitation indicating jets/outflows. Six objects exhibit photospheric absorption features, with almost no outflow activity, and no detection of the accretion-related \Brgamma emission line.  Comparing these results with archival sample of Class I $K$-band spectra, we find that the CO and \Brgamma emission lines are systematically more luminous in Class 0s, suggesting the accretion is on average more vigorous in the Class 0 phase. Typically associated with the heated inner accretion disk, the much higher detection rate of CO overtone emission in Class 0s indicate also that episodes of high accretion activity are more frequent in Class 0 systems. The kinematics of the Class 0 CO overtone emission suggest either an accretion-heated inner disk, or material directly infalling onto the central region. 
This could point toward an accretion mechanism of different nature in Class 0 systems than the typical picture of magnetospheric accretion.
\end{abstract}

\section{Introduction}
\label{sec:intro}

Along the star formation evolutionary path, protostars originate from pre-stellar cores and will ultimately lead to pre-main-sequence (or T-Tauri) stars \citep{Adams1987,Kenyon1993a}. Class I protostars were first characterized by their infrared (IR) spectral energy distribution (SED) slope \citep{Wilking1989} in association with dense molecular gas \citep{Myers1987}. However, they have been interpreted as being already relatively evolved, with typical ages of $\sim$ 10$^{5}$ years \citep{Greene1994,Kenyon1995,Kristensen2018}, consisting of a central proto-stellar embryo surrounded by both a diffuse envelope and circumstellar disk \citep{Andre1994}. The Class 0 protostellar phase has thus been defined as the short life-time main accretion phase succeeding the formation of the second Larson core \citep{Maury2011b}, during which most of the final star's mass that initially is in the form of a dense and cold circumstellar envelope, is accreted on to the central nascent protostellar embryo \citep{Andre1993,Andre2000,Andre2002}. The spectral energy distribution (SED) of Class 0 protostars is mainly governed by the dust thermal emission emanating from a circumstellar envelope, characterized by ground based submillimeter observations (\eg see the continuum surveys with Bolocam \citealt{Enoch2009b}, MAMBO \citealt{Maury2011}, and SCUBA \citet{Mottram2017,Karska2018}), and space based (far-) IR telescopes (surveys by \textit{Spitzer} and \textit{Herschel}; \eg \citealt{Evans2009,Andre2010,Megeath2012,Dunham2015,Marsh2016,Ladjelate2020,Fiorellino2021a,Pezzuto2021}).

Protostars derive a substantial fraction of their luminosity from the accretion processes, that liberate a significant fraction of energy in the form of radiative energy. Several surveys \citep{Kenyon1990b,Kenyon1994,Evans2009,Dunham2010a,Dunham2013} reported the existence of a ``luminosity problem'' comparing the observed distribution of protostellar luminosities with the expected average accretion luminosity that should emanate from these objects during their typical lifetime \citep{Myers1998,Young2005}. 
Later on, different models implementing time-dependent accretion rates, episodic in nature or not, found reasonable match with observed protostellar luminosities \citep[\eg][]{Myers2009,Myers2010,Offner2011,Fischer2017}.
Indeed, several observational works suggested that protostars actually accrete at a variable rate with time, on timescales of months/years \citep{Fischer2019,LeeYH2021,Park2021,Zakri2022}. This is also supported by indirect evidences of past episodes of high envelope temperature \citep{Visser2012,Jorgensen2013,Visser2015,Anderl2016,Frimann2016}, and traces of variability in the mass ejection rate detected in molecular outflows \citep{Arce2001,Tafalla2004,Plunkett2015,Dutta2023}. While the accretion scenario occurring in the youngest embedded objects remains uncertain, these recent improvements now led the field to consider a ``protostellar luminosity spread'' \citep{Fischer2023}, for which the physical mechanisms occurring at small scales (\ie $\lesssim$ 1 au) must be characterized.

Protostellar accretion is a fundamental problem that is related to a variety of physical mechanisms.
In this embedded phase, a complete understanding of the status of disk formation and its properties (magnetization, turbulence; see recent review by \citealt{Maury2022,Tsukamoto2022}), the extraction of angular momentum by the launching of jets and outflows \citep{Bally2016}, the magnetization of the central embryo, the coupling of the disk with the infalling envelope, is still lacking. For example, recent observations have shown how complex the inner envelope infalling structures can be, with potentially strong impacts on the subsequent accretion onto the central object(s) (see \citealt{Pineda2023} and references there in). In the much less embedded T-Tauri stars, the accretion is thought to occur via magnetospheric accretion (see review by \citealt{Hartmann2016}).  Within $\sim$0.1 au typically, the disk is truncated by the stellar magnetosphere at a few stellar radii, and the magnetic field lines guide the matter through funnel flows to the star. In protostars, the dominant accretion mechanism remains to be fully characterized.

There are two accepted ways to estimate the mass accretion rate in protostars, whose emission in the UV and visible is shielded by the dust in the disk and envelope. One is the measurement of the hot continuum excess  that originates from the accretion shock where disk material infalling along magnetic field lines impacts the stellar surface \citep{Bertout1988,Gullbring1998,Greene1996,White2004,White2007}. This blue excess peaks in the UV, whose tail can be detected in the optical and NIR.
The second technique is to analyze emission lines produced toward the accretion shock, \eg the H {\footnotesize{I}} emission lines such as Balmer, Paschen, and Brackett series, whose emission has been calibrated to the accretion luminosity directly measured from the UV continuum excess \citep{Najita1996a,Muzerolle1998,Alcala2017}. A few NIR studies have explored accretion properties of Class I protostars, revealing H {\footnotesize{I}} \Brgamma emission line profile consistent with infall \citep{Nisini2005,Doppmann2005,Antoniucci2008,Connelley2010,ContrerasPena2017}. More recently, \citet{Fiorellino2021,Fiorellino2023} derived the mass accretion rate for a sample of Class Is via a semi-empirical method based on the luminosity of \Brgamma line, the bolometric luminosity, the veiling and using either the birthline as defined by \citet{Palla1993} or the 1 Myr isochrone of the \citet{Siess2000} evolutionary tracks. They found that their sample shows larger mass accretion rates compared to Class II stars, but not high enough for the systems to build up the inferred stellar mass. This suggests that most of the accretion must occur during the Class 0 phase, or that the accretion process proceeds in a non-steady framework such that a single epoch of accretion rate measurements may not properly reflect the final stellar mass build up.

During the earlier, relatively short Class 0 stage, it is not clear what is the dominant mode of accretion. The typical size of Class 0 disks suggest that the disk formation is affected by the magnetic fields that redistribute the angular momentum of the collapsing and rotating inner envelope material \citep{Maury2019}. This suggests that the inner envelope magnetization could be a strong regulator of the actual accretion onto the protostellar embryo. In their numerical work, \citet{LeeYN2021} have found that the mid-plane of embedded disks is initially  lacking turbulence, and that the highly turbulent disk upper layers would drive the accretion. In this picture, the accretion is not purely disk mediated but results from a strong coupling between the large scale dynamics of the magnetized infalling envelope and the inner disk (see also \citealt{Kuffmeier2018}). Other non-ideal MHD simulations implementing ambipolar and ohmic diffusion predict an efficient redistribution of the magnetic flux preventing the amplification of the magnetic field beyond 0.1 G in the first Larson core, resulting in a initial fully thermally supported proto-stellar core \citep{Vaytet2018,Wurster2018b,Wurster2020b}. 

It is not clear if magnetospheric accretion is occurring at the Class 0 stage, \ie if the internal dynamo of the protostellar embryo started to truncate the inner disk and funnel the accretion flows. Recent numerical studies found that the turbulent motion emerging at the birth of the protostellar embryo could drive a dynamo process much earlier on than what was previously thought \citep{Bhandare2020,Ahmad2023}. In a Class I object, \citet{JohnsKrull2009} derived a strong magnetic field of $\sim$ 3 kG using Zeeman broadening observations, suggesting that magnetospheric accretion can be established in the protostellar phase. One would need to also determine what magnetic field strength would be required to sustain the mass accretion rate expected in the Class 0 phase via magnetospheric accretion. Such predictions need to be challenged by NIR observations of Class 0s that can constrain the kinematics of the accretion material and inner disk.

It remains difficult to characterize the accretion onto central objects in the Class 0 phase due to the high extinction, given that their NIR counterparts are difficult to detect. Characteristics of Class 0 protostars' accretion activity have always been performed with indirect measurements, such as measurement of the mass ejection rate \citep{Bontemps1996}, the effect of accretion luminosity on the chemistry of the envelope \citep{Visser2012b}, or the detailed kinematics of the infalling envelope material   \citep{Mottram2013,Pineda2020,Cabedo2021,ValdiviaMena2022}. Recently, \citet{Laos2021} conducted a pilot study toward a few protostars that are bright enough in the NIR to be observed with spectroscopy. They exhibited a few detections of \Brgamma and CO overtone emission lines, suggesting that the quantitative characterization of Class 0 accretion properties is possible with  moderate-resolution NIR spectroscopy. The present work expands this initial study to a larger sample of Class 0 sources, allowing us to perform relevant comparisons with the line characteristics of the same accretion tracers observed toward more evolved objects. 

This paper is structured as follows. In Section \ref{sec:obs}, we present our new sample of Class 0 protostars observed in the NIR with Keck, and detail the data reduction processes. The spectral line analysis of this sample and other previously-observed Class 0 objects is presented in Section \ref{sec:results}, where we run different diagnostics in light of the physical mechanisms expected to be responsible for these spectral features. Section \ref{sec:comp_ClassI} is dedicated to quantitatively compare the results of the spectral line analysis performed on our Class 0 sample, with those obtained on a sample of archival NIR spectroscopic observations of Class I objects. We discuss our results in Section \ref{sec:disc} in the context of the specific accretion properties constrained with this work. We draw our conclusions in Section \ref{sec:ccl}.

\section{Observations}
\label{sec:obs}

\subsection{Selection of Class 0 Protostars}
\label{sec:Class0_sample}

In order to enable significant comparisons between the NIR spectral properties of Class 0 and I protostars, we had to increase the sample of Class 0 observed with NIR spectroscopy from the 8 objects targets presented in \citet{Greene2018} and \citet{Laos2021}. In order to build this new sample, we evaluated different source catalogs that identified YSOs via IR observations \citep[\ie the c2d Spitzer legacy project, the Gould Belt legacy survey, and Herschel Orion Protostar Survey][respectively]{Evans2009,Dunham2015,Furlan2016}, focusing on nearby star forming molecular clouds (\ie $\leq$ 500 pc from us). 

We first selected sources that met the usual Class 0 criteria, \ie that differentiate Class 0 from Class I protostars by having a bolometric temperature $T_{\textrm{bol}}\,\leq\,70$, and a massive and cold circumstellar envelope as a major contributor of the SED with $L_{\textrm{submm}}$/$L_{\textrm{bol}}\,\geq\,0.05\%$ \citep{Andre2000,Andre2002}. Then, we looked for Class 0 sources that have a bright and compact enough NIR counterpart, using the 2MASS \citet{Skrutskie2006}, UKIRT Infrared Deep Sky Survey (UKIDSS; \citealt{Lawrence2007}), and VISTA Orion survey \citep{Meingast2016,Grobschedl2019} catalogs. We required a $K$-band magnitude less than 16, and sources less extended than $\sim\,2\,^{\prime\prime}$ in the $K$-band images. Class 0 protostars are often totally extincted in the NIR due to the significant dust mass in the surrounding envelope. However, in some cases, the blueshifted cavity of the bipolar outflow clears enough envelope material out, such that NIR (scattered) light can escape from the inner regions. 

The process described above resulted in the selection of 18 new sources as bona fide low-mass Class 0 protostellar objects having $K$-band brightness $14-16$ mag: 1 in Perseus, 7 in Orion A, 9 in Aquila/Serpens, and 1 in Cepheus L1174. The characteristics of all 26 (18 new and 6 previously-observed) objects are presented in Table \ref{t.C0_source_charac}. We report the bolometric temperature ($T_\textrm{bol}$) and bolometric luminosity ($L\textrm{bol}$, that we scale with the clouds' distance we adopt) values taken from the literature \citep{Maury2011,Dunham2015,Furlan2016}. The $T_\textrm{bol}$ and $L\textrm{bol}$ uncertainties are typically estimated to be  $\sim\,30-40$ \% in these references. 
None of our protostars have a measurement of their parallax from \textit{Gaia} data DR2 due to their embedded nature, frequently resulting in extinction values >\,10 mag at visual wavelengths. We report the distance measurements of the local dust clouds from \citet{Herczeg2019,Zucker2019,Zucker2020} for the Serpens, Perseus, and Cepheus sources. For the Orion Sources, we used the distance estimates from \citet{Tobin2020}, who used the distances of neighboring stars in Orion A form \citet{McBride2019}. Typical uncertainties are $\pm$ 25 pc for the Serpens sources, $\pm$ 15 pc for the Cepheus and Perseus sources, and $\pm$ 10 pc for Orion sources.

\begin{table*}
\centering
\small
\caption[]{ Sample of 26 Class 0 objects}
\label{t.C0_source_charac}
\setlength{\tabcolsep}{0.6em} 
\begin{tabular}{p{0.14\linewidth}ccccccccc}
\hline \hline \noalign{\smallskip}
 Source & near-IR $\alpha_{\textrm{J2000}}$ & near-IR $\delta_{\textrm{J2000}}$ & UT Date & Int. time & $L_{\textrm{bol}}$ $^{(a)}$ & $T_{\textrm{bol}}$ $^{(a)}$ & Distance $^{(b)}$ & $v_{\textrm{lsr}}$ \\ 
 & & & & minutes & $L_\odot$ & K & pc & km s$^{-1}$\\
\noalign{\smallskip}  \hline 
\noalign{\smallskip} 
S68N$^*$ & 18:29:48.14 & +01:16:44.92 & 2014 Jun 18 & 248.7 &  18.6 & 30 & 495 & 8.45\\  
\noalign{\smallskip}  \hline \noalign{\smallskip} 
HOPS 32$^*$ &  05:34:35.4 & -05:39:59.0  & 2019 Oct 13 & 20 &  1.7 & 59 & 391 & 8.8\\ 
\noalign{\smallskip}  \hline \noalign{\smallskip}
HOPS 44$^*$ &  05:35:10.6 & -05:35:06.3  & 2019 Oct 14 & 12 &  1.5 & 44 & 392 & 9.0\\ 
\noalign{\smallskip}  \hline \noalign{\smallskip}
Per emb 8$^*$ &  03:44:44.0 & +32:01:36.2   & 2019 Oct 13 & 40 &  4.9 & 41 &294 & 9.8\\ 
\noalign{\smallskip}  \hline \noalign{\smallskip}
Per emb 25$^*$$^{(e)}$ & 03:26:37.4 & +30:15:28.4 & 2019 Oct 12 & 66 &  1.3 & 73 & 294 & 5.2\\ 
\noalign{\smallskip}  \hline \noalign{\smallskip}
Per emb 26$^*$ &   03:25:38.8 & +30:44:06.2    & 2019 Oct 14 & 48 &  9.7 & 48 & 294 & 4.6\\ 
\noalign{\smallskip}  \hline \noalign{\smallskip}
Per emb 28$^*$ &  03:43:51.0 & +32:03:08.1  & 2019 Oct 14 & 28 &  0.6 & 57 & 294 & 8.6\\ 
\noalign{\smallskip}  \hline \noalign{\smallskip}
Per emb 21$^*$$^{(c)}$ &  03:29:10.7 & +31:18:20.6  & 2019 Oct 14 & 32 &  4.6 & 58 & 294 & 8.6\\ 
\noalign{\smallskip}  \hline \noalign{\smallskip}
Ser SMM3 &  18:29:59.31 & +01:14:01.47  & 2022 Aug 16 & 45 &  28.6 & 44 & 495 & 8.0\\ 
\noalign{\smallskip}  \hline \noalign{\smallskip}
Aqu MM4 &  18:29:08.47 & -01:30:39.08  & 2022 Aug 16 & 35 &  8.6 & 32 & 455 & 7.3\\ 
\noalign{\smallskip}  \hline \noalign{\smallskip}
HOPS 50 &  05:34:40.90 & -05:31:44.50  & 2022 Nov 11 & 40 &  3.6 & 51 & 392 & 8.5\\ 
\noalign{\smallskip}  \hline \noalign{\smallskip}
HOPS 60 &  05:35:23.36 & -05:12:03.30  & 2022 Nov 10 & 20 & 19.2  & 54 & 393 & 8.0\\ 
\noalign{\smallskip}  \hline \noalign{\smallskip}
HOPS 87$^{(c)}$ &  05:35:23.41 & -05:01:28.63  & 2022 Nov 11 & 25 &  31.9 & 38 & 393 & 10.5 \\ 
\noalign{\smallskip}  \hline \noalign{\smallskip}
HOPS 164 &  05:37:00.50 & -06:37:10.23  & 2022 Nov 10 & 75 &  0.5 & 50 & 385 & 6.7\\ 
\noalign{\smallskip}  \hline \noalign{\smallskip}
HOPS 171 &  05:36:17.23 & -06:38:01.30  & 2022 Nov 11 & 45 &  1.5 & 62 & 383 & 8.0\\ 
\noalign{\smallskip}  \hline \noalign{\smallskip}
HOPS 203 &  05:36:22.76 & -06:46:03.30 & 2022 Nov 10 & 60 &  17.0 & 44 & 383 & 8.2\\ 
\noalign{\smallskip}  \hline \noalign{\smallskip} 
HOPS 250 &  05:40:48.82 & -08:06:57.39  & 2022 Nov 11  & 30 & 7.1 & 70 & 429 & 5.5\\ 
\noalign{\smallskip}  \hline \noalign{\smallskip} 
Per emb 24$^{(e)}$ &  03:28:45.30 & +31:05:42.02  & 2022 Nov 10  & 45 & 0.6 & 72 & 294 & 7.4\\ 
\noalign{\smallskip}  \hline \noalign{\smallskip}
Aqu MM11 &  18:30:46.94 & -01:56:45.48  &2023 Jun 07  & 45 &  0.8 & 54 & 455 & 6.4\\ 
\noalign{\smallskip}  \hline \noalign{\smallskip}
Ceph mm &  21:01:32.83	& +68:11:20.78  & 2023 Jun 07 & 10 &  5.8 & 26 & 341 & 2.8\\ 
\noalign{\smallskip}  \hline \noalign{\smallskip}
Ser emb 2  &  18:29:52.51 & +00:36:11.16  & 2023 Jun 07 & 45 &  100.7 & 68 & 495 & 7.9\\ 
\noalign{\smallskip}  \hline \noalign{\smallskip}
Ser emb 15  &  18:29:54.35 &  +00:36:01.51  & 2023 Jun 07 & 45 & 2.3 & 59 & 495 & 7.8\\ 
\noalign{\smallskip}  \hline \noalign{\smallskip}
Aqu MM5$^{(d)}$  &  18:29:23.41 & -01:38:55.59  & 2023 Jun 08 & 15 & 1.5 & 188 & 455 & 7.4\\ 
\noalign{\smallskip}  \hline \noalign{\smallskip}
Aqu MM8  &  18:30:28.95 & -01:56:02.643  &2023 Jun 08 & 35 & 1.7  & 34 & 455 & 8.6\\ 
\noalign{\smallskip}  \hline \noalign{\smallskip}
Ser emb 22  & 18:29:57.59 & +01:13:00.40  & 2023 Jun 08 & 15 & 38.6 & 59 & 495 & 6.9\\ 
\noalign{\smallskip}  \hline \noalign{\smallskip}
SerS MM16  &  18:30:02.45 & -2:02:46.083  & 2023 Jun 08 & 45 & 33.7 & 45 & 455 & 7.6\\ 
\noalign{\smallskip} 
\hline
\noalign{\smallskip}
\end{tabular}
\tablecomments{\small We scale the luminosities found in the literature to the distances we adopt for each source. 
$^{(a)}$ Bolometric temperature and bolometric luminosity values from the literature. \citet{Furlan2016} for the Orion sources. \citet{Dunham2015} for the Perseus and Serpens Main sources. \citet{Maury2011} for the Serpens South sources.
$^{(b)}$ Distance of the cloud using \citet{Herczeg2019,Zucker2019,Zucker2020}. For the Orion Sources, we used the distance estimates from \citet{Tobin2020}.
$^{(c)}$ HOPS 87 and Per emb 21 did not have continuum detection.
$^{(d)}$ In spite of its high bolometric temperature value (320 K in \citealt{Dunham2015}, and 188 K in \citealt{Maury2011}), Aqu MM5 has been classified as a Class 0 by \citet{Maury2011} due to its position in the $M_{\textrm{env}}\,-\,L_{\textrm{bol}}$ diagram and the $L^{\lambda>350 \mu \textrm{m}}_{\textrm{sub-mm}}/L_{\textrm{bol}} \,>\,1\%$ criteria.
$^{(e)}$ These two Perseus sources have $T_{\textrm{bol}}\,>\,70$ K in \citet{Dunham2015}. However, \citet{Enoch2009} calculated 67 K for Per emb 24 and 68 K for Per emb 25 (using 2MASS, BOLOCAM, and \textit{Spitzer}), and \citet{Mottram2017,Karska2018} calculated 60 K for Per emb 25 (using 2MASS, SCUBA on the JCMT, \textit{Spitzer}, and \textit{Herschel}).
$^*$ From the existing sample of Class 0 previously presented in \citet{Greene2018,Laos2021}.}
\end{table*}

\subsection{Data Collection and Reduction}
\label{sec:data_reduc}

The 18 new Class 0 objects were observed with the MOSFIRE instrument \citet{McLean2010,McLean2012} in its long-slit mode, on the Keck I telescope on Maunakea, Hawaii. A spectroscopic resolving power of $R\,=\,\lambda/ \delta \lambda$ of $\sim\,3000-3400$ was measured in our spectra, depending on the seeing conditions.
The plate scale was 0.1798 pixel$^{-1}$ along the 46$^{\prime\prime}$ slit length, and a 1$^{\prime\prime}$ slit width was used. The order-sorting MOSFIRE K filter was used to record the $\lambda\,=\,1.95 - 2.4\,\mu$m wavelength range. 
We used an initial long offset along the slit in order to maximize the wavelength coverage at the red end.
Data were acquired in ABBA pairs to enable efficient background subtraction, with the telescope nodding $\sim\,20^{\prime\prime}$ along the slit between each integration that were 150 seconds. 

The data reduction was performed using the PypeIt package \citep{Prochaska2020}. 
PypeIt first uses a dome flat to identify and trace the slit edges. Then, we used the OH atmospheric lines present in the 2D science frames for the wavelength calibration instead of the arc lamps exposures. The 2D spectrum images are processed with a cosmic-ray flagging routine, flat-fielded, and sky-subtracted. We extract the 1d spectra from the 2D images using a constant extraction width as a function of wavelength with an object finding algorithm. The extraction width is calculated based on a PypeIt-based optimal aperture mask maximizing the signal-to-noise ratio (S/N). The 1d spectra of each exposure are then co-added. The A1 dwarf GD 71 and the G2V P330E star were observed to build a MOSFIRE $K$-band sensitivity function to perform the flux calibration of the protostar spectra. Finally, we perform the telluric correction with the telluric model fitting routine of PypeIt.

In a few sources, we notice that the  large signals of the brightest H$_2$ lines causes the PypeIt extraction algorithm to flag out the corresponding brightest pixels in the detector image, likely because it erroneously identifies these as invalid pixels. We deactivate this option for these sources, to recover the total flux of these H$_2$ lines. However, sometimes this flagging occurs when the H$_2$ emission lines are very extended, causing a poor definition of the local sky around the impacted spectra. Also, for very extended H$_2$ emission, the B nod position would sometimes move a counterpart onto the A position in the 2D spectra. In these two cases, using the 2D model mask was more efficient to recover the most H$_2$ flux.

Two Class 0 sources, Per emb 21 (from the \citealt{Laos2021} sample) and HOPS 87, did not have any detected continuum in the detector image but only H$_2$ lines, precluding us from using the PypeIt source finding algorithm. We performed a PypeIt-based manual extraction of these two objects, based on the location of the H$_2$ lines. For HOPS 87, we then performed the same steps of flux calibration, co-adding and telluric correction. For Per emb 21, we repeated the reduction procedure performed by \citet{Laos2021}.

\begin{figure*}[tbh!]
  \centering
  \savebox{\myimage}{\includegraphics[scale=0.55,clip,trim= 0cm 0cm 0cm 0cm]{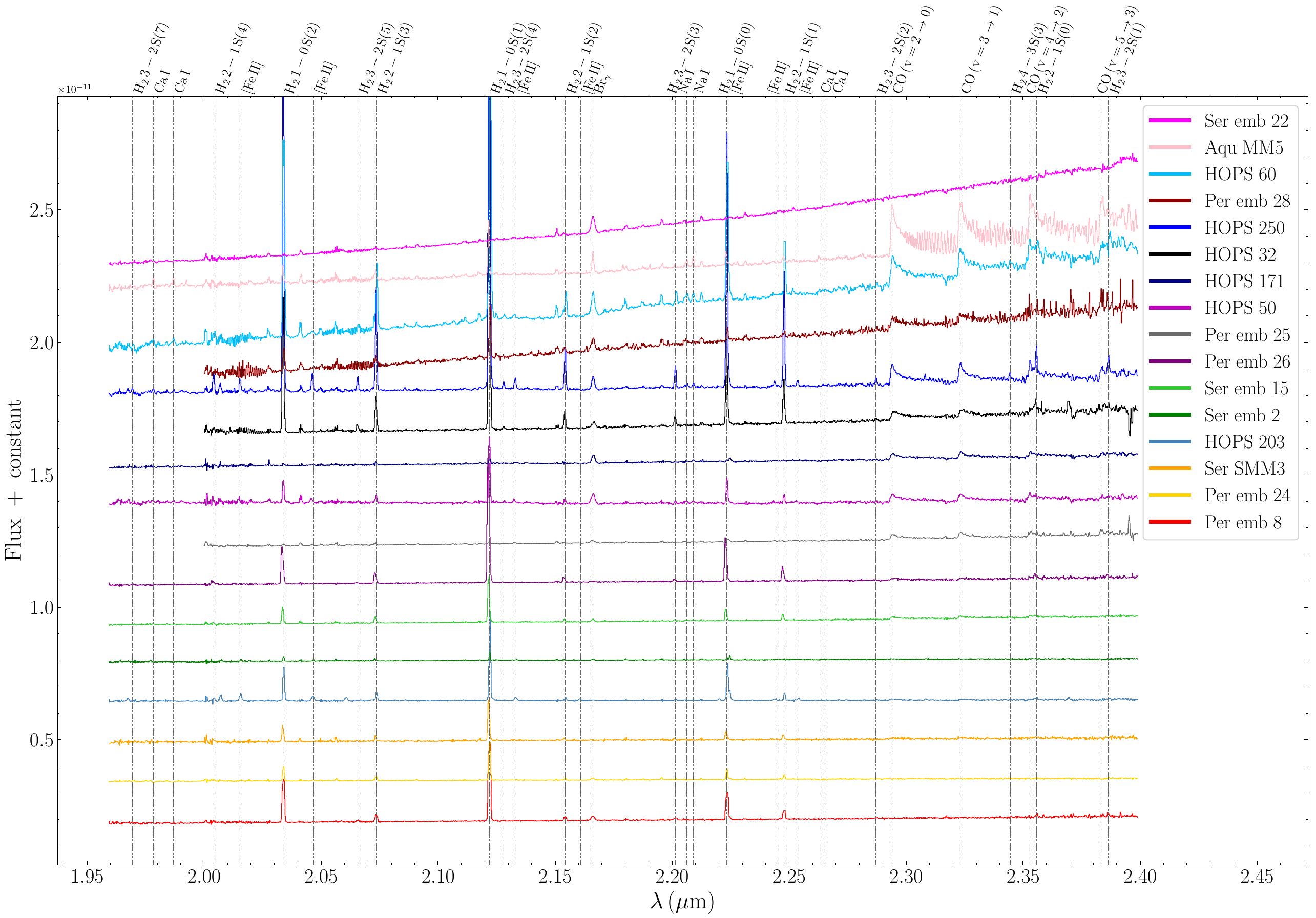}}
  \rotatebox{90}{
    \begin{minipage}{\wd\myimage}
      \usebox{\myimage}
      \caption{Extracted 1D $K$ band spectra of our sample of Class 0 protostars showing clear \Brgamma or CO overtone emission. We sorted the spectra such as the reddest spectra and the CO overtone emitting sources are at the top, while the sources exhibiting photospheric features are at the bottom. These spectra are not corrected for extinction. The grey vertical dashed line report all the detected lines in the sample. Per emb 21 and HOPS 87 are not shown here because no continuum has been detected in these sources. However, their spectra show bright emission lines for nearly all of the H$_2$ lines outlined in this Figure.}
      \label{fig:all_specs_1}
    \end{minipage}}
\end{figure*}

\begin{figure*}[tbh!]
  \centering
  \savebox{\myimage}{\includegraphics[scale=0.55,clip,trim= 0cm 0cm 0cm 0cm]{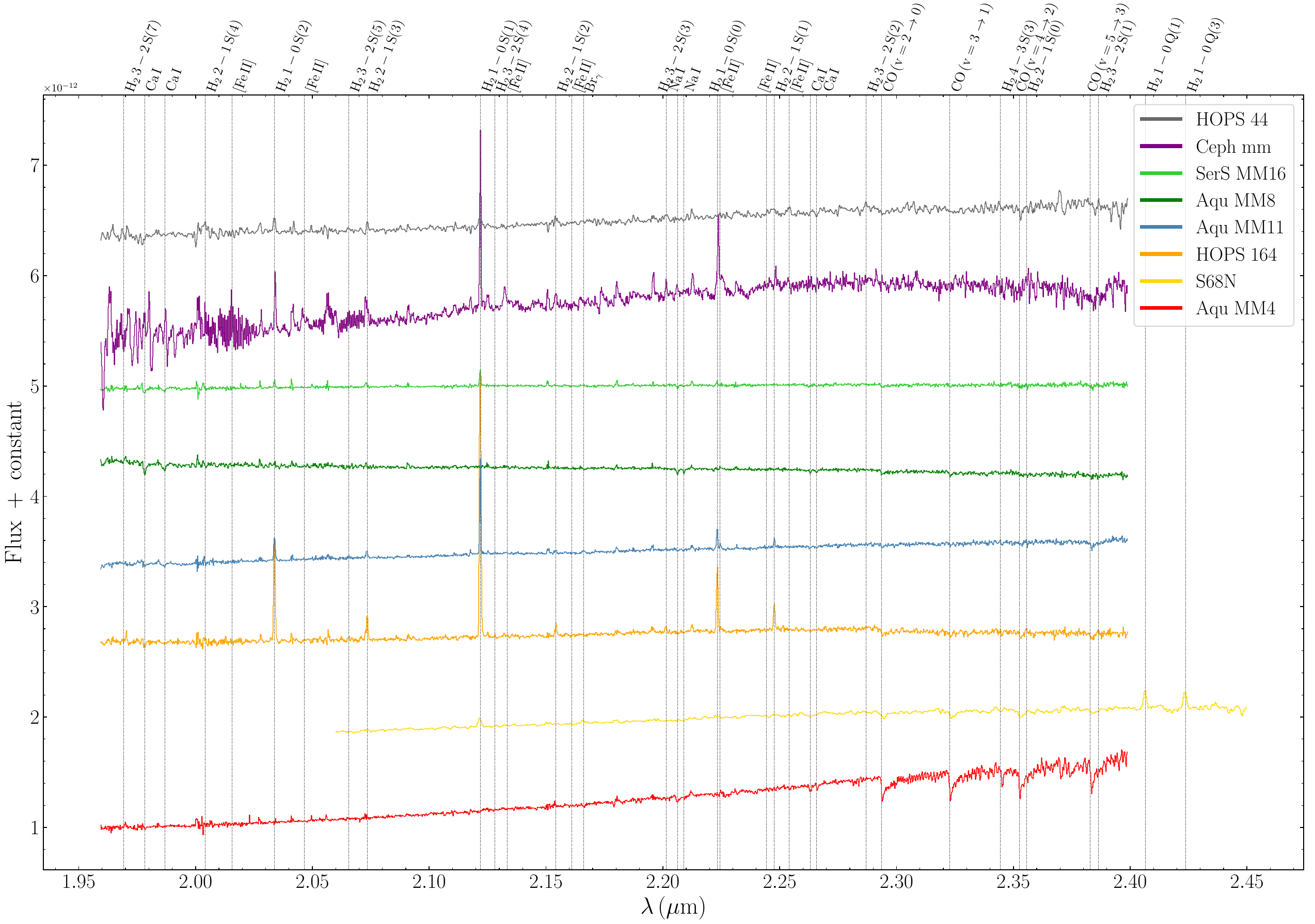}}
  \rotatebox{90}{
    \begin{minipage}{\wd\myimage}
      \usebox{\myimage}
      \caption{Same as Figure \ref{fig:all_specs_1} for the Class 0 sources showing photospheric features or absence of strong \Brgamma or CO overtone emission features.}
      \label{fig:all_specs_2}
    \end{minipage}}
\end{figure*}

\section{Spectral analysis and results}
\label{sec:results}

\subsection{Analysis techniques}
\label{sec:analysis_tech}

For ultimate wavelength calibration of the spectral analysis, we correct for the velocity shift induced by Earth's orbit around the Sun for each observation. In addition, we correct for the velocity of the local standard of rest (LSR) for each protostellar object. These values are generally known thanks to high spectral resolution radio observations of molecular gas emission lines known to be good tracers of protostellar dense gaseous envelopes, or of the cold gas clumps surrounding protostars (\eg C$^{18}$O, NH$_3$, N$_2$H$^+$). We retrieve the $v_{\textrm{lsr}}$ values of our Class 0 sources in \citealt{Grobschedl2021} for the Orion sources, in \citet{Kun2008} for the Cepheus source, in \citet{Levshakov2013,Lee2014,FernandezLopez2014} for the Serpens/Aquila sources, and in \citet{Carney2016} for the Perseus sources. Typical uncertainties for these $v_{\textrm{lsr}}$ values are $\pm$ 1 km s$^{-1}$.

For each identified spectral line, the continuum level was determined via a multi-order polynomial fitting on a spectral region centered on the expected line's wavelength, that excludes any identified lines. 
Within this line-free spectral sub-region, the noise level is set by the root mean square (\textit{rms}) calculated on the continuum-subtracted spectra. We attempted to perform a 1D Gaussian fitting and use two parameters to assess whether the line is detected or not. The first one is the ratio of the peak over the \textit{rms}. The second one is the ratio between the peak of the fitted Gaussian profile and its fitting uncertainty. Both signal to noise ratios (S/N) are taken into account throughout this work. We typically require the two S/N values to be $\geq\,3$ to declare a line detected. If a line is detected, we calculate the flux and equivalent width by integrating the observed spectra over the Gaussian-fitted profile (\ie we integrate over $[-3/2\times\textrm{FWHM};3/2\times\textrm{FWHM}]$). If a line is not detected, we calculate a line flux upper limit with $3\times rms\times \lambda_{\textrm{line}}/R$, where R is the resolution of the spectra (and accordingly for the equivalent width). To calculate the flux and equivalent width uncertainties, we propagate the noise (\ie the \textit{rms} mentioned above) in the line integration. The uncertainty of the gaussian fitting provides the error on the line centroid, and full width half max. We note that in this work we do not subtract the photospheric template to compute the flux and EW of emission lines (\eg see the impact of such subtraction in \citealt{Tofflemire2019}).

Several telluric lines are not perfectly subtracted from the spectra by the PypeIt pipeline, which led us to carefully check the 2D spectra to identify those and enable an accurate identification of the protostellar spectral lines. These imperfectly subtracted lines are illustrated relative to an extracted object spectrum in Figure \ref{fig:tell_lines_hops250} in Appendix \ref{app:tell_line}.

\subsection{Emission lines}
\label{sec:results_emission}

All the extracted 1D $K$-band spectra of our Class 0 sources (18 new and 6 previously published) are shown in Figures \ref{fig:all_specs_1} and \ref{fig:all_specs_2}. A large variety of spectral features are detected. While a sub-sample of 6 sources show several absorption features (Ca I, Na I, and CO overtone), 17 sources sources exhibit an emission-line-rich spectrum with notably \Brgamma, CO overtone emission, and sometimes atomic line emission (Ca I, Na I, and [Fe II]). H$_2$ lines are seen in emission in nearly all of the sources. 

Table \ref{t.line_detection} presents whether emission lines are detected or not for each source,  based on the S/N criteria given in Section \ref{sec:analysis_tech}. These include the strongest H$_2$ lines, \Brgamma, and the first two CO overtone emission bands. We detect the brightest H$_2$ lines in $\sim$ $80-90$,\%, and \Brgamma in 65\,\% of the sources.
The first and second CO overtone bands are detected in emission in 50\,\%, and 54\,\% of the sources, respectively. The detection rates of \Brgamma and CO overtone bands may be underestimations of the true occurrence of these emission lines because of the high extinction inherent to Class 0 sources. Indeed, while the H$_2$ lines can emanate from the shocked outflow cavities, \Brgamma, CO overtone likely originate from regions deeper within the protostar, where the extinction is much higher compared to cavities. For example, the strong H$_2$ lines in Per emb 21 and HOPS 87, indicative of strong ejection, do not come along with strong accretion spectral features because of the strong extinction that these two sources' inner region likely suffer from, explaining the non-detection of their NIR continuum. We are thus only sensitive to the shocked cavities in these two sources. However, we note that the H$_2$ lines have also much higher peak flux than the CO bands, making them easier to detect. That is partially due to extinction, but also due to larger intrinsic brightness for some extended shocked regions, exciting H$_2$ over a larger area.

\begin{figure*}[!tbh]
\centering
\subfigure{\includegraphics[scale=0.4,clip,trim= 4cm 0.9cm 4cm 0cm]{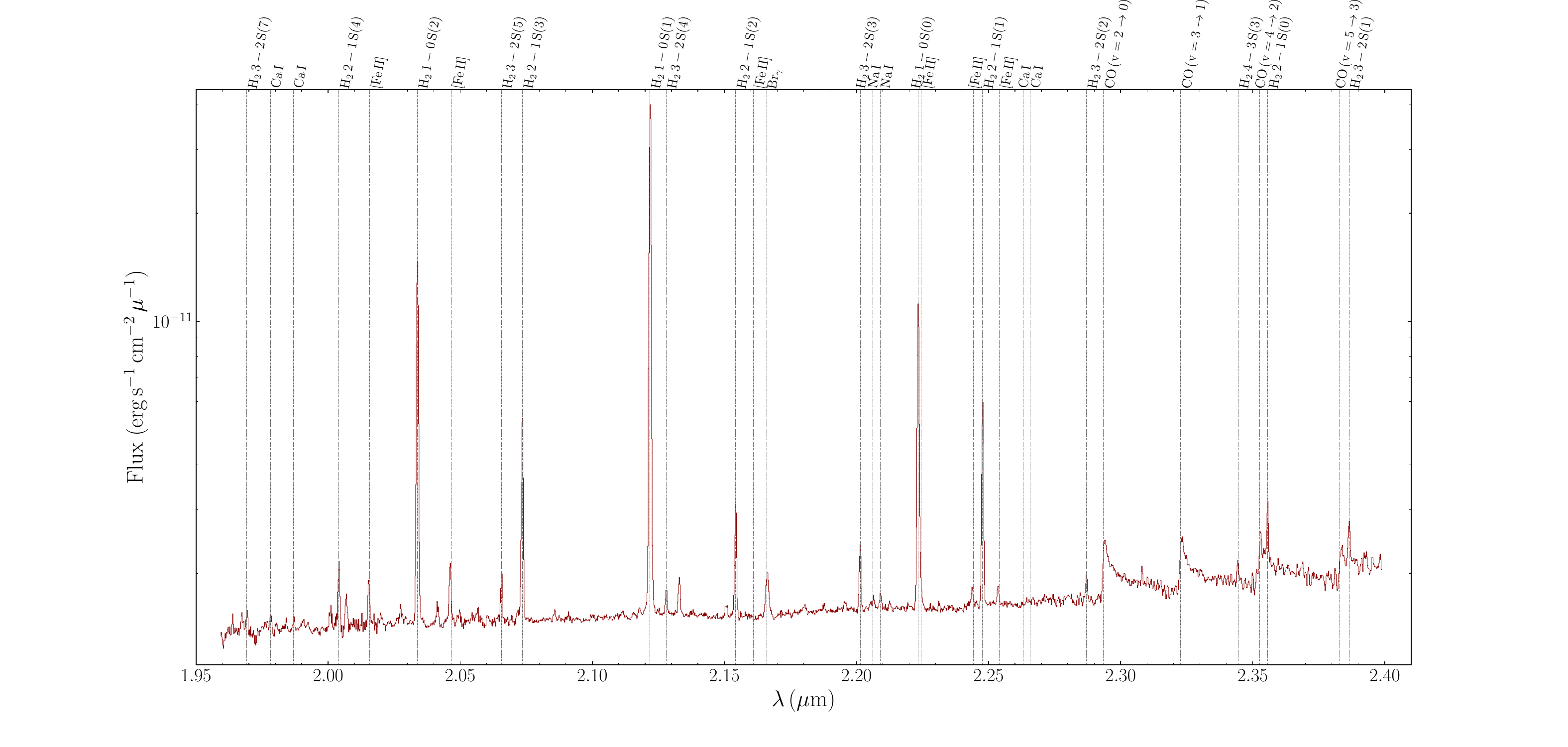}}
\subfigure{\includegraphics[scale=0.40,clip,trim= 5.6cm 0cm 4.5cm 0cm]{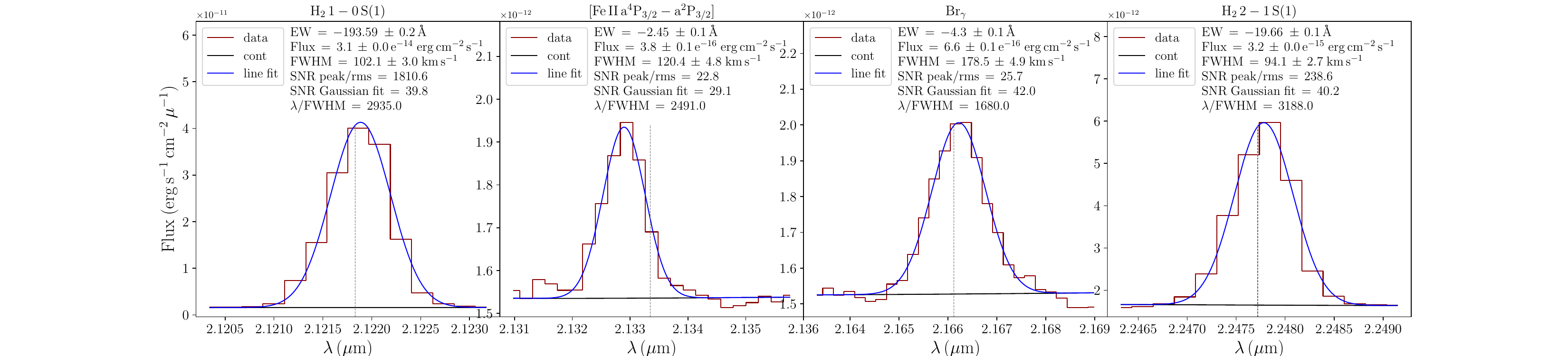}}
\vspace{-0.4cm}
\caption{\small $K$-band spectra example of the HOPS 250 Class 0 source. \textit{Top:} Same as Figure \ref{fig:all_specs_1} for HOPS 250. \textit{Bottom:} Line fitting results for four spectral lines. In each panel, the solid red line is the observed spectra, the solid black line is the fitted continuum, and the solid blue line is the resulting Gaussian best fit. The grey vertical dashed line indicates the line reference wavelength.}
\label{fig:ex_spec_hops250}
\end{figure*}

Figure \ref{fig:ex_spec_hops250} presents an example of our spectral line analysis as performed on the Class 0 source HOPS 250. The fit of four lines are shown, alongside the derived values of equivalent width (EW), line flux, line full-width-half-max (FWHM) velocity, velocity shift with the respect to the line reference wavelength, and the two S/N criteria introduced above. The values of $\lambda_{\textrm{line}}/\textrm{FWHM}$ are also shown, to outline how resolved are the lines, compared to the spectral resolution of the observations (\ie $R$ = 3300 in the case of HOPS 250). Throughout this paper, we report the measured line fluxes uncorrected for extinction. However, the line luminosities are corrected for extinction. We mainly discuss the implication on extinction of the Class 0s' emission lines in Section \ref{sec:comp_ClassI}.

\begin{table*}
\small
\centering
\caption[]{Emission line detections}
\label{t.line_detection}
\setlength{\tabcolsep}{0.35em} 
\begin{tabular}{p{0.22\linewidth}cccccccc}
\hline \hline \noalign{\smallskip}
 Source &$\rm H_{2}\,1-0\,S(2) $&$\rm H_{2}\,1-0\,S(1) $&$\rm H_{2}\,1-0\,S(0) $&$\rm H_{2}\,2-1\,S(1) $&$\rm Br_{\gamma}$&$\rm CO\,(v = 2\rightarrow 0)$&$\rm CO\,(v = 3\rightarrow 1)$\\ 
\noalign{\smallskip}  \hline 
\noalign{\smallskip} 
HOPS 50 & $\checkmark$ & $\checkmark$ & $\checkmark$ & $\checkmark$ & $\checkmark$ & $\checkmark$ & $\checkmark$\\ 
\noalign{\smallskip}  \hline \noalign{\smallskip}  
HOPS 60 & $\checkmark$ & $\checkmark$ & $\checkmark$ & $\checkmark$ & $\checkmark$ & $\checkmark$ & $\checkmark$\\ 
\noalign{\smallskip}  \hline \noalign{\smallskip}  
HOPS 87 & $\checkmark$ & $\checkmark$ & $\checkmark$ & $\checkmark$ & X & X & X\\ 
\noalign{\smallskip}  \hline \noalign{\smallskip}  
HOPS 164 & $\checkmark$ & $\checkmark$ & $\checkmark$ & $\checkmark$ & X & X & X\\ 
\noalign{\smallskip}  \hline \noalign{\smallskip}  
HOPS 171 & $\checkmark$ & $\checkmark$ & X & X & $\checkmark$ & $\checkmark$ & $\checkmark$\\ 
\noalign{\smallskip}  \hline \noalign{\smallskip}  
HOPS 203 & $\checkmark$ & $\checkmark$ & $\checkmark$ & $\checkmark$ & $\checkmark$ & $\checkmark$ & $\checkmark$\\ 
\noalign{\smallskip}  \hline \noalign{\smallskip}  
HOPS 250 & $\checkmark$ & $\checkmark$ & $\checkmark$ & $\checkmark$ & $\checkmark$ & $\checkmark$ & $\checkmark$\\ 
\noalign{\smallskip}  \hline \noalign{\smallskip}  
Per emb 24 & $\checkmark$ & $\checkmark$ & $\checkmark$ & $\checkmark$ & $\checkmark$ & X & $\checkmark$\\ 
\noalign{\smallskip}  \hline \noalign{\smallskip}  
Ser SMM3 & $\checkmark$ & $\checkmark$ & $\checkmark$ & $\checkmark$ & X & $\checkmark$ & $\checkmark$\\ 
\noalign{\smallskip}  \hline \noalign{\smallskip}  
Aqu MM4 & X & X & X & X & X & X & X\\ 
\noalign{\smallskip}  \hline \noalign{\smallskip}  
S68N & --- & $\checkmark$ & X & X & X & X & X\\ 
\noalign{\smallskip}  \hline \noalign{\smallskip}  
HOPS 32 & $\checkmark$ & $\checkmark$ & $\checkmark$ & $\checkmark$ & $\checkmark$ & $\checkmark$ & $\checkmark$\\ 
\noalign{\smallskip}  \hline \noalign{\smallskip}  
HOPS 44 & $\checkmark$ & $\checkmark$ & X & X & X & X & X\\ 
\noalign{\smallskip}  \hline \noalign{\smallskip}  
Per emb 8 & $\checkmark$ & $\checkmark$ & $\checkmark$ & $\checkmark$ & $\checkmark$ & X & X\\ 
\noalign{\smallskip}  \hline \noalign{\smallskip}  
Per emb 25 & $\checkmark$ & $\checkmark$ & $\checkmark$ & X & $\checkmark$ & $\checkmark$ & $\checkmark$\\ 
\noalign{\smallskip}  \hline \noalign{\smallskip}  
Per emb 26 & $\checkmark$ & $\checkmark$ & $\checkmark$ & $\checkmark$ & $\checkmark$ & $\checkmark$ & $\checkmark$\\ 
\noalign{\smallskip}  \hline \noalign{\smallskip}  
Per emb 28 & $\checkmark$ & $\checkmark$ & $\checkmark$ & $\checkmark$ & $\checkmark$ & $\checkmark$ & $\checkmark$\\ 
\noalign{\smallskip}  \hline \noalign{\smallskip}  
Per emb 21 & $\checkmark$ & $\checkmark$ & $\checkmark$ & $\checkmark$ & X & X & X\\ 
\noalign{\smallskip}  \hline \noalign{\smallskip}  
Aqu MM11 & $\checkmark$ & $\checkmark$ & $\checkmark$ & $\checkmark$ & $\checkmark$ & X & X\\ 
\noalign{\smallskip}  \hline \noalign{\smallskip}  
Ceph mm & $\checkmark$ & $\checkmark$ & $\checkmark$ & $\checkmark$ & X & X & X\\ 
\noalign{\smallskip}  \hline \noalign{\smallskip}  
Ser emb 2 & $\checkmark$ & $\checkmark$ & X & $\checkmark$ & $\checkmark$ & $\checkmark$ & $\checkmark$\\ 
\noalign{\smallskip}  \hline \noalign{\smallskip}  
Ser emb 15 & $\checkmark$ & $\checkmark$ & $\checkmark$ & $\checkmark$ & $\checkmark$ & $\checkmark$ & $\checkmark$\\ 
\noalign{\smallskip}  \hline \noalign{\smallskip}  
Aqu MM5 & $\checkmark$ & $\checkmark$ & $\checkmark$ & $\checkmark$ & $\checkmark$ & $\checkmark$ & $\checkmark$\\ 
\noalign{\smallskip}  \hline \noalign{\smallskip}  
Aqu MM8 & X & X & $\checkmark$ & X & X & X & X\\ 
\noalign{\smallskip}  \hline \noalign{\smallskip}  
Ser emb 22 & X & X & X & $\checkmark$ & $\checkmark$ & X & X\\ 
\noalign{\smallskip}  \hline \noalign{\smallskip}  
SerS MM16 & $\checkmark$ & $\checkmark$ & $\checkmark$ & X & X & X & X\\ 
\noalign{\smallskip}  \hline \noalign{\smallskip}  
Detection Rate Class 0s & 88.0 & 88.5 & 76.9 & 73.1 & 61.5 & 50.0 & 53.8\\ 
Detection Rate Class 1s & 29.4 & 77.1 & 64.8 & 45.7 & 77.9 & 14.1 & 5.9\\ 
\noalign{\smallskip} 
\hline
\end{tabular}
\vspace{0.03cm}
\tablecomments{\small Emission line detections for \HtonetoOStwo, \HtonetoOSone, \HtonetoOSO, \HttwotooneSone, \Brgamma, and the two CO first  overtone bands. Check-marks indicate a detection, crosses indicate a non-detection, and dashes indicate that the spectrum does not cover the line. The last two lines show the total detection rates ( \#detections/total \# in percent) for our Class 0s, and the Class Is sample presented in Section \ref{sec:comp_ClassI}.}
\vspace{-0.05cm}
\end{table*}

The Class 0 line parameters (flux, equivalent width, and FWHM) of all the H$_2$ and \Brgamma emission lines are shown in Tables \ref{t.C0_line_obs1}, \ref{t.C0_line_obs2}, and \ref{t.C0_line_obs3} of Appendix \ref{app:line_param}. The line parameters of CO are presented in Section \ref{sec:results_CO_overtone} and \ref{sec:results_absorption}.
We present in the following subsections the main results considering each group of spectral features, sorted by emitting species, and if they are detected in emission or absorption.
The results of this Class 0 spectral line parameter analysis are quantitatively presented in Section \ref{sec:comp_ClassI}, where we compare the Class 0s with a sample Class I archival observations.

\subsubsection{Molecular Hydrogen Emission}
\label{sec:results_H2}

Among our sample of $K$-band spectra of 26 Class 0 objects, we detect H$_2$ emission in 23 sources, and we detect in total up to 14 different H$_2$ emission lines. In spite of its widely spaced rotational levels,  H$_2$ is especially bright in excited media, and visible thanks to its high abundance. It is commonly seen toward protostellar outflow cavity shocks, revealing the molecular content and momentum of outflowing gas \citep{CarattiOGaratti2006}. At NIR wavelengths, several ro-vibrational transitions can be observed, whose $v$ = 1--0 S1 line at 2.12 $\mu$m is the brightest. The highest upper-energy level transition we detect is $v$ = 4--3 S(3).

The working surface of outflow shocks is large toward the base of protostellar outflows compared to the $1-1.5^{\prime\prime}$ slit width we used. This explains the general 2D extent we see in the raw detector images for the brightest four H$_2$ lines, \ie 1--0 S2, 1--0 S1, 1--0 S0, and 2--1 S1. Also, scattering of outflow cavity walls is expected to be a strong contributor of NIR light in outflow cavities, which challenges the determination of the H$_2$ emission spatial origin. The H$_2$ lines are on average only marginally resolved, with a mean and standard deviation of 69 $\pm$ 49 km s$^{-1}$, for the $v$ = 1--0 S1 line FWHM (where the instrument resolution of $\sim$100  km s$^{-1}$ has been subtracted in quadrature). 
The velocity shift of the H$_2$ lines is not predominantly red- or blueshifted, with on average a shift of -2 $\pm$ 28 km s$^{-1}$ for 1--0 S1. This indicates that we are not necessarily detecting only the blueshifted outflow cavity shocks, but rather a variety of shock working surfaces in both bipolar cavities. 

Unlike the CO overtone or \Brgamma emission lines, which are thought to originate from deep within the protostellar core, the H$_2$ emission lines most likely originate from the outflow cavity shocks. 
These regions should suffer from a variable amount of foreground extinction in the different star forming clouds hosting our sources.
Evaluating all detected H$_2$ lines together for each source, \ie building a rotational diagram and computing the ortho-to-para ($o/p$) ratio, can provide simple tools to broadly estimate the amount of extinction affecting them. For sources with enough H$_2$ detections, we compute the extinction (using the \citealt{Cardelli1989} extinction law) that minimizes the standard deviation of the set of $o/p$ ratios computed within each rotational ladder (\ie $v = 1-0$, $v = 2-1$, and $v = 3-2$). We use the standard definition of $o/p$ ratio provided in \citet{Wilgenbus2000}, where a rotational temperature and a LTE equilibrium value of $o/p$ ratio need to be assumed. We use $o/p_{\textrm{LTE}}\,=\,3$, which is a reasonable assumptions given the typical $T_{\textrm{rot}, v\geq1}\,\geq \,2000$ K we derive in the observations.
We use $T_{\textrm{rot}} = 2500$ for $v = 1-0$ lines, $T_{\textrm{rot}} = 3000$ for $v = 2-1$ for , and $T_{\textrm{rot}} = 4000$ for $v = 3-2$ to then compute the $o/p$ ratio in each given ladder. Indeed, the $o/p$ ratio is expected to increase with rotational temperature, but to exhibit some homogeneity with a given rotational ladder \citep{Neufeld2006}. This technique does not provide extinction values for 5 sources, \ie Aqu MM4, Aqu MM8, Ser emb 2, HOPS 44, and HOPS 171. Aqu MM4, Aqu MM8 does not have H$_2$ detection at all, so these two sources does not take any part in the analysis of the H$_2$ lines. 
The number of H$_2$ detections and/or S/N of the spectra are too low in Ser emb 2, HOPS 44, and HOPS 171. For these three sources, we assume a value of $A_v\,=\,30$ mag because it corresponds to the mean of the extinction values found toward the others sources in our sample.

Adopting, these extinction values, the resulting mean H$_2$ $o/p$ ratio using \HtonetoOSone and \HtonetoOSO is 2.36 $\pm$ 0.70 in our sample of Class 0 spectra, indicative of somewhat hot gas temperature, \ie $\gtrsim\,1000$ K \citep{Neufeld1998,Wilgenbus2000,Neufeld2006}.  Per emb 25 has a low $o/p$ ratio (0.33) compared to the rest of the sample, suggesting a cooler shock and a lower pre-shock $o/p$ ratio. However, the uncertainty on the $A_v$ determination is relatively high for this source. 
The bulk of the $o/p$ ratio population we derive is within 2 and 3, for which the $A_v$ determination is more reliable. Comparing with the shock models of \citet{Wilgenbus2000}, this indicates a pre-shock $o/p$ ratio $\geq 2$ in the case of $J$ type shocks, and $\geq 1$ in the case of $C$ type shocks, assuming a pre-shock density of $n_{\textrm{H}}\,\sim\,10^{5}-10^{6}$ cm$^{-3}$. This suggest the pre-shock medium is cold, \ie $\leq$ 300 K \citep{Neufeld2006}. 
Our method to determine the extinction of the H$_2$ emitting gas is quite uncertain, which affects the accuracy of our line ratio measurements. Among the 19 values of $o/p$ ratio we are able to derive in our sample of Class 0 spectra, two are $>3$ (HOPS 32 and HOPS 60, with $o/p\,=\,3.3$ and 3.1, respectively), which is not physical. 
The H$_2$ lines of these two sources are well detected, so overestimated extinction could explain their $o/p$ ratio being > 3.
Indeed, the \HtonetoOSone and \HtonetoOSO are distant enough in wavelength for their ratio to be sensitive to the extinction. 
Our approach only provides a first approximation of the extinction of the H$_2$ emitting gas. More extended wavelength coverage would be necessary to derive the extinction with higher accuracy, using pairs of H$_2$ lines sharing the same upper energy level.
We also don't obtain any correlation between our values of $o/p$ ratio taken against the \HtonetoOSone or \HtonetoOSO line fluxes which would have been expected from shocked gas. Again, this suggests that we are limited by our extinction determination, or that we lack in dynamic range of shock conditions.

The population of the H$_2$ vibrational levels would also vary for different excitation mechanisms, \ie between a dominant contribution of UV, against gas collisions, or X-ray. The question of the dominant H$_2$ excitation mechanism is relevant in Class 0s, as the expected vigorous accretion/ejection activity during this protostellar evolutionary phase would also produce specific ionization/irradiation conditions in the inner envelope \citep{Cabedo2022,LeGouellec2023a}. \citet{Black1987,Gredel1995} argued that the $v$ = 2--1 S1/$v$ = 1--0 S1 H$_2$ line ratio can be used to identify the major H$_2$ excitation mechanism. Fluorescent UV excitation of H$_2$ significantly populates the vibrational levels $v_{\textrm{up}}\,\geq\,2$ and thus enhance the emission of the $v$ = 2--1 S1 line, producing 2--1 S1/1--0 S1 $\lesssim$ 4. However, the high-density environment encountered in shocks would thermalize the H$_2$ emitting gas to vibrational temperature $\sim$ 2000 -- 4000 K, producing 2--1 S1/1--0 S1 $\gtrsim$ 5. X-rays would produce a wide range of 2--1 S1/1--0 S1 ratio of $\sim\,1.9-16.7$, depending of the fractional ionization of the emitting gas \citet{Gredel1995}. Using the $A_v$ values for the visual extinction discussed above and the extinction law of \citealt{Cardelli1989} with a visual extinction to reddening ratio of 5 (in order to into account the significant level of grain growth occurring in protostellar envelopes and disks; \citealt{Weingartner2001}), we find that the mean and standard deviation of Class 0s' 2--1 S1/1--0 S1 ratio is 18.3 $\pm$ 4.1, thus consistent with collisional excitation via most likely a shocked gas in a wind, or X-ray excitation, but pure UV excitation is discarded.

\subsubsection{CO Overtone Emission}
\label{sec:results_CO_overtone}

The ro-vibrational transitions of the CO molecule produce multiple bands in the $K$-band that are seen in emission in 13 sources of our Class 0 sample. Four of these overtone (\ie $\Delta v = 2$) bands are within our spectral coverage. This emission has previously been observed in several YSOs, from low- to high-mass objects \citep{Connelley2010,Martins2010,RamirezTannus2017,Laos2021}, and originates from hot (2000-5000 K) and dense gas ($N_{\textrm{CO}}\,\sim\,10^{20}-10^{22}$ cm$^{-2}$) of the disk inner regions \citep{Carr1993,Najita1996b,Bik2004,Aspin2010}, or the base of a stellar wind \citep{Carr1989}. The analysis of these bands can thus reveal important information about the gas distribution in the inner regions of Class 0s.

We also employ a Gaussian fitting as a detection criterion (results are in Table \ref{t.line_detection}), and we characterize the bands with the following method. In each spectrum, because no continuum can be evaluated longward to the CO overtone bandheads in our $K$-band spectra, we determine the continuum level by computing a moving local minimum whose window size is $\sim$ 0.01 micron (on the order of the size of a CO band head) that is then fit by a low-order polynomial, to which we add the standard deviation of the continuum shortward to the CO bandhead. Line flux and equivalent width are computed accordingly for each CO band, and their uncertainties are calculated by propagating the noise level (the \textit{rms} shortward of the first CO band) in the integration of the CO bands.

We now aim to determine the kinematics information of the CO emitting regions by constraining the velocity shift and the non-thermal broadening of the bandheads. To do so, we use the PHOENIX high resolution synthetic spectra library \citep{Husser2013}. A set of high resolution ($R$ = 10,000) spectra is built for $T_{\textrm{eff}} = 2300-5500$ K, log $g$ = 1.5, and [Fe/H] = 0. We chose to use this relatively low value of disk-like surface gravity, and to explore the full relevant range of temperature (2300 K is the minimum available temperature of the PHOENIX library and CO dissociates above 5000 K). These spectra show clear CO absorption features, that we turn into emission for our comparisons with the Class 0s CO emission spectra. For each comparison with an observed Class 0 spectrum, we perform a $\chi^2$ test with the synthetic spectra that have been spectrally shifted, binned and convolved to a given velocity shift and Gaussian broadening kernel. \citet{Laos2021} applied this technique iterating manually on the parameters to find the model that best match the data. In our work on this extended sample, we chose to perform three parameter (\ie $T_{\textrm{eff}}$, velocity shift, and FWHM of the broadening) Markov Chain Monte Carlo (MCMC) exploration with the \textit{emcee} package \citep{ForemanMackey2013} in order to assess the accuracy and degeneracy of the parameter estimation. The assigned priors to the model parameter values before performing the MCMC analysis are the following: 2500 -- 4500 K, 40 -- 200  km s$^{-1}$, and -100 -- 100 km s$^{-1}$, for the temperature, FWHM of the broadening kernel, and velocity shift, respectively. These ranges are sufficiently large compared to the full grid parameter exploration, such that they don't affect the resulting posterior distribution. As mentioned above, the PHOENIX grid starts at 2300 K and CO dissociates above 5000 K. The lower value of the FWHM prior range of 40 km $s^{-1}$ is reasonable given that our spectra have a spectral resolution of $\sim$ 80-100 km $s^{-1}$. The upper value is also reasonable, given that CO would not be able to emanate from region close enough to the protostar to generate a broadening of 200 km $s^{-1}$, as it would be dissociated before. This ensures that the FWHM prior range encompasses all reasonable values and does not limit the posterior distribution. Finally, the velocity shift prior range is large enough compared to the expected posteriors.

While, the broadening and velocity shift are efficiently constrained with this approach, we highlight that the determination of the effective temperature of the CO emitting area appears to not be reliable given the resulting ranges of posterior temperature values. This was expected given our simplistic approach. Indeed, determining the impact of the $K$-band continuum on the normalized CO bands, the emitting area, column density and thus optical thickness would be necessary to accurately model the CO overtone emission. 
In addition, we compare and fit the normalized CO spectra in this approach, whereas the continuum level of the Class 0 observations is not only composed of the central proto-stellar embryo, but also inner disk and scattered light (so-called veiling excess).
We thus chose to focus on the broadening and velocity shift results obtained with this simple approach, and not to consider the temperature that is only used for fitting purposes (as the temperature is the parameter most responsible for the relative strength of the normalized CO bandhead in the PHOENIX spectra). The detailed modeling of the Class 0 CO overtone emission will be the focus on a forthcoming paper, and is beyond the scope of this work.

\begin{figure*}[!tbh]
\centering
\includegraphics[scale=0.45,clip,trim= 3.5cm 0.8cm 4cm 2cm]{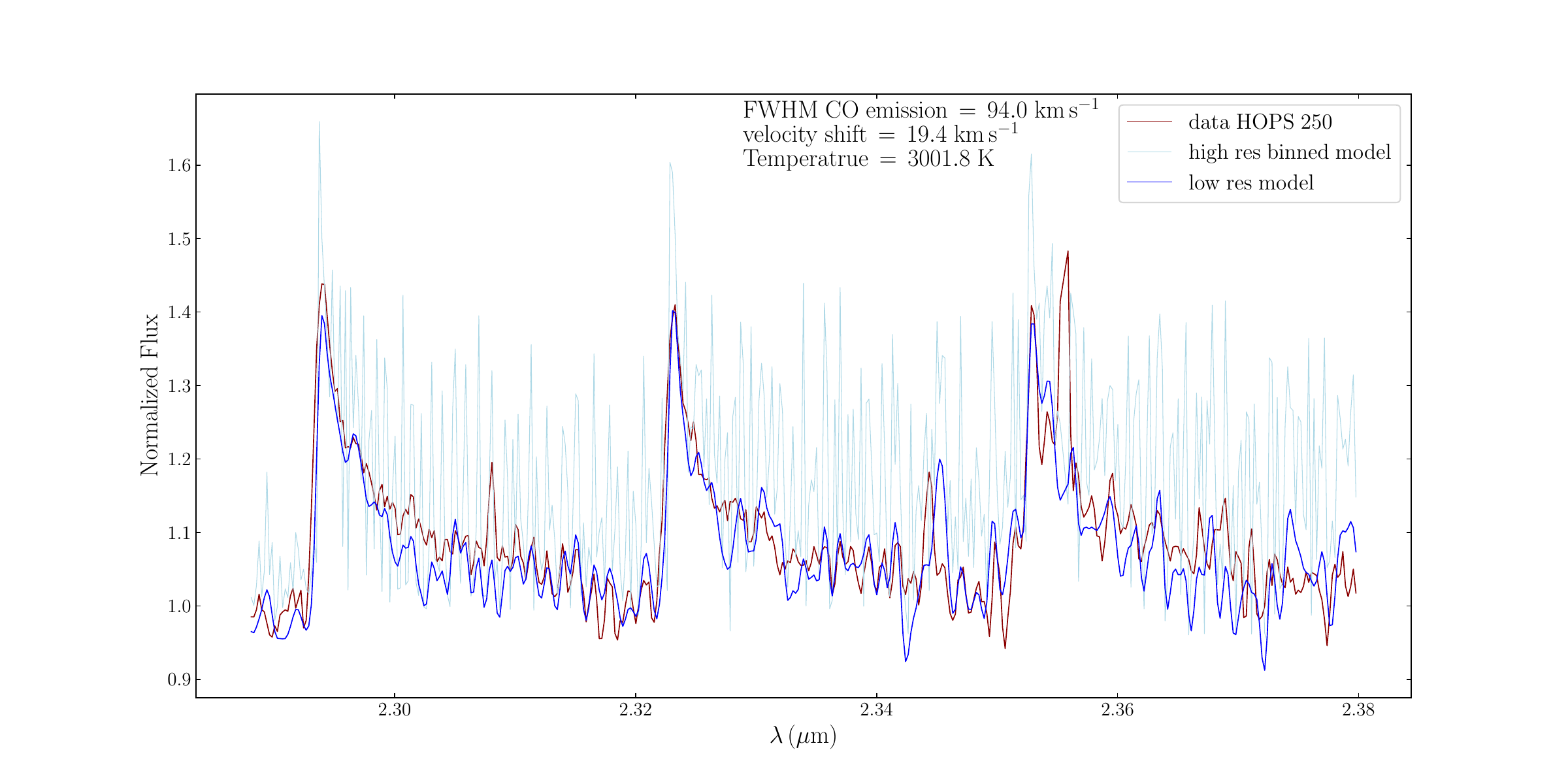}
\vspace{-0.1cm}
\caption{\small Example of our CO emission modeling of the first three CO overtone emission bands of HOPS 250. The diagram shows the normalized observed spectrum of HOPS 250 in solid red. The solid light blue line shows the best modeled spectrum binned to the resolution of the Class 0 data, and the dark blue solid line shows this best model (\ie the median posterior values of the full MCMC fitting) convolved with a 1D Gaussian kernel. In this case the best model has a temperature of 3000 K, a velocity red shift of 19.4 km\,s$^{-1}$, and a broadening kernel of 94 km\,s$^{-1}$.
}
\label{fig:CO_HOPS250_model}    
\end{figure*}

Table \ref{t.CO_em_results} shows the results of the CO analysis for all the CO-emitting Class 0 sources. We note that given the proximity of the resulting broadening kernel FHWM value to the spectral resolution of the observations and its uncertainty, we cannot determine the deconvolved FWHM with high precision. An example of a fit is shown for HOPS 250 in Figure \ref{fig:CO_HOPS250_model}. Upper limits for the broadening kernels' FWHM are provided in Table \ref{t.CO_em_results} when 85\% of the resulting distribution of FWHM values obtained from the MCMC fall below the spectral resolution of the observations (same criteria is used to determine if a source has spectrally CO overtone emission). The MCMC explorations don't converge for HOPS 203, Ser emb 2 and Per emb 28, due to poor S/N of the observations. For these sources, we determine the velocity shift using a cross-correlation with a reference synthetic CO emission spectra ($T_{\textrm{eff}} = 2300$ K, log $g$ = 1.5, and [Fe/H] = 0), where the uncertainty is the velocity shift corresponding to 2\% of the peak of the cross-correlation function.

\begin{figure}[!tbh]
\centering
\includegraphics[scale=0.32,clip,trim= 1.5cm 0.3cm 1.5cm 1.5cm]{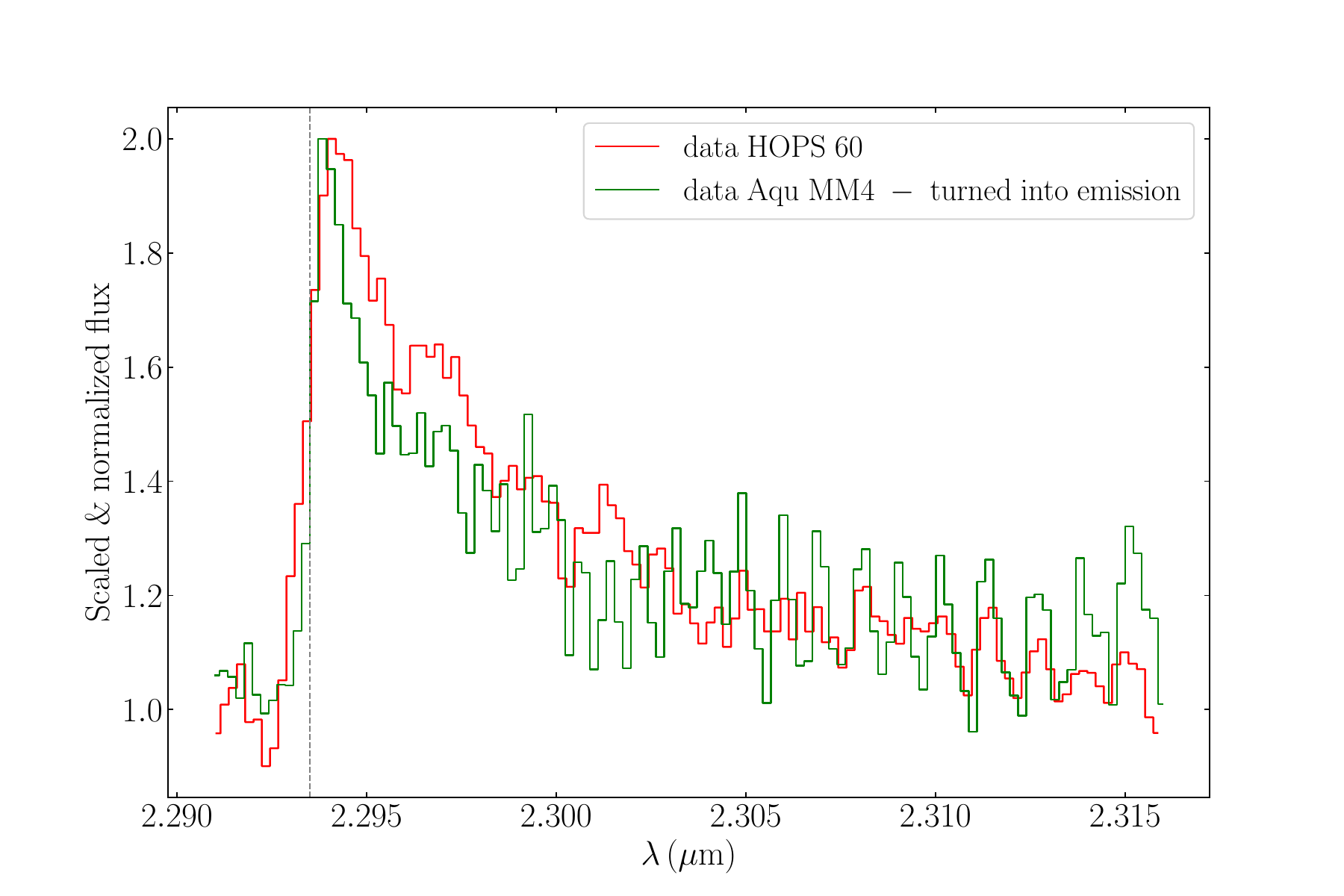}
\vspace{-0.3cm}
\caption{\small CO $v = 2-0$ band of the HOPS 60 (red solid line) and Aqu MM4 (green solid line) Class 0s. The spectra have been normalized by the continuum level, and linearly scaled between 1 and 2 for display purposes. The CO absorption spectra of Aqu MM4 has been turned into emission. The vertical dashed line denotes the reference wavelength of the first CO overtone bandhead. We can notice the apparent broader CO profile of HOPS 60. The modest spectral resolution of the MOSFIRE data preclude us from distinguishing between Gaussian and Keplerian broadening mechanism.}
\label{fig:CO_ab_vs_em}    
\end{figure}

We find that the observed high S/N CO emission bands in our sample of Class 0s are on average unresolved, or only slightly resolved. While the CO overtone emission bands of HOPS 32, Per emb 25, Per emb 26, and Aqu MM5 are clearly unresolved, the rest of the sources have a broader distribution of resulting FHWM broadening values, precluding us from firmly conclude whether these CO bands are resolved or not. Only HOPS 60 exhibits clear spectrally resolved CO emission.
Figure \ref{fig:CO_ab_vs_em} shows the spectra of the first CO overtone bandhead of HOPS 60, alongside the spectra of Aqu MM4 which exhibit a clear photosphere signature (a high S/N unresolved CO overtone spectrum, see Section \ref{sec:results_absorption}), where the spectrum has been turned into emission for comparison. This is why we also cannot attempt to determine whether or not the bandhead profile is consistent with material in Keplerian rotation. Higher spectral resolution observations are required for this. The distribution of velocity shift values of the CO overtone emission is centered on $\sim 0$ km s$^{-1}$. HOPS 32, HOPS 60, HOPS 171, Per emb 26, and Ser emb 15 are consistent with an emission redshifted more than 10 km s$^{-1}$, while no high S/N source has significantly blueshifted CO emission.

\begin{table*}
\small
\centering
\caption[]{CO Emission line parameters}
\label{t.CO_em_results}
\setlength{\tabcolsep}{0.5em} 
\begin{tabular}{p{0.2\linewidth}ccccc}
\hline \hline \noalign{\smallskip}
 Source  & $\rm Flux $ $\rm CO\,(v = 2\rightarrow 0)$& $\rm EW $ $\rm CO\,(v = 2\rightarrow 0)$& $\rm FWHM $& $\rm velocity\,shift $\\ 
  & $10^{-19}\,\textrm{W}\,\textrm{m}^{-2}$ &  \AA  & $\textrm{km}\,\textrm{s}^{-1}$ & $\textrm{km}\,\textrm{s}^{-1}$\\ 
\noalign{\smallskip}  \hline \noalign{\smallskip} 
HOPS 32 & 23.8$\pm 0.3$ & -17.6$\pm 0.2$ & $ \leq\,$125 & 48$_{-11}^{+52} $\\ 
\noalign{\smallskip} 
HOPS 50 & 15.4$\pm 0.1$ & -12.4$\pm 0.1$ & 87$_{-2}^{+4} $ & 0$_{-1}^{+1} $\\ 
\noalign{\smallskip} 
HOPS 60 & 52.9$\pm 0.2$ & -10.2$\pm 0.0$ & 228$_{-130}^{+91} $ & 23$_{-12}^{+67} $\\ 
\noalign{\smallskip} 
HOPS 171 & 11.6$\pm 0.1$ & -14.9$\pm 0.1$ & 81$_{-3}^{+29} $ & 18$_{-2}^{+2} $\\ 
\noalign{\smallskip} 
HOPS 203$^{*}$ & 3.5$\pm 0.1$ & -16.7$\pm 0.3$ & --- & -28$\pm 53$\\ 
\noalign{\smallskip} 
HOPS 250 & 34.5$\pm 0.1$ & -19.4$\pm 0.1$ & 94$_{-22}^{+5} $ & 19$_{-17}^{+2} $\\ 
\noalign{\smallskip} 
Per emb 25 & 11.9$\pm 0.2$ & -17.6$\pm 0.2$ & $ \leq\,$125 & 5$_{-3}^{+6} $\\ 
\noalign{\smallskip} 
Per emb 26 & 6.0$\pm 0.1$ & -23.5$\pm 0.5$ & 156$_{-41}^{+81} $ & 63$_{-6}^{+29} $\\ 
\noalign{\smallskip} 
Per emb 28$^{*}$ & 33.2$\pm 0.5$ & -8.9$\pm 0.1$ & --- & 28$\pm 63$\\ 
\noalign{\smallskip} 
Ser SMM3$^{*}$ & 4.6$\pm 0.1$ & -11.9$\pm 0.3$ & 209$_{-164}^{+110} $ & -62$_{-88}^{+102} $\\ 
\noalign{\smallskip} 
Ser emb 2$^{*}$ & 1.8$\pm 0.1$ & -5.0$\pm 0.2$ & --- & -170$\pm 124$\\ 
\noalign{\smallskip} 
Ser emb 15 & 5.3$\pm 0.1$ & -12.6$\pm 0.2$ & 92$_{-3}^{+4} $ & 16$_{-2}^{+3} $\\ 
\noalign{\smallskip} 
Aqu MM5 & 119.3$\pm 0.1$ & -40.8$\pm 0.0$ & $ \leq\,$91 & 2$_{-6}^{+4} $\\ 
\noalign{\smallskip} 
\hline
\noalign{\smallskip}
\end{tabular}
\vspace{0.03cm}
\tablecomments{\small Derived line parameters of the CO overtone emission bands. The continuum-subtracted flux and EW of the first band $\rm CO\,(v = 2\rightarrow 0)$ are shown alongside the results of the MCMC velocity broadening (``FWHM'' in the Table) and velocity shift analysis fitting the first three CO bands simultaneously. 
Median values and 16\% - 85\% of the distribution are given for each MCMC parameter.
Upper limits are shown when the MCMC converges broadening kernel whose FWHM is lower than the spectral resolution of the observations.
Sources with a $^{*}$ have poor SNR toward the CO overtone bands, meaning that the results are poorly constrained.  The MCMCs don't converge for HOPS 203, Ser emb 2 and Per emb 28. For these sources, we determine the velocity shift using a cross-correlation with a reference synthetic CO emission spectra.
}
\end{table*}

\subsubsection{\Brgamma Emission}
\label{sec:results_Brgamma}

\Brgamma emission is detected in 17 sources of our Class 0 sample by our spectral analysis. The emission of this H\small{I} recombination line usually emanates from high density region confined toward the source center, tracing gas exited either in the accretion region (accretion funnel flows) or at the base of stellar/disk winds. We note that all Class 0 sources exhibiting CO overtone emission also have \Brgamma detected in emission, expect for Per emb 26,
and Ser SMM3. Among the sources that have CO overtone in absorption, only Aqu MM1 has \Brgamma detected in emission. Also, Aqu MM8 and HOPS 164 exhibit \Brgamma in absorption, with a low S/N though (S/N of 4.1 and 4.5, respectively).

All the \Brgamma profiles are spectrally resolved by our observations, the mean and standard deviation of their FWHM values are 182.4 $\pm$ 48.4 km s$^{-1}$. We do not find any systemic velocity shift of the \Brgamma lines, given the mean line centroid value of 2.2 $\pm$ 22.0 km s$^{-1}$. The profiles are on average symmetric, and no sign of redshifted absorption features are detected. 
Finally, no 2D extent in the raw detector images have been found for \Brgamma, meaning that the emitting area (or absorbing) are confined within the seeing disk of $0.4-0.7^{\prime\prime}$. 

\Brgamma is commonly used in the literature to compute the accretion luminosity related to the energy of the accretion funnel flow where \Brgamma is supposed to originate \citep{Muzerolle1998,Calvet2004,Alcala2014,Alcala2017}. However, such methods requires good knowledge of the stellar properties (radius and mass) and the extinction of the inner regions to accurately derive a mass accretion rate. The determination of these properties are currently highly degenerate for Class 0 protostars (which are generally not visible in the $J$ and $H$ bands where stellar parameters can be derived; see the method of \citealt{Fiorellino2021}), preventing us from quantitatively discuss the mass accretion rate of these objects.

\subsubsection{[\ion{Fe}{2}] Emission}
\label{sec:results_atomic_em}

Similar to H$_2$, [Fe \small{II}] traces outflowing material. However, while H$_2$ traces the shocked molecular gas, [Fe \small{II}] traces completely different material emanating from regions confined toward the high velocity collimated jet, where the fast ($\geq\,30$ km s$^{-1}$) dissociative $J$-type shocks induce the destruction of grains liberating iron in the gas phase \citep[\eg][]{Stapelfeldt1991,Gredel1994}. This corresponds to conditions of moderate ionization, high temperatures (8000 -- 15 000 K), and electron density of $\sim$ 10 $^5$ cm$^{-3}$ \citep{Nisini2002}. We detect 6 different [Fe \small{II}] lines, for which the brightest ones are detected in 20\% of the sources. Most of these sources have \Brgamma or CO overtone bands detected in emission. The [Fe \small{II}] lines are usually resolved, with FWHM of $\sim\,80-100$ km s$^{-1}$. All of them exhibit a systematic blue velocity shift of $\sim$ -10 $-$ -80 km s$^{-1}$, indicating that only the jets from the blueshifted cavities are usually detected.

\subsubsection{Other Atomic Emission}

Na\small{I} and Ca\small{I} are seen in emission in 20 $-$ 40 \% of the sources. These lines, when seen in emission are expected to emanate from hot high density regions that are related to the accretion material or to the base of the stellar/disk winds \citep{McGregor1988}. They are most of the time of low S/N, precluding us from performing robust kinematic analyses. However, we find that these lines are marginally resolved and do not exhibit a clear systematic velocity shift. 
`
No correlation can be measured between the relative velocity shift of the H$_2$, [Fe \small{II}], or CO overtone emission lines with the shift of the Na\small{I} and Ca\small{I} emission lines.

\subsection{Absorption Lines in Protostellar Photospheres}
\label{sec:results_absorption}

\begin{figure*}[!tbh]
\centering
\includegraphics[scale=0.5,clip,trim= 3cm 1.2cm 3cm 0cm]{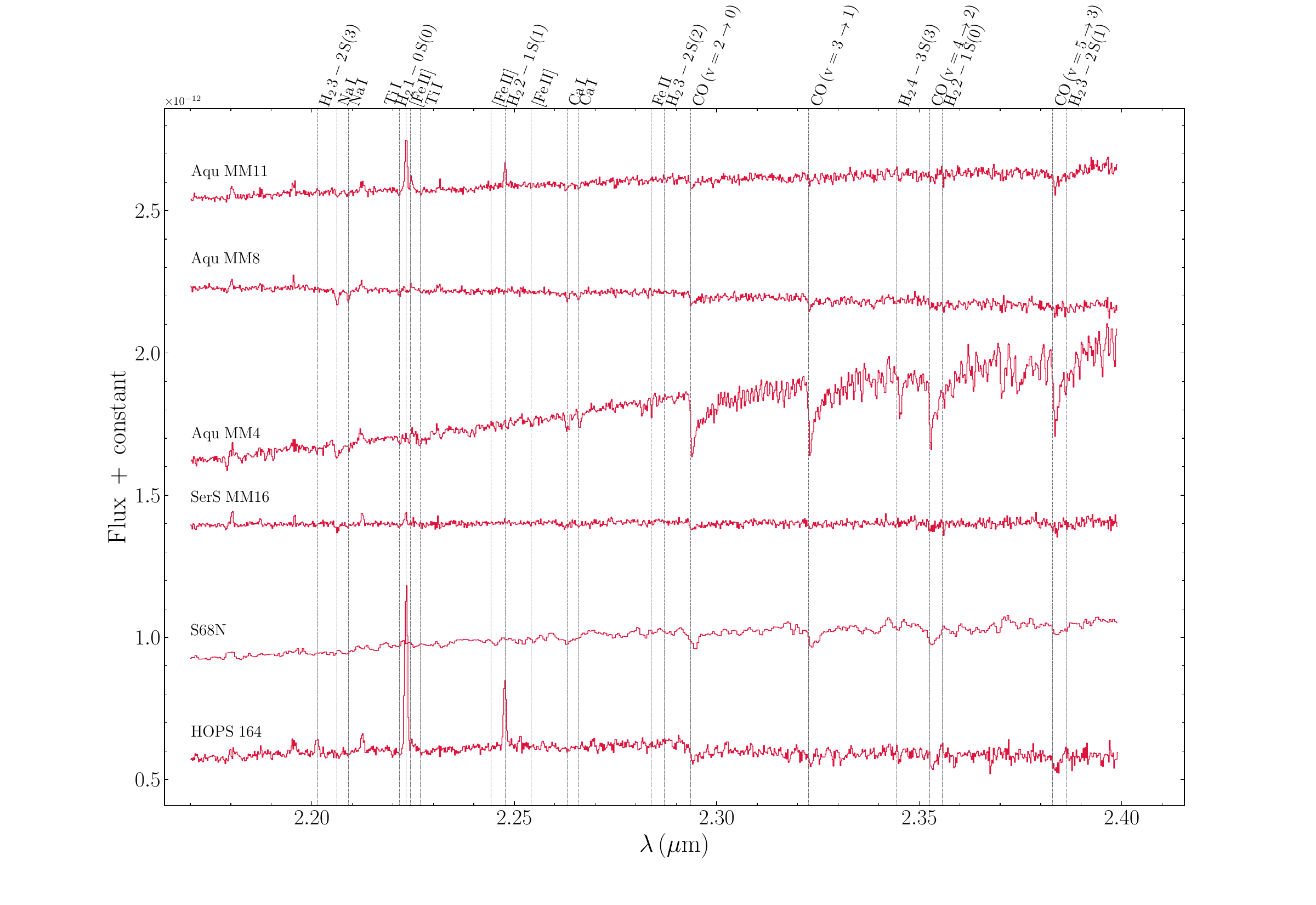}
\caption{\small Extracted 1D $K$ band spectra of our the six Class 0 protostars of our sample showing photospheric absorption features. These spectra are not corrected for extinction. The grey vertical dashed lines show all the lines tentatively detected in this sub-sample. While the CO overtone bands, Na I, and Ca I lines are clearly detected in absorption, several Ti and [Fe II] are tentatively detected, except in Aqu MM4, where the absorption features are strong.
}
\label{fig:all_specs_C0_abs}
\end{figure*}

We detect the CO overtone bands in absorption instead of emission in 6 sources of our sample (see Figure \ref{fig:all_specs_C0_abs}). Weak absorption lines of Na \small{I} and Ca \small{I} at 2.21 $\mu$m and 2.26 $\mu$m are present in five of them. 
These features are consistent with absorption from the photosphere of the nascent protostellar embryo \citep{Greene2018}, generally detected in Class I protostars with low veiling and low accretion activity \citep{Doppmann2005,Connelley2010}. 
These 6 sources are the first photospheres detected in Class 0 protostellar cores.
We list further details about the Class 0 nature of these 6 protostars in Appendix \ref{app:photospheres_disc}. We highlight that the Aqu MM4 NIR source may not be associated with the Class 0 system we thought it was, given the coordinate shift between the ALMA high angular resolution submillimeter dust continuum peak and the NIR continuum. The Class 0 nature of this NIR object is thus unclear.

We have constrained the FWHM velocity broadening, velocity shifts and temperatures the same way as the emission with PHOENIX models, but using a surface gravity value of log $g$ = 2.5 \citep{Greene2018}. First modeling attempts with Starfish \citep{Czekala2015} of the Aqu MM4 spectra also revealed a low surface gravity (\ie $\sim 2.8$). However, further detailed modeling of the photospheres will be performed in an upcoming paper. The results of our first fitting approach are shown in Table \ref{t.CO_abs_results}. The absorption features of Aqu MM11 are too weak to be fitted by our MCMC process. We determine its shift via a cross-correlation with a reference PHOENIX spectrum. Only Aqu MM4 has redshifted CO absorption, while the other five sources are consistent with no velocity shift. We do not spectrally resolve any absorption features (neither CO, Na, or Ca), precluding us from determining the rotational broadening of the photospheres. The velocity shift of the Na \small{I} and Ca \small{I} lines are consistent with the shift of the CO absorption features. The temperature estimations are not conclusive in SerS MM16 and Aqu MM8 due to poor S/N of the CO absorption bands. In these six cases where the photosphere is detected we also notice that the H$_2$ emission lines are weaker when compared to the CO emitting sources. Indeed, within the \HtonetoOSone (the strongest H$_2$ line in our spectral coverage) luminosity values of our full sample, SerS MM16 and HOPS 164 are in the first quantile, and S68N and Aqu MM11 are in the first half with values less than 0.1 times the mean. Aqu MM4 and Aqu MM8 have no emission lines detected at all. Also, no \Brgamma or atomic emission lines are detected in these objects.

\begin{table*}
\small
\centering
\caption[]{CO Absorption line parameters}
\label{t.CO_abs_results}
\setlength{\tabcolsep}{0.35em} 
\begin{tabular}{p{0.2\linewidth}ccccc}
\hline \hline \noalign{\smallskip}
 Source  & $\rm EW $ $\rm CO\,(v = 2\rightarrow 0)$& $\rm FWHM $& $\rm velocity\,shift $& $T_{\textrm{eff,\,CO}}$\\ 
  &  \AA  & $\textrm{km}\,\textrm{s}^{-1}$ & $\textrm{km}\,\textrm{s}^{-1}$ & $ \rm K $\\ 
\noalign{\smallskip}  \hline \noalign{\smallskip} 
Aqu MM4 & 21.8$\pm 0.1$ & $ \leq\,$91 & 21$_{-8}^{+1} $ & 3313$_{-913}^{+143} $\\ 
\noalign{\smallskip} 
HOPS 164 & 12.0$\pm 0.2$ & 91$_{-3}^{+7} $ & 14$_{-1}^{+3} $ & 2500$_{-100}^{+22} $\\ 
\noalign{\smallskip} 
S68N & 6.3$\pm 1.2$ & $ \leq\,$300 & 9$_{-1}^{+1} $ & 4745$_{-95}^{+51} $\\ 
\noalign{\smallskip} 
SerS MM16 & 10.8$\pm 0.3$ & 85$_{-7}^{+8} $ & -5$_{-6}^{+2} $ & 2498$_{-98}^{+47} $\\ 
\noalign{\smallskip} 
Aqu MM8 & 11.2$\pm 0.2$ & $ \leq\,$91 & -4$_{-1}^{+1} $ & 2537$_{-137}^{+47} $\\ 
\noalign{\smallskip} 
Aqu MM11 & 6.2$\pm 0.1$ & --- & 0$\pm 50$ & ---\\ 
\noalign{\smallskip} 
\hline
\end{tabular}
\vspace{0.03cm}
\tablecomments{\small Derived line parameters of the CO overtone absorption bands. The EW of the first band $\rm CO\,(\nu = 2\rightarrow 0)$ is shown alongside the results of the MCMC analysis fitting the first three CO bands at the same time. 
Median values and 16\% - 85\% of the distribution are given for each of the three MCMC parameters.
Upper limits are shown when the MCMC converges to the lower limit of the parameter grid, \ie the lowest temperature of the PHOENIX grid, or to broadening kernel whose FWHM is lower to the spectral resolution of the observations.
The MCMC does not converge for Aqu MM11. For this source, we thus determine the velocity shift using a cross-correlation with a reference synthetic CO emission spectra.
}
\end{table*}

\section{Comparisons with Class I objects}
\label{sec:comp_ClassI}

In order to quantitatively evaluate the evolution of the accretion properties of protostars with time, we have gathered a sample of archival NIR spectroscopic observations of Class I and Flat-spectrum low-mass protostars. The goal is to perform a spectral line analysis using the same tools as those used in Section \ref{sec:results} on the sample of more evolved Class I objects and to statistically compare these results with those obtained on the Class 0 objects. 

\subsection{Archival Sample of Class I spectra}
\label{sec:arch_sample_CIs}

We use the samples of Class I/Flat Spectrum (FS) sources presented in \citet{Greene2010}, \citet{Doppmann2005}, and \citet{Fiorellino2021}. \citet{Fiorellino2021} observed 17 Class I/FS objects in the NGC1333 region of the Perseus cloud with VLT KMOS (see their Table 1 for all the relevant information about the source sample). $K$-band spectra were obtained with a spectral resolution of $\sim$ 4200. The flux calibrated spectra were made available by the authors at \citet{Fiorellino2021Vizier}. We also use the sample of 52 Class I/FS objects presented in \citet{Doppmann2005}. They used Keck NIRSPEC to observe protostars in nearby star forming clouds (\ie $\rho$ Ophiuchus, Taurus-Auriga, Serpens, Perseus, and Corona Australis) with a spectral resolution of 18,000. We also use the work of \citet{Greene2010} who re-observed 18 sources of this sample with Keck NIRSPEC to study a few H$_2$ lines in the $K$-band. 
The idea was to gather a large and  homogeneous sample of low-mass Class I/FS objects within local star forming clouds, \ie regions with similar metallicity and not directly exposed to the feedback of massive stars.
Other spectral studies of Class I /FS sources that we do not reanalyze here include those of 
\citet{Nisini2005,Antoniucci2008,Connelley2010,Martins2010,Antoniucci2011,Connelley2014,Itrich2023}.
We discuss the potential impact of environment in Section \ref{sec:disc_enviroment}.

We had to perform manual flux calibration of the \citet{Doppmann2005} and \citet{Greene2010} observations. To do so, we have retrieved the $K$-band magnitudes of the objects, and their 2D-spatial extent from the 2MASS (UKIDSS when available) archive. Then, taking into account the slit width and 2MASS angular resolution, we can predict the amount of light that should correspond to the absolute flux in the $K$-band spectra. 
The same technique was realized in \citet{Fiorellino2021}. We note that the 2MASS and UKIDSS surveys are not exactly contemporary to these Class Is NIR spectroscopic observations, and that this represents a caveat of our approach since protostars can be variable. Recent variability surveys reported a typical infrared variability of $\sim$0.1-1 mag for YSOs in local star forming clouds \citep{AlvesdeOliveira2008,MoralesCalderon2011,Park2021,Fischer2023}.
In addition, while equivalent widths are independent of extinction, the distribution of continuum levels can be different between Class 0 and I (different intrinsic veiling due to different inner disk and NIR scattered light contributions), which can affect our line strength comparisons to the continuum.

Therefore, we chose to also compute the line luminosities. In order to compare them among the two samples, we correct the line flux for extinction and take into account the distance of the objects. Similarly to the Class 0 objects, we use the work by \citet{Herczeg2019,Zucker2019,Zucker2020} to find the distance of the Class I objects. 
As for the \citet{Doppmann2005,Greene2010} objects, extinction values were calculated by de-reddening each object’s $JHK$ to the CTTS locus presented in \citet{Meyer1997}. 
We used preferentially 2MASS magnitudes. However, for a few sources the $J$ flux is not detected. For these sources, we thus retrieve the photometry from UKIDSS, from the VISIONS survey (for Ophiuchus sources; \citealt{Meingast2023a}), from \citet{Gorlova2010} (for Serpens sources), or from \citet{Haas2008} (for Corona Autralis sources), depending on the respective surveys' sky coverage.
We use the extinction law from \citet{Cardelli1989} to obtain the extinction in the $K$-band, with a total to selective extinction ratio $R_V$ of 5. 
Given that young stars have a lot of circumstellar material, to what specific region is the extinction computed on is uncertain, and the extinction values are likely underestimates \citep{Doppmann2005}.
The visual extinction of the VLT KMOS NGC1333 objects were reported in Table 5 and 6 in \citet{Fiorellino2021}. The Perseus sources being more embedded, many of them don't have a reliable $J$-band photometry. Assuming a veiling and an age for the objects (\eg the birthline from \citealt{Palla1993}), \citeauthor{Fiorellino2021} chose to constrain the $K$-band extinction of their targets by calculating the expected bolometric magnitude in the $K$-band given an estimate of the stellar photosphere luminosity. This estimate is obtained by subtracting the accretion luminosity from the bolometric luminosity using the relation from \citet{Muzerolle1998} between the accretion luminosity and the luminosity of the \Brgamma line \citep{Alcala2017}.
Determining the extinction of these embedded sources remains a difficult challenge. 
We have also attempted an alternative method in order to uniformly correct all the Class Is for extinction. We de-redden the ($H$,$K$) photometry of the objects to match the $H-K$ color of the bluest spectra of our sample, \ie DG Tau A. However, this method is not necessarily better, because Class I objects have some non-uniform level of NIR excess which is not possible to distinguish from extinction. 
This gives extinction values in reasonable agreement with the methods described above, with the distribution of differences having a mean and standard deviation of $A_v = -0.6 \pm 9.6 $ mag. The results of the statistical comparisons of Section \ref{sec:stats_comp} are also unchanged. From now on, we use the methods initially described, \ie the \citet{Fiorellino2021} values and the values obtained with the \citet{Meyer1997} method for the \citet{Doppmann2005,Greene2010} observations.

The extinction values of Class 0 protostars are extremely difficult to precisely constrain because of several intrinsic observational limitations. SED modeling of unresolved FIR and submillimeter counterparts of these objects generally produce very uncertain extinction values due to poor constraints obtained on the dense substructures, outflow cavity geometry, etc. One would need to model the disk and envelope with ALMA and ACA observations, with scattering along with dust thermal emission, to model both the SED and the sub-millimeter interferometric images in order to precisely constrain the small-scale dense substructures, which are responsible for most of the extinction. In these sources, NIR and sub-millimeter peaks usually have different spatial locations, which make such spatially resolved modeling essential in order to constraint the extinction of the NIR light. Such work is beyond the scope of this paper. To treat the extinction of Class 0 sources, we start by  applying a uniform value of $A_v$ of 50 mag in most of Section \ref{sec:stats_comp} to compare the \Brgamma and CO overtone emission line luminosities, while we use the extinction found in Section \ref{sec:results_H2} for the H$_2$ lines. 
The value $A_v$ value of 50 mag is motivated by the extinction measurement of $A_v$ $\simeq$ 50 mag for S68N from its H$_2$ line ratio in \citet{Greene2018}.
At the end of Section \ref{sec:stats_comp}, we perform Monte Carlo simulations to account for a dispersion around this assumed value of Class 0 visual extinction. 

Finally, in order to compare velocity shifts from the source reference frames, we manually correct the Class I observations for Earth orbit motions with respect to the Sun, and for the $v_{\textrm{lsr}}$ of the sources using \citet{Carney2016,Stephens2018,Dhabal2019} for the Perseus sources, \citet{Covey2006,Friesen2017} for $\rho$ Ophiuchus and Taurus-Auriga sources, \citet{Covey2006,Levshakov2013,Lee2014} for the  Serpens sources, and \citet{Lindberg2012} for the Corona Australis sources.

\subsection{Statistical comparisons}
\label{sec:stats_comp}

Table \ref{t.line_detection} provides the emission lines detection rates of the Class I sample.
In this section, we compare the different spectral line detection rates between our newly observed and literature Class 0 sample, and the Class I objects.

All of the strongest H$_2$ lines in the $K$-band (\ie \HtonetoOStwo, \HtonetoOSone, \HtonetoOSO, and \HttwotooneSone) clearly have higher detection rates in the Class 0 sample (from 62 to 89 \%) than in the Class I sample (from 35 to 77 \%). Similarly, CO overtone emission is detected much more frequently in Class 0 objects, with 42 \% and 54 \% of detections for the first and second band, against 14 \% and 6 \% in the Class I cases. Conversely, the CO overtone bands are detected in absorption much more frequently in the Class I sources, with 69 \% of detections, against 23 \% in Class 0s.
Finally, \Brgamma is less frequently detected in Class 0 than in Class I (58 \% versus 79 \%). 

\begin{figure*}[!tbh]
\centering
\subfigure{\includegraphics[scale=0.5,clip,trim= 1cm 0cm 2cm 0.8cm]{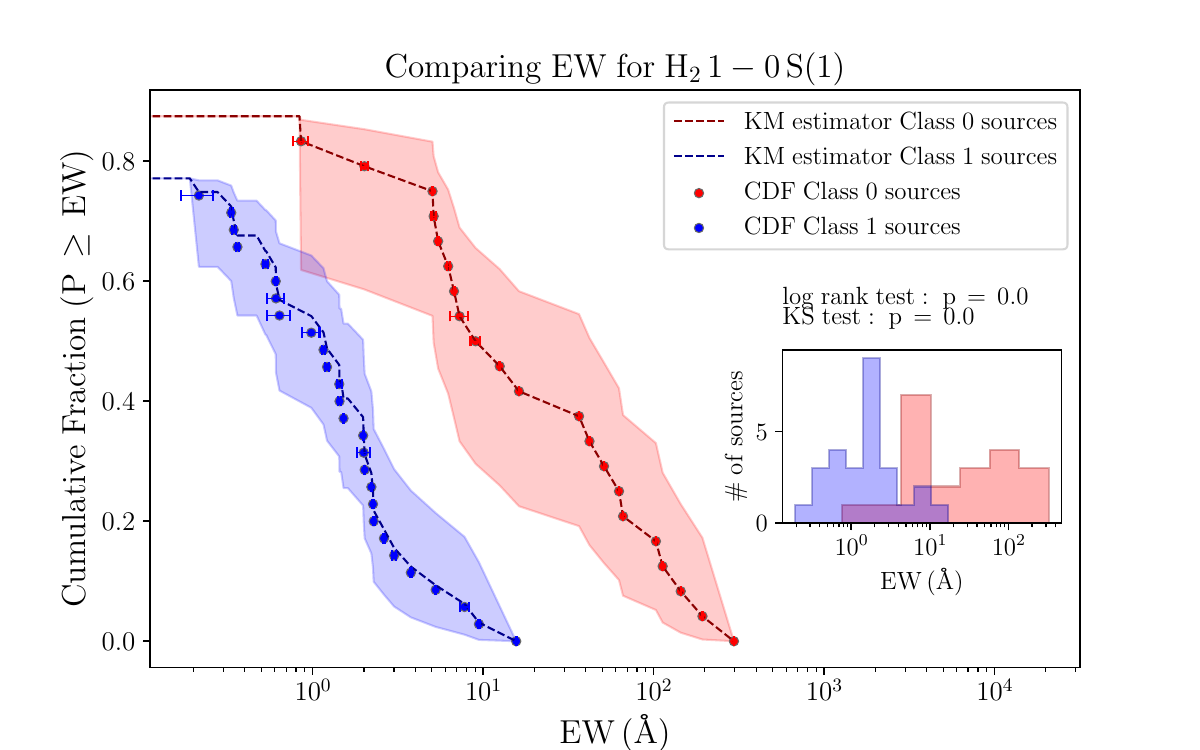}}
\subfigure{\includegraphics[scale=0.5,clip,trim= 1cm 0cm 2cm 0.8cm]{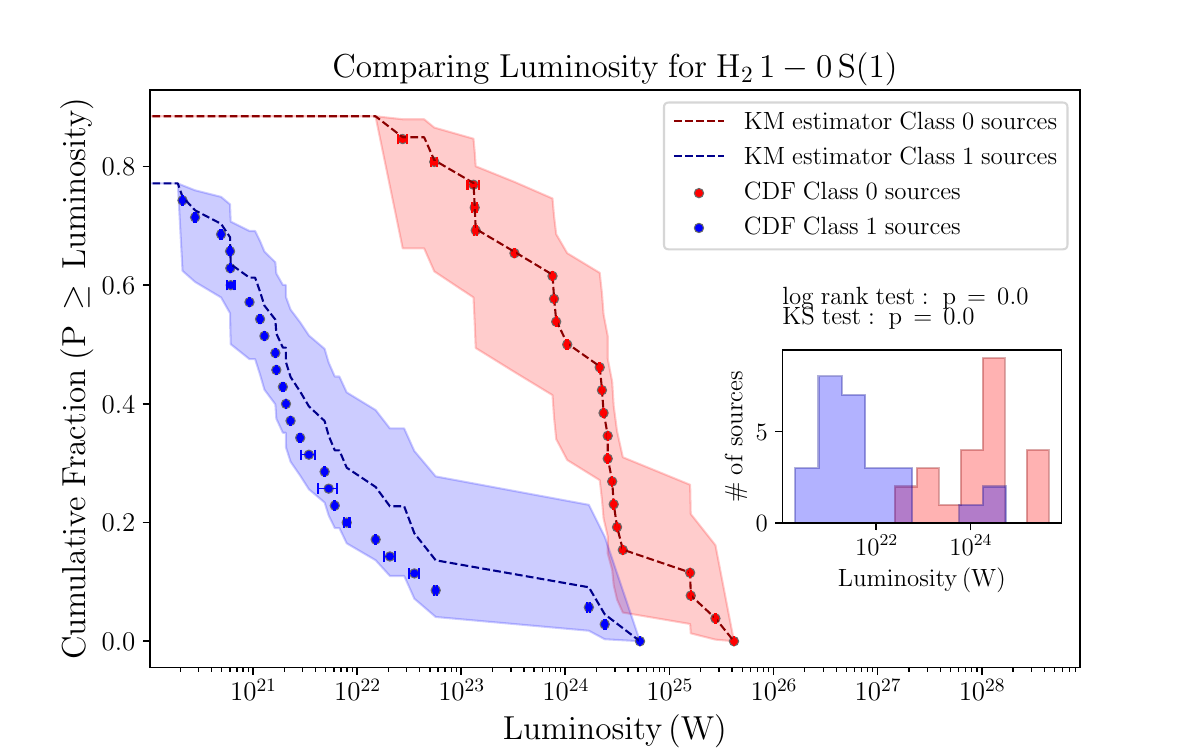}}
\subfigure{\includegraphics[scale=0.5,clip,trim= 1cm 0cm 2cm 0.8cm]{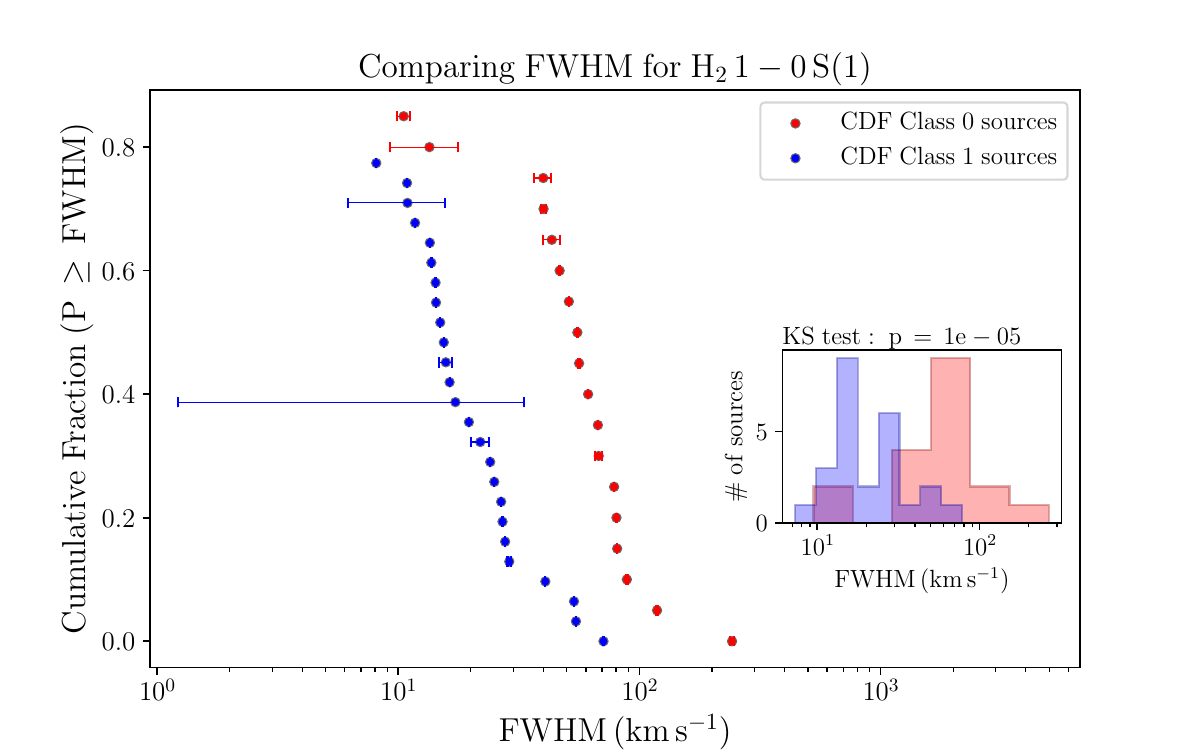}}
\subfigure{\includegraphics[scale=0.5,clip,trim= 1cm 0cm 2cm 0.8cm]{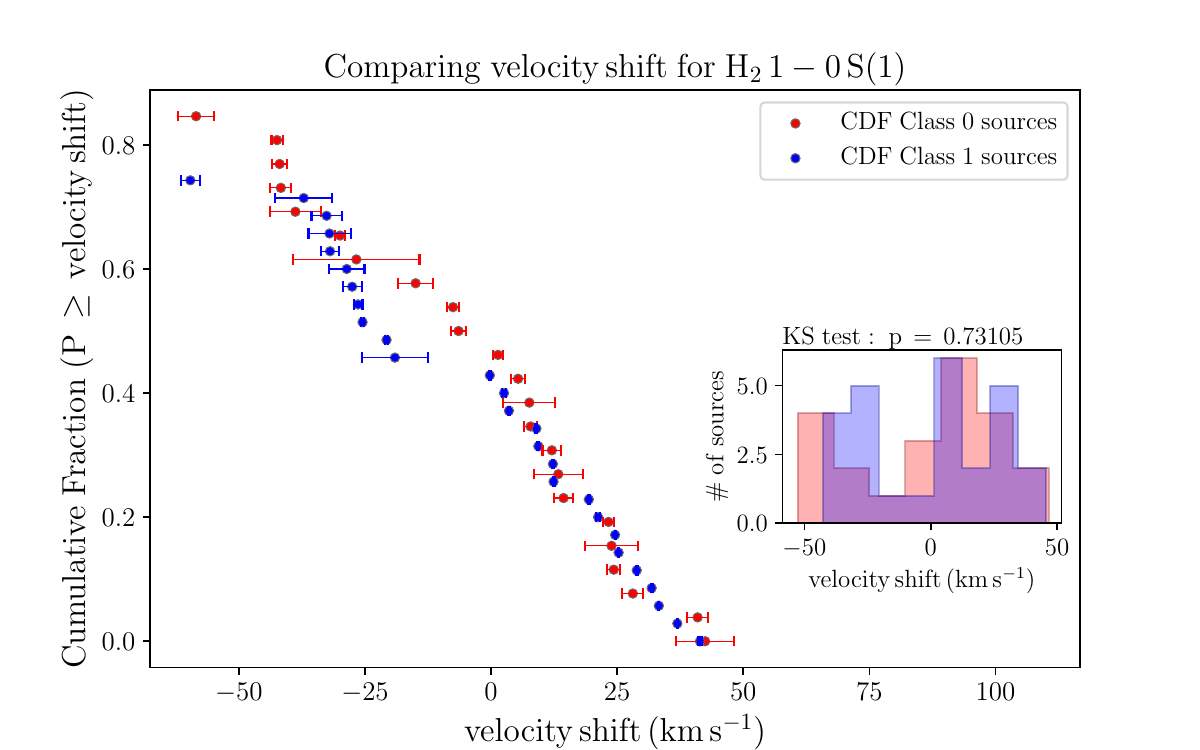}}
\caption{\small Comparisons of the \HtonetoOSone emission line parameters between our sample of Class 0 protostars and the sample of archival Class I protostars. Each panel focuses on one line parameter, \ie line equivalent width (\textit{top left} panel; we show EW values as positive), line luminosity (\textit{top right} panel), line FWHM (\textit{bottom left} panel), and velocity shift from the reference frame (\textit{bottom right} panel). In each panel, the distribution of line parameter values is shown in the form of a cumulative distribution function (CDF; normalized by the detection rate of the line in a given sample), and a histogram (shown in the sub-box on the right hand side of the plot). Each point of the CDFs correspond to one object's measurement and uncertainty of the given line parameter.  The red color is used for Class 0 objects, and the blue color is used for Class Is. The $p$-value of the two-sided Kolmogorov-Smirnov (KS) test are shown above the histograms on the right. The $p$-value is the probability that the two populations come from the same parent population.
In the cases where upper-limits can be derived from non-detections (\ie for line equivalent width and luminosity), we show the resulting Kaplan-Meier estimator function for the Class 0 and I distributions with the dashed line and shaded area, which indicate the 95\,\% confidence interval of the survival function. In these cases, the resulting $p$-value of the log-rank statistical test, that takes into account non-detections (unlike the KS test), is shown on the right.}
\label{fig:C0_C1_comp_H2_1-0_S1}
\end{figure*}

\begin{figure*}[!tbh]
\centering
\subfigure{\includegraphics[scale=0.5,clip,trim= 1cm 0cm 2cm 0.8cm]{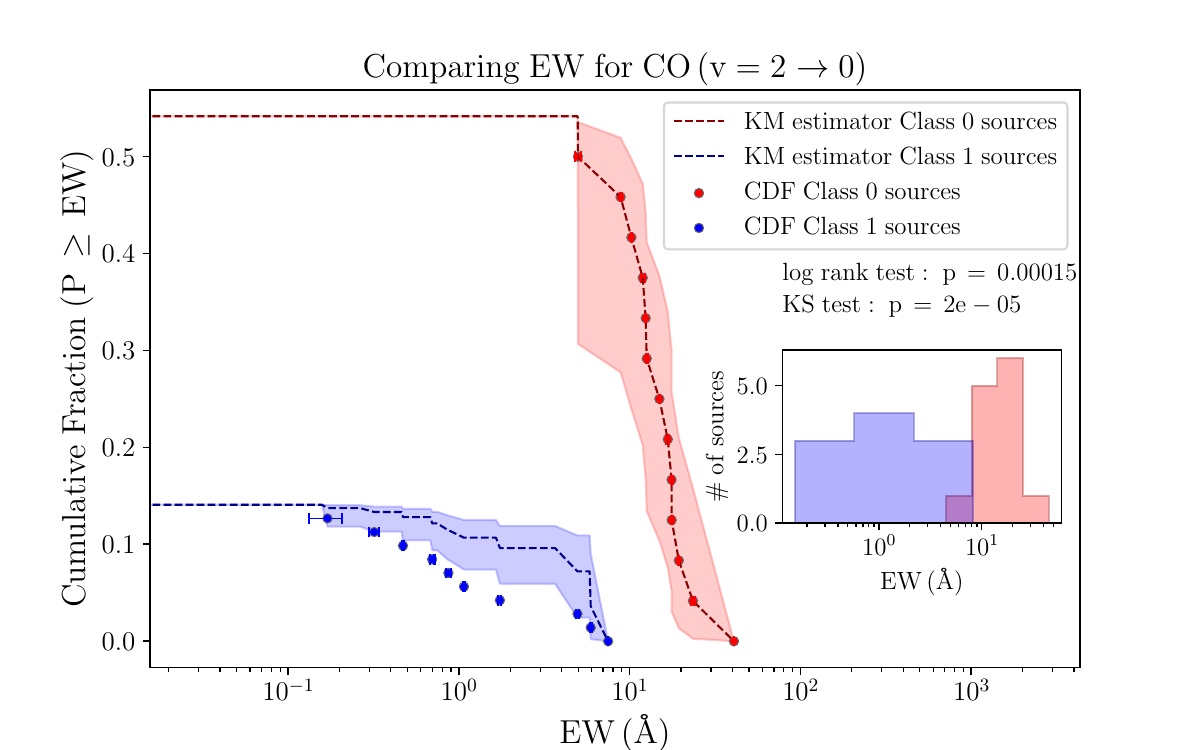}}
\subfigure{\includegraphics[scale=0.5,clip,trim= 1cm 0cm 2cm 0.8cm]{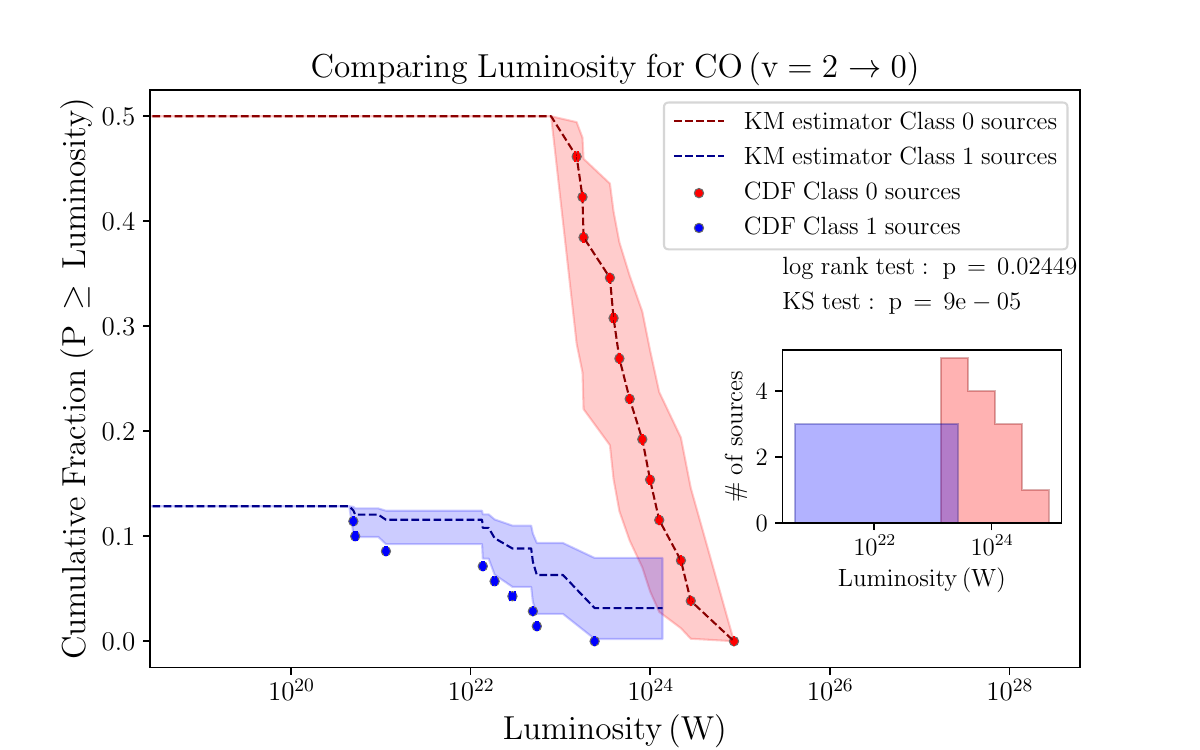}}
\subfigure{\includegraphics[scale=0.5,clip,trim= 1cm 0cm 2cm 0.8cm]{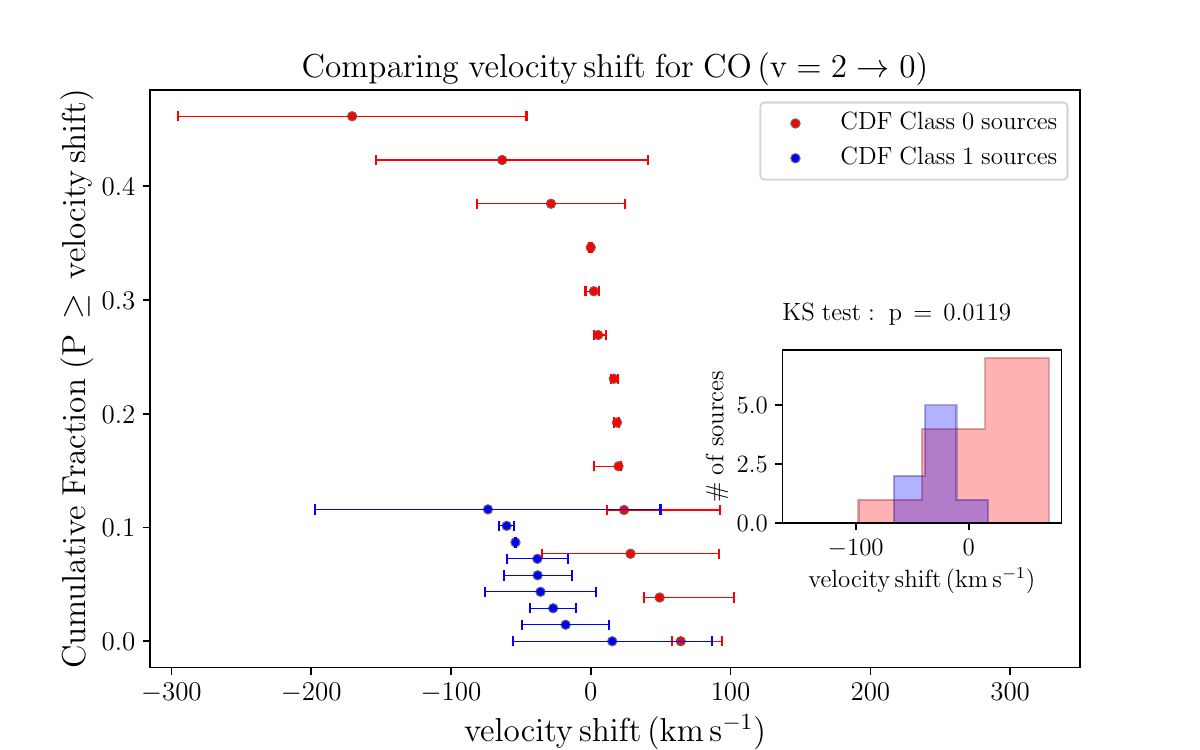}}
\caption{\small Same as Figure \ref{fig:C0_C1_comp_H2_1-0_S1} for the first CO overtone band emission. Only the equivalent width, luminosity, and velocity shift results are shown. Determining the broadening of the CO overtone emission requires modeling, which has only been done for the Class 0 sources (see Section \ref{sec:results_atomic_em}).}
\label{fig:C0_C1_comp_CO}
\vspace{0.2cm}
\end{figure*}

We now aim to quantify the differences in line parameters (\ie equivalent width, luminosity, velocity broadening FWHM, and velocity shift) between the Class 0 and I samples. To do so, we build cumulative distributions functions (CDFs) for each line and line parameter. We use the Kaplan-Meier (KM) estimator from the lifelines Python package\footnote{See \href{http://lifelines.readthedocs.io/}{http://lifelines.readthedocs.io/}} that takes into account upper limits. We note that our analysis assumes that each value is precisely known (\ie it does not consider the uncertainties of the line parameters). Figures \ref{fig:C0_C1_comp_H2_1-0_S1} (see also Figures \ref{fig:C0_C1_comp_H2_1-0_S0}, \ref{fig:C0_C1_comp_H2_1-0_S2}, and \ref{fig:C0_C1_comp_H2_2-1_S1} in Appendix \ref{app:add_comp_plots}), \ref{fig:C0_C1_comp_Bry}, and \ref{fig:C0_C1_comp_CO}, present these statistical comparisons for \HtonetoOSone, \Brgamma, and the \COfirstband band, respectively. In these Figures, the KM estimators departs from the CDF where upper limits of non-detections are present at a given value of the line parameters. We used and compare absolute values of the emission line equivalent widths to work with positive values. 

To test the statistical significance of the observed shifts in the Class 0 and Class I line parameter distributions, we perform two statistical tests: the two-sided Kolmogorov-Smirnov (KS) test and the log-rank test. The KS test is a non-parametric test that quantitatively evaluates the difference between the cumulative distribution of two data sets. The log-rank test is a non parametric method that compares the survival distributions of two samples, taking into account non-detections. In both tests, the null hypothesis is that distributions are identical at all line parameter values. Each test produces a $p$-value ($p$), that evaluates if the data sets are drawn from the same parent distribution, with low values of $p$ indicating different parent populations. We typically consider that a $p$-values lower than 0.001 corresponds to distributions statistically distinct, $p$-values between 0.001 and 0.05 to distributions that are marginally different, and finally that $p$-values higher than 0.05 correspond to distributions statistically similar. These tests are less reliable for small numbers of detections (like some H$_2$ lines). Also, a big difference between log-rank and KS tests $p$-values can be seen for a large number of non-detections (e.g.,  the CO overtone emission case).

The H$_2$ emission lines exhibit larger EW values in Class 0s than Class Is with statistically significant differences; the line luminosities are also larger in the Class 0s but the statistical difference is marginal when taking account non-detections (otherwise the KS tests results in high statistical difference; see Figures \ref{fig:C0_C1_comp_H2_1-0_S1}, \ref{fig:C0_C1_comp_H2_1-0_S0}, \ref{fig:C0_C1_comp_H2_1-0_S2}, and \ref{fig:C0_C1_comp_H2_2-1_S1}). The H$_2$ line profiles are broader in the Class 0 stage, with strong to marginal statistical significance, depending on the H$_2$ lines ($p$ values ranging from 10$^{-5}$ to 0.02; the case of \HtonetoOStwo is not conclusive due to too few Class I detections). However, we note that because the MOSFIRE instrument spectral resolution is lower than the archival Class I observations (Keck NIRSPEC and VLT KMOS), this FWHM analysis may suffer from systematics even if we have subtracted in quadrature the instrument resolution (variable seeing conditions may have degraded the spectral resolution of our Class 0 observations). The velocity shifts appear to be similar between Class 0 and I objects, with mean values within $\pm\,5$ km s$^{-1}$.

The EW and luminosities of the \COfirstband band are clearly higher in Class 0s with statistical significance, given results of the KS tests. However, the log-rank tests are less conclusive for the CO luminosities given the high number of non-detections in the Class I sample. The CO emission bandheads are not resolved in the Class 0s, which preclude us from comparing the broadening between the two samples. The Class 0s have more redshifted CO overtone emission compared to the Class I objects, whose CO emission bands are on average consistent with a slight blueshift or are not shifted in velocity.

The \Brgamma emission lines exhibit higher luminosity and EW values in the Class 0s with a very significant statistical difference (Figure \ref{fig:C0_C1_comp_Bry}). The distributions of line FWHMs are very similar between Class 0s and Class Is, with mean values of $\sim\,190-195$ km s$^{-1}$. Interestingly, the distributions of line velocity shifts are statistically different, with a mean and standard deviation values of 1 $\pm\,22$ km s$^{-1}$ in the Class 0s, and $-55$ $\pm\,38$ km s$^{-1}$ in the Class Is, with Class I values down to -130 km s$^{-1}$.

\begin{figure*}[!tbh]
\centering
\subfigure{\includegraphics[scale=0.5,clip,trim= 1cm 0cm 2cm 0.8cm]{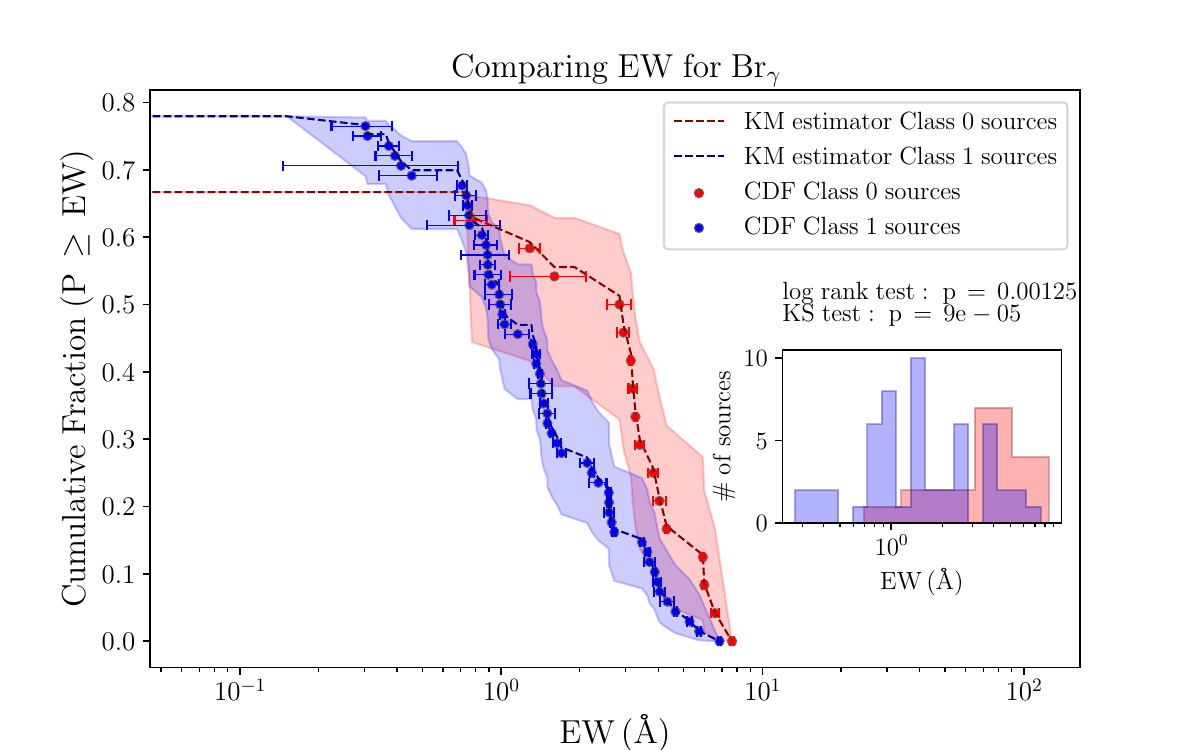}}
\subfigure{\includegraphics[scale=0.5,clip,trim= 1cm 0cm 2cm 0.8cm]{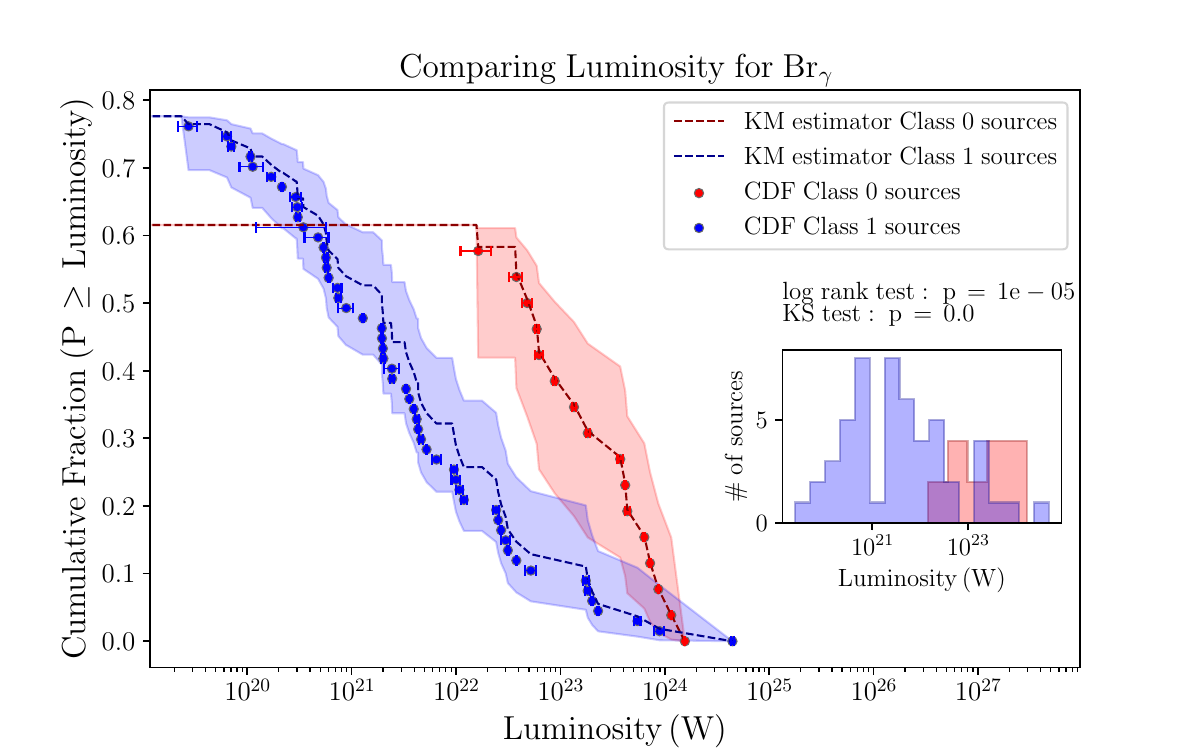}}
\subfigure{\includegraphics[scale=0.5,clip,trim= 1cm 0cm 2cm 0.8cm]{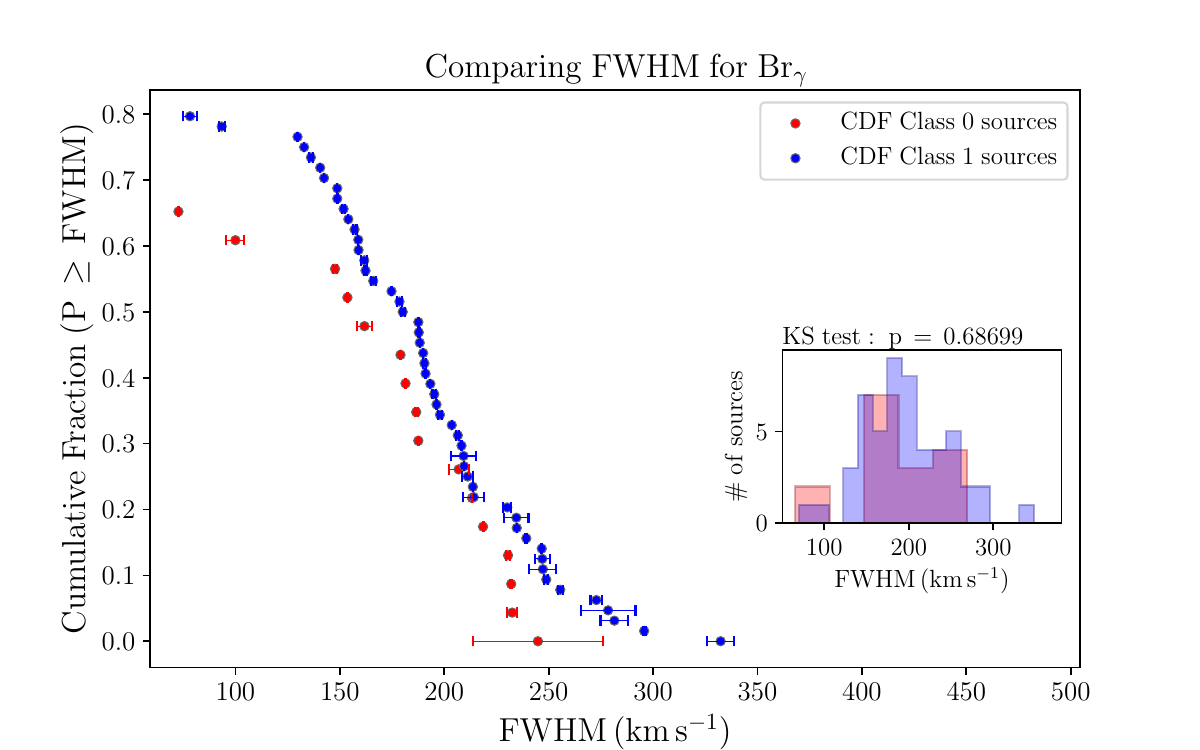}}
\subfigure{\includegraphics[scale=0.5,clip,trim= 1cm 0cm 2cm 0.8cm]{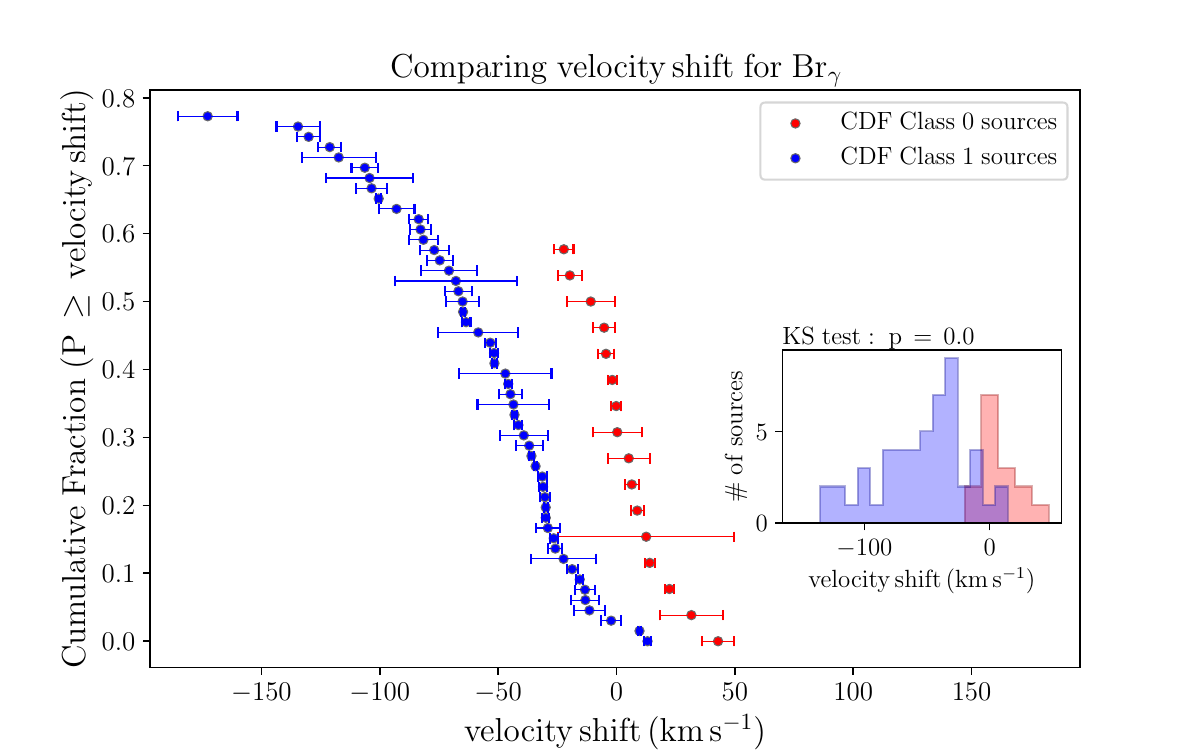}}
\caption{\small Same as Figure \ref{fig:C0_C1_comp_H2_1-0_S1} for the \Brgamma emission lines.}
\label{fig:C0_C1_comp_Bry}
\vspace{0.2cm}
\end{figure*}

\begin{figure*}[!tbh]
\centering
\subfigure{\includegraphics[scale=0.5,clip,trim= 1cm 0.3cm 1.5cm 1cm]{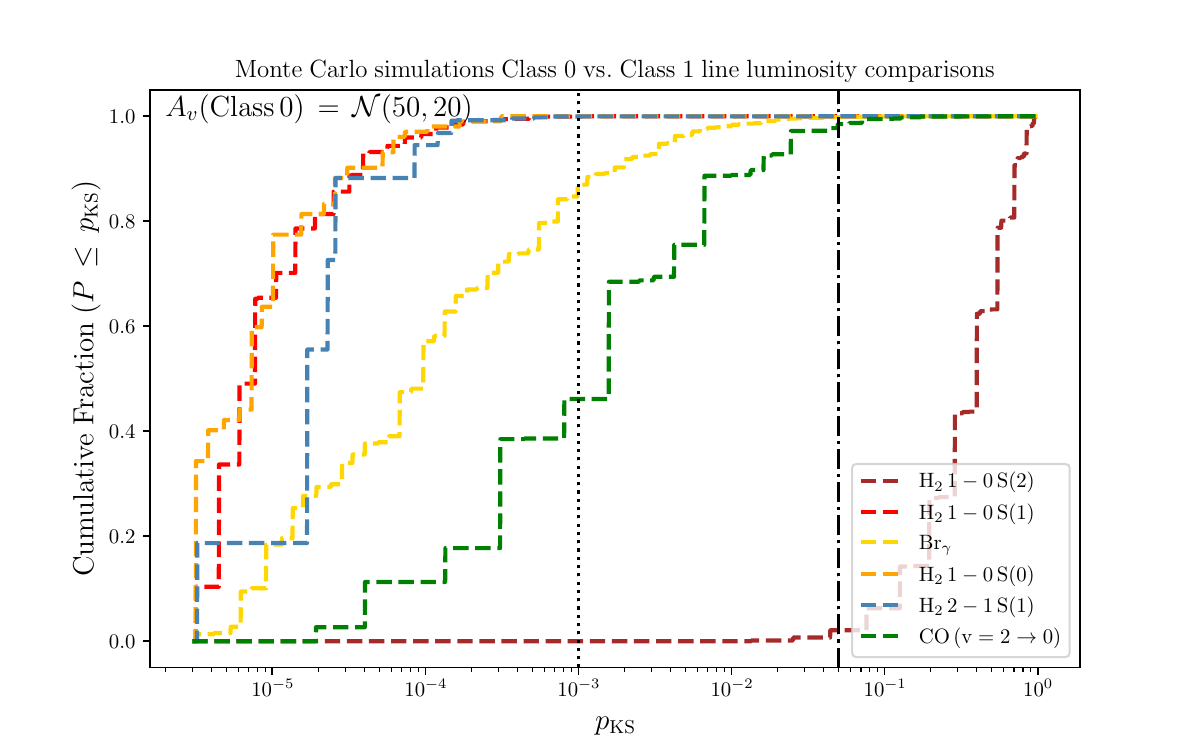}}
\subfigure{\includegraphics[scale=0.5,clip,trim= 1cm 0.3cm 1.5cm 1cm]{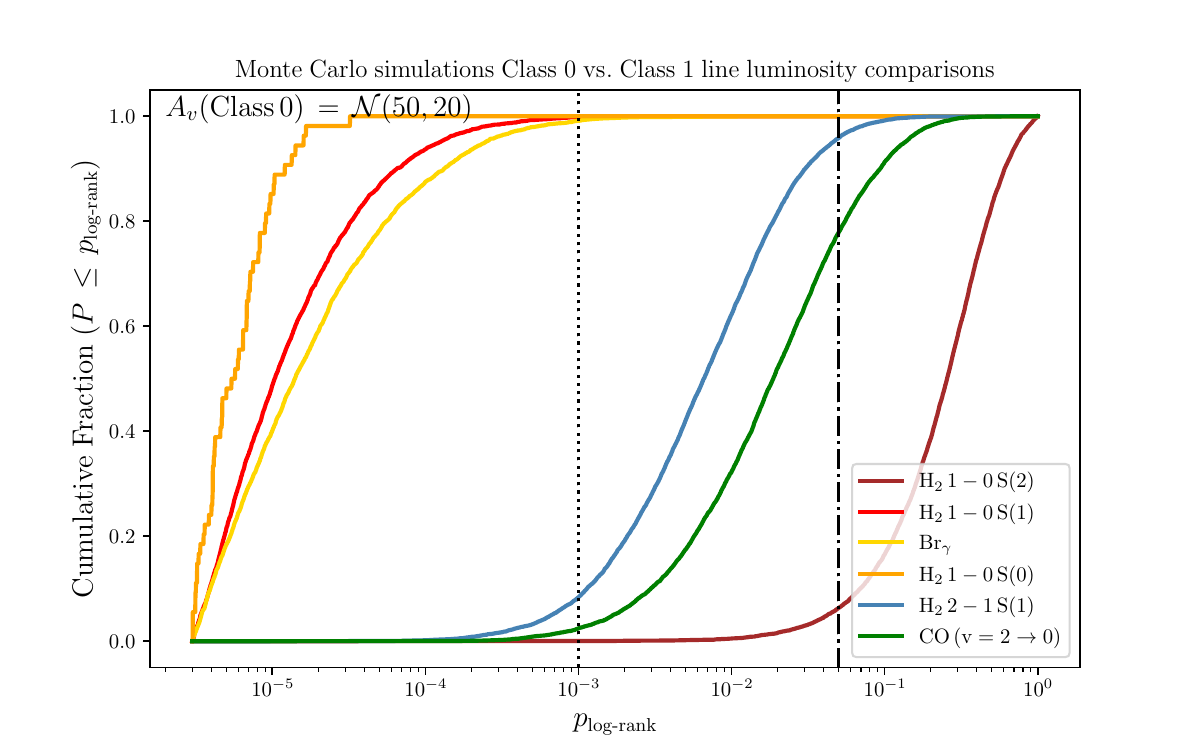}}
\caption{\small Comparisons of the line luminosity distributions for different emission lines between the samples of Class 0 and I protostars. Given the large uncertainty of the Class 0 extinction values, the two plots show the results of Monte Carlo simulations. For each run out of 10$^4$ tries, the extinction values of the Class 0 sources is randomly selected to follow a normal distribution of $A_v$ centered on 50 mag, with a dispersion of 20 mag. Then, for a given spectral emission line, the distribution of synthetic line luminosity for the Class 0 sample is compared with the distribution of Class I line luminosity via a statistical test (KS for dashed lines in the left panel, log-rank test for solid lines in the right panel). The cumulative distribution function of the resulting $p$-values from the two-sided KS or log-rank tests are shown for each spectral line by a different color. The vertical dotted and dot-dotted dashed lines correspond to $p$ values of 0.05 and 0.001, respectively.
}

\label{fig:C0_C1_comp_MC_simu_20_50}
\end{figure*}

\begin{figure}[!tbh]
\centering
\includegraphics[scale=0.5,clip,trim= 0cm 0cm 0cm 0cm]{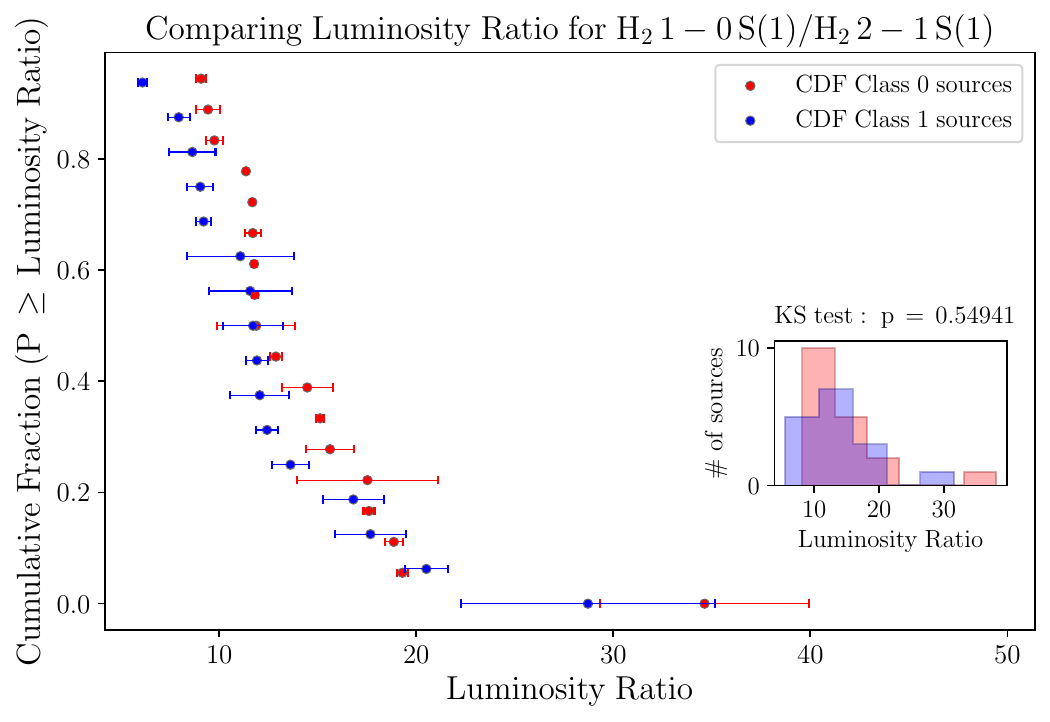}
\caption{\small Same as Figure \ref{fig:C0_C1_comp_H2_1-0_S1} for the \HtonetoOSone to \HttwotooneSone line luminosity ratio.}
\label{fig:C0_C1_comp_H_2_ratio}
\end{figure}

\begin{figure}[!tbh]
\centering
\includegraphics[scale=0.5,clip,trim= 0cm 0cm 0cm 0cm]{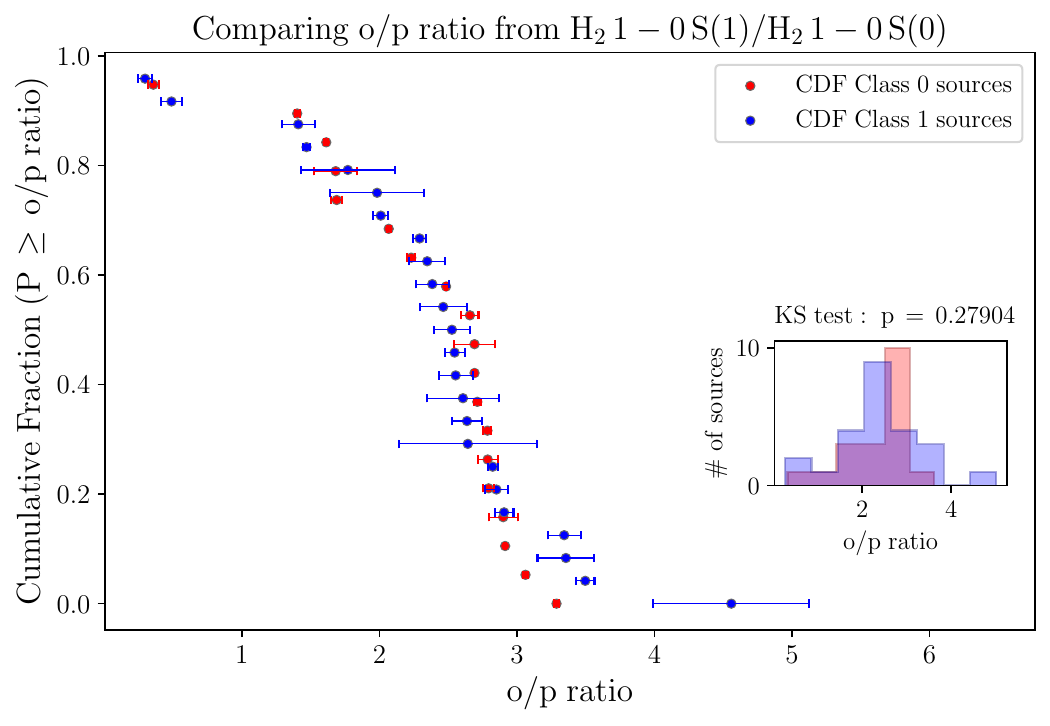}
\caption{\small Same as Figure \ref{fig:C0_C1_comp_H2_1-0_S1} for the H$_2$ $o/p$ ratio using the \HtonetoOSone to \HtonetoOSO ratio.}
\label{fig:C0_C1_comp_H_2_ratio_op}
\end{figure}

This line luminosity analysis is not completely accurate since the extinction values of the Class 0s are not strongly constrained. Given the expected depletion of the envelope with time, higher extinction values are expected in Class 0 objects. However, we must account for the non-uniformity of the distribution of extinction values of our Class 0 objects. We perform a series of 10$^4$ Monte Carlo simulations where each run implements a randomly-selected normal distribution of extinction values $A_v$ for the Class 0s centered on 50 mag with a standard deviation of 20 mag. For each run, we store the $p$-value of the log-rank and KS tests, and we present the cumulative luminosity distributions in Figure \ref{fig:C0_C1_comp_MC_simu_20_50} for the main emission lines we discuss. The distribution of $A_v$ extinction values of our sample of Class Is objects has a mean, median, and standard deviation of 23.3, 24.5, and 10.0, respectively. Thus we find it is reasonable to chose normal distributions peaking at $A_v$ of 50 mag for the Class 0 objects \citep{Greene2018,Laos2021}, given their embedded nature and the complexity of the inner envelope density structures (see also Figures \ref{fig:C0_C1_comp_MC_simu_20_30} and \ref{fig:C0_C1_comp_MC_simu_20_70} in Appendix \ref{app:add_comp_plots}, where the normal distributions are taken centered on 30 and 70 mag). With this approach, we find that the larger luminosity values of the Class 0 \Brgamma, \HtonetoOSO, and \HtonetoOSone lines are statistically different compared to the values encountered in Class I objects. \COfirstband, \HttwotooneSone, and \HtonetoOStwo suffer from a large number of non-detections in the Class I sample, which makes the comparison tests less straightforward and likely less accurate. The distributions of \HttwotooneSone luminosities appear statistically distinct in the Class 0s versus the Class Is considering the KS test, but only marginally different with the log-rank test. The distributions of \COfirstband luminosities appears to be marginally different between Class 0s and Is. Finally, the distribution of \HtonetoOStwo luminosities does not present any statistical difference between Class 0s and Is.

We also compare the distribution of \HtonetoOSone to \HttwotooneSone line luminosity ratio in Figure \ref{fig:C0_C1_comp_H_2_ratio} using the extinction values of Section \ref{sec:results_H2} for the Class 0s. 
Class 0s clearly exhibit very similar \HtonetoOSone/\HttwotooneSone luminosity ratio, compared to the Class Is.
The KS test suggests that one cannot differentiate between the two distributions.
We note that this ratio is less sensitive to the extinction, and that our results (\ie this H$_2$ line ratio being similar Class 0s and in Class Is) still holds applying a distribution of extinction for the Class 0s as we do above. We also compute the distribution of ortho:para ratios of H$_2$ using the \HtonetoOSone to \HtonetoOSO line ratio in the Class I sample, which is compared to the Class 0 results in Figure \ref{fig:C0_C1_comp_H_2_ratio_op}. We find that the mean ortho:para ratio for our Class I object sample is $\langle o/p\rangle \,=\, 2.42 \pm 0.9$. Again, we don't notice any differences between the two samples, as Class 0 objects have similar $ o/p$ ratio with a KS $p$-value of 0.46.

Finally, we do not seek to quantitatively compare the faint atomic emission lines detected in the Class 0 spectra, \ie [Fe II], Ca I, and Na I. The \citet{Doppmann2005} and \citet{Greene2010} NIRSPEC spectra do not necessarily have the wavelength coverage to allow these detections, and when it does these atomic lines are seen in absorption in most of the sources. Also, most of the \citet{Fiorellino2021} spectra of Class I would lack sensitivity to detect these atomic lines.

\section{Discussion}
\label{sec:disc}

\subsection{Physical origin of the CO overtone and \Brgamma emission in Class 0/I objects}
\label{sec:disc_physical_origin_Co_Brgamma}

The spatial origin of the CO overtone and \Brgamma emitting areas have long been debated. The hydrogen ionization and excitation structure in winds of T-Tauri stars have been found to primarily correspond to neutral gas \citep{Natta1988,Hartmann1990}, conditions that can favor CO overtone and/or \Brgamma emission (\eg \citealt{Carr1989}). 
The analysis and modeling of the line profiles have then revealed that \Brgamma could be more consistent with infalling gas within the innermost regions of circumstellar disks pointing toward a magnetospheric accretion origin \citep{Najita1996a,Muzerolle1998,Hartmann2016}. 
Because a wide range of excitation conditions can contribute to the overtone emission, the spatial emission of CO seems to be less constrained, and different modeling attempts explored contributions from a neutral wind \citep{Carr1989,Chandler1995}, or the inner edge of a heated accretion disk atmosphere \citep{Calvet1991,Carr1993,Glassgold2004,Aspin2007,Berthoud2007}. More recently, several NIR interferometric observations constrained the spatial scale of the CO overtone and \Brgamma emitting area in several T Tauri and Herbig Ae/Be stars \citep{Tatulli2008,Eisner2014,CarattioGaratti2020,Koutoulaki2021,Soulain2023}. These studies found that the \Brgamma emission can be consistent with magnetospheric accretion, but in several Herbig Ae/Be stars the emitting region extends well beyond the magnetospheric accretion region (located typically below 0.1 au), where the line profiles suggest an additional contribution from material in Keplerian rotation at the base of the disk wind and/or from a inner hot gaseous disk \citep{CarattioGaratti2015b,Mendigutia2015,Hone2019}. CO overtone emission is generally attributed to a hot ($T\,\sim\,2000-3000$ K) and dense ($N_\textrm{CO}\,\sim\,10^{21}-10^{22}$ cm$^{-2}$) inner disk inside the dust sublimation radius (\eg \citealt{Ilee2014,Lyo2017,Poorta2023}).

In our sample of low-mass Class 0 protostellar objects, most of the sources are consistent with little or no CO velocity shift, which is expected if the CO overtone emitting regions are homogeneously distributed along the inner edges of a rotation disk, heated by the accretion luminosity that must be high enough to heat the inner disk neutral material to temperatures $\geq$ 2000K. This would suggest that the episodes of high accretion activity probed by the CO overtone emission is disk mediated. However, the CO emission is redshifted by at least 20 km s$^{-1}$ in a few Class 0s sources, suggesting that CO emission can also trace an accretion of high column density infalling gas, in a case of a highly inclined source (\ie close to pole-on projection). This has been proposed in MHD simulations of embedded disks where accretion onto the proto-stellar embryo can directly happen from envelope streams of infalling material \citep{LeeYN2021}.

\begin{figure}[!tbh]
\centering
\includegraphics[scale=0.52,clip,trim= 0cm 0cm 0cm 0cm]{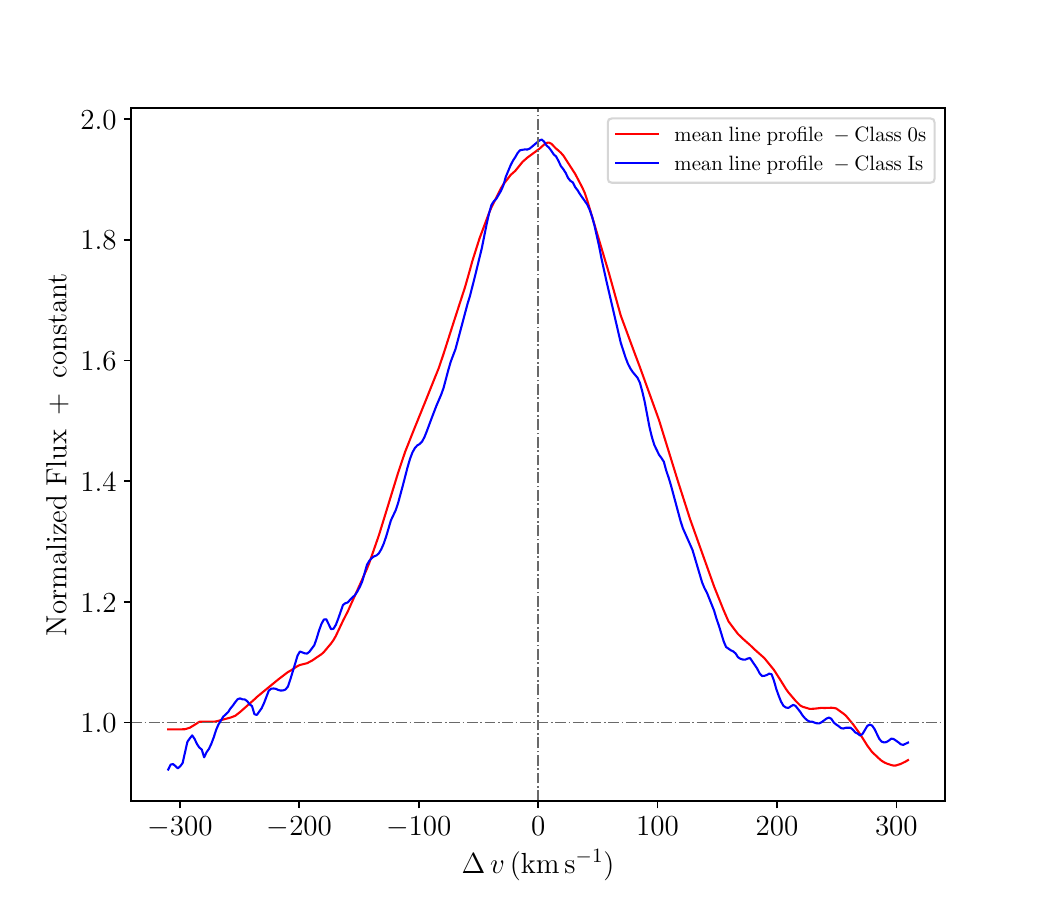}
\caption{\small Mean \Brgamma profile in the Class I sample, and in the Class 0 sample. The profiles were normalized, and linearly scaled between 1.0 and 2.0.
}
\label{fig:Bry_profile}
\end{figure}

While \Brgamma is clearly blueshifted in Class Is, with velocity shift values corresponding to high velocity protostellar jets, \Brgamma does not exhibit a clearly preference of velocity shift in the Class 0s. Looking at the mean \Brgamma line profiles, no prominent self-absorption features are detected (see Figure \ref{fig:Bry_profile}). However, looking at the individual cases, a redshifted absorption feature is clearly seen in several Class I spectra, which sometimes can bias the determination of the line velocity shift from the Gaussian fit toward artificial blueshifted values. None of the Class 0 spectra show strong absorption, but this diagnostic is limited by the lower spectral resolution of the MOSFIRE observations. \citet{Edwards1994,Hartmann1994,Najita1996a} argued that infalling gas in a magnetosphere is characterized by redshifted absorptions and blueshifted emission centroids of the hydrogen Balmer series, in sources with high inclinations, producing similar \Brgamma line profiles compared to the Class I spectra used in our study. 
This suggests that the \Brgamma emission mechanism is similar in the Class Is. The Class 0 objects do not exhibit a blueshifted distribution of \Brgamma profiles, but have very similar FWHM compared to Class Is. The fact that Class 0s need to have an inclination close to pole-on for the scattered light in the blueshifted cavity to be detected can explain the observed lack of blueshifted \Brgamma emission if it does not originate in a jet. This velocity shift analysis means that we cannot conclude that the \Brgamma emission emanates from a different region compared to more evolved objects. In addition, no one-to-one correlation is seen between the velocity shifts of the \Brgamma lines with the jet tracers [Fe\small{II}] lines in our Class 0 spectra, suggesting that \Brgamma may not have a jet origin.

If the blueshifted \Brgamma emission seen in the Class I sources has a jet origin, the direct link between the \Brgamma luminosity and the accretion luminosity in protostars launching jets may not be straightforward (\eg see \citealt{Beck2010}). Jets and stellar winds are linked to accretion activity, but the accretion luminosity may not be directly linked with the physical conditions responsible for the \Brgamma if it emanates from the jet. The EW and luminosity of \Brgamma is well correlated with the CO overtone emission (see Figure \ref{fig:C0_C1_Bry_CO}). This suggests these two lines share the same physical link to the current accretion activity. However, the correlation between H$_2$ and \Brgamma line EW is much poorer (see Figure \ref{fig:H2_Bry}), both in Class 0s and Is, suggesting that the link between \Brgamma and ejection is less strong. In addition, the large differences in the FWHM of \Brgamma and CO overtone emission profiles in both Class 0s and Is agree with the picture of CO originating from an inner disk, and \Brgamma from an accretion funnel flow.

\subsection{Intrinsic differences between the accretion and ejection properties of Class 0 and Class I protostars}
\label{sec:disc_accr_diff_C0_CIs}

Seeing on average larger CO overtone and \Brgamma luminosity values in our Class 0 protostars compared to the Class Is in the statistical analysis of Section \ref{sec:stats_comp} suggests that the accretion luminosity is on average higher in the Class 0 phase. 
Given our lack of quantitative constraints on the protostellar embryo parameters (mass and radius) and inner disk radius, we cannot derive mass accretion rates for our sample of protostars.
However, as we expect a somewhat larger radius of the protostellar embryo (\eg \citealt{Greene2018} derived a low surface gravity for S86N and proposed a large protostellar radius), and a smaller inner disk radius (in order to explain the higher CO overtone emission detection rate in the Class 0 phase) in the Class 0 phase, a prototypical larger accretion luminosity would translate into an average larger mass accretion rate in the Class 0 phase (given the dependence of the accretion luminosity on these parameters; \citealt{Gullbring1998}).
In addition, while the detection rate of \Brgamma emission is similar between Class 0s and Is, the much higher detection rate of CO overtone emission suggests that episodes of high accretion activity is more frequent in the Class 0 phase (see also the recent survey from \citealt{Itrich2023}, who detected CO overtone emission in 12.5\% of their Class Is in CMa-$l$224). Detecting photospheres (and thus sources with low-veiling indicative of small mass accretion rate; \citealt{Connelley2010}) much more frequently in the Class I sample supports this hypothesis.

Molecular hydrogen is also brighter, more luminous, and detected more frequently in the Class 0 objects. This likely implies stronger shocks, \ie large pre-shock density and shock velocity. However, it is interesting to note that the H$_2$ excitation and distribution of $o/p$ ratio appear to be similar in the two samples. The higher accretion luminosity would produce more UV, that could in theory affect the properties of the shocks in the base of outflow cavities. However, such photons are easily extincted, rapidly resulting in IR-FIR irradiation counterpart that have much less effect on the shock properties. Detection of UV driven chemistry, like via the dissociation of water into OH \citep{Tabone2021}, would a stronger probe of the accretion-induced UV field in protostars. X-rays could also represent a counterpart in the excitation of H$_2$, but constraints on the ionization of the wind/jet system are required to quantity this. In our sample, only one source (Per emb 8) has been detected by Chandra in \citet{Wang2016}. However, this may just be a detection/extinction effect. \citet{Grosso2020} also reported a X-ray flare for the HOPS 383, probing an intense magnetic activity in the vicinity of this Class 0. Detailed analysis of the NIR and MIR targeted by JWST shall further constrain the physical conditions of the gas toward the base of protostellar cavities.

\begin{figure*}[!tbh]
\centering
\subfigure{\includegraphics[scale=0.4,clip,trim= 0cm 0cm 0cm 0cm]{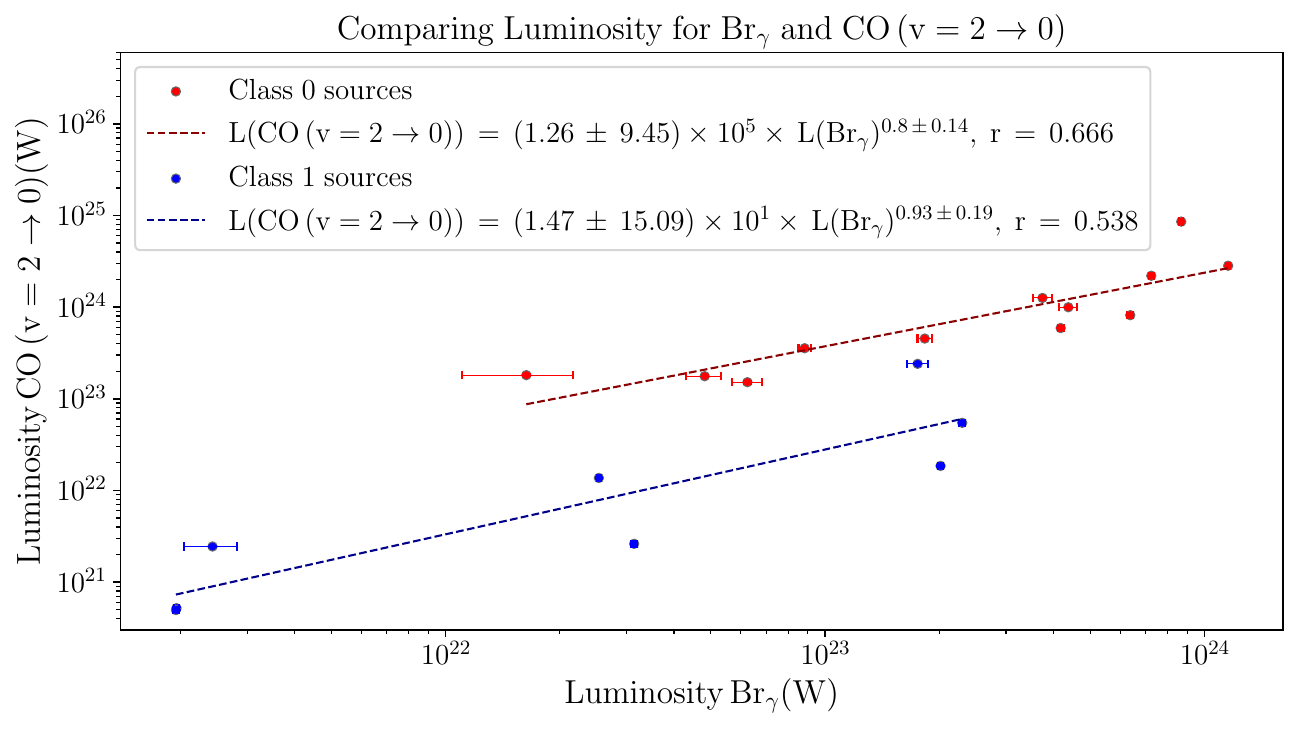}}
\subfigure{\includegraphics[scale=0.4,clip,trim= 0cm 0cm 0cm 0cm]{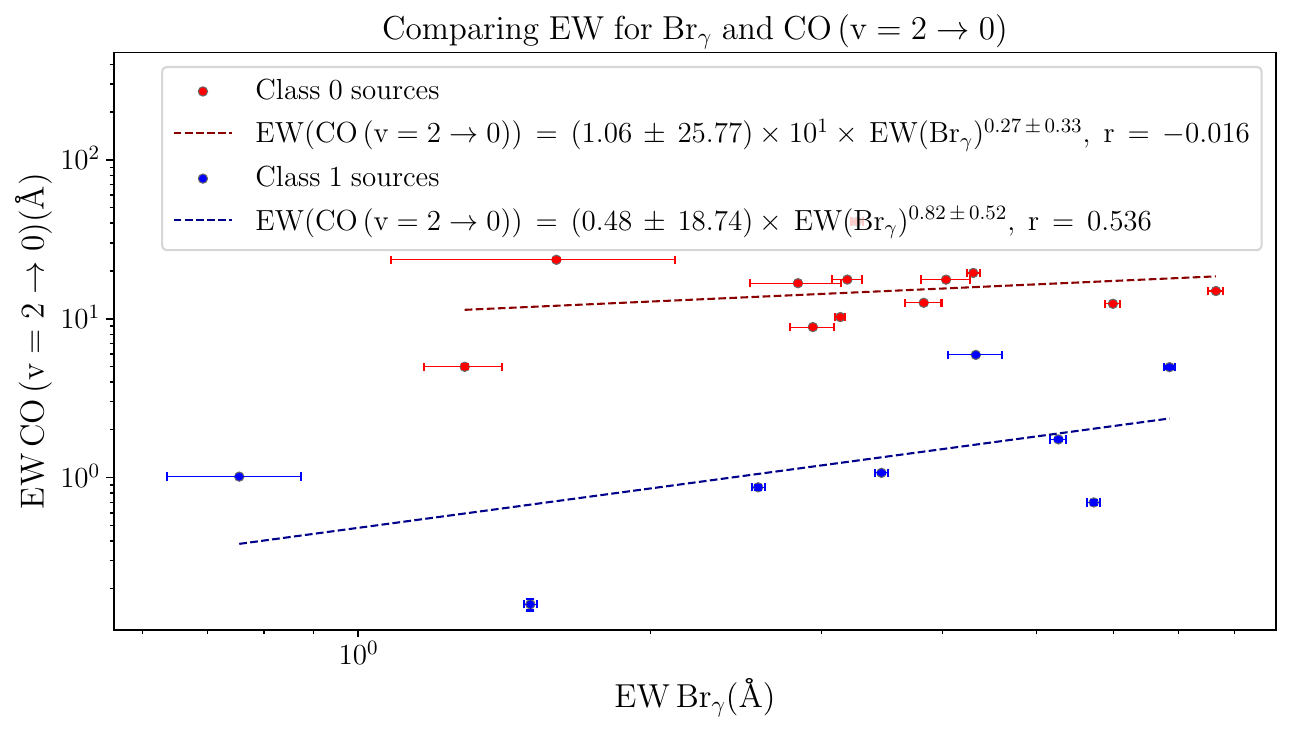}}
\caption{\small 2D correlations of \COfirstband versus \Brgamma emission lines. The left panel compares the EWs, and the right panel compares the luminosities. A power law (dash lines) is fitted in each case. The Pearson r coefficient is given each time. While the distribution of EW and luminosities values appear to differ between Class Is and Class 0s, the correlation parameters seem to be similar between the two samples.}
\label{fig:C0_C1_Bry_CO}
\end{figure*}

\begin{figure*}[!tbh]
\centering
\subfigure{\includegraphics[scale=0.4,clip,trim= 0cm 0cm 0cm 0cm]{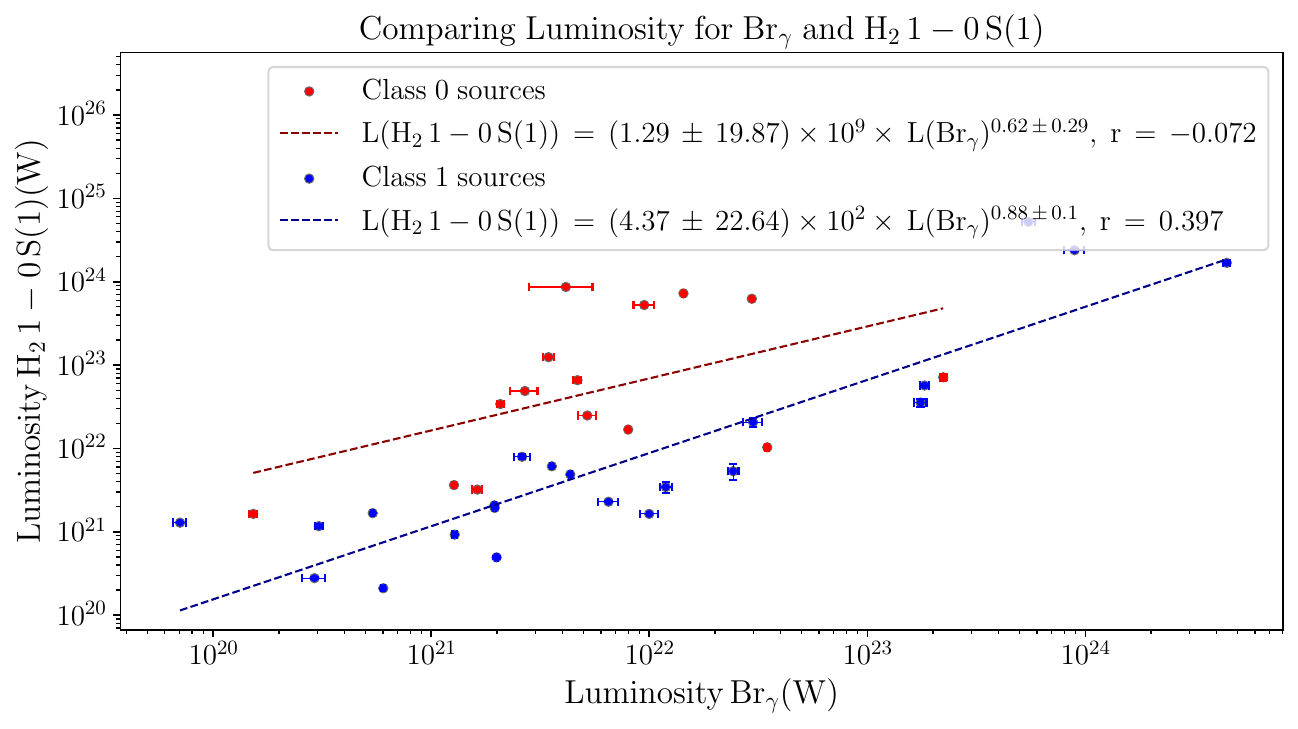}}
\subfigure{\includegraphics[scale=0.4,clip,trim= 0cm 0cm 0cm 0cm]{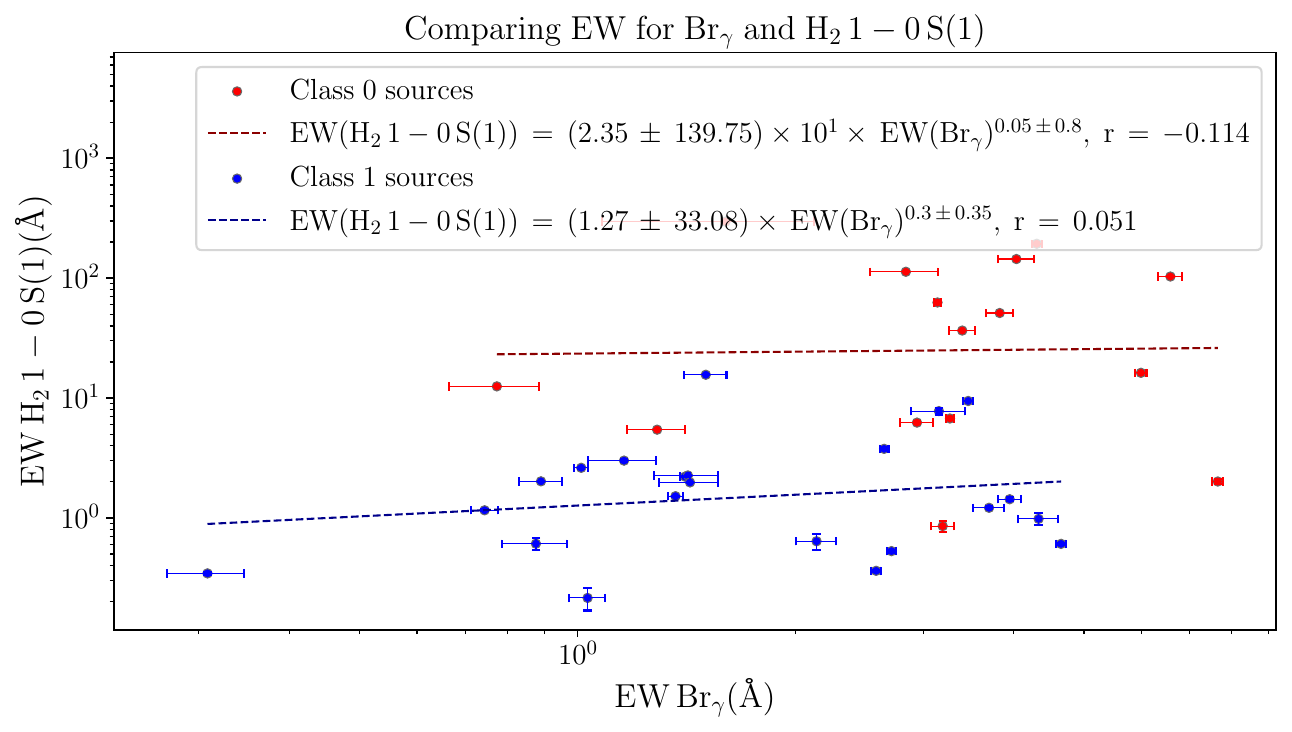}}
\caption{\small Same as Figure \ref{fig:C0_C1_Bry_CO} for \HtonetoOSone versus \Brgamma.}
\label{fig:H2_Bry}
\end{figure*}

The \Brgamma versus CO overtone emission plots point toward a continuous evolution between Class 0s and Is, but obviously shifted to higher values of EW and luminosities in the Class 0s (see Figure \ref{fig:C0_C1_Bry_CO}). We note that the correlation between CO overtone and \Brgamma emission is similar in Class 0s and Is. The statistical differences we see in this study suggests that Class 0 as identified in our sample correspond to a truly different evolutionary stage than Class I/FS, where the accretion properties are different.

\subsection{Which accretion mode for Class 0 objects?}
\label{sec:disc_C0_accretion}

A variety of explanations have been proposed to explain the origin of CO overtone emission in YSOs, including circumstellar inner disks, stellar or disk winds, magnetic accretion mechanisms such as funnel flows, inner disk instabilities. Time variability of the CO overtone emission is expected with any of these models (see review by \citealt{Fischer2023}). A clear dichotomy is seen between the CO overtone and \Brgamma emitting sources versus the photospheric spectra where no accretion tracers are detected. We can relate this to period of high accretion activity with strong emission lines, versus a quiescent phase where the surface gravity of the photosphere is low \citep{Greene2018}, due to thermal expansion of the protostellar embryo subsequently to the recent high accretion \citep{Baraffe2012}. However, an average higher mass accretion rate in the Class 0s would have implications on the time variability of the NIR accretion tracers. One would need to investigate the required magnetic field strength to sustain the accretion with the timescale of inner disk depletion and replenishment. The 50 \% of CO detections in emission tells us that on average Class 0s spend half of their lifetime in such high accretion mode, but we cannot determine what is the duration of such cycles, typically made of a high accretion episode and a quiescent phase where low veiling and the photosphere are seen. 
These Class 0 spectra are similar to ExOri-like systems (given the CO band and Na lines in emission). However, one would need different epochs to detect variability of the CO overtone emission, as it was previously seen in ExOri-like systems (\eg \citealt{Lorenzetti2009,Kospal2011}, and \citet{CarattioGaratti2017} for a high-mass outbursting system), thus determining the duration of the CO emission-related high accretion state. 
IR and sub-millimeter variability in Class 0/I systems \citep{Yoon2021,Yoon2022, Guo2020, LeeYH2021, Park2021, Zakri2022} has been attributed to variability in the accretion and ejection processes, exhibiting variability on timescales ranging from a few months to several years. However, the sub-millimeter emission is expected to be delayed and smoothed out with respect to the NIR activity \citep{Johnstone2013}. 

Archival WISE+NEOWISE NIR and MIR photometry of our sample of CO emitting Class 0s does not exhibit systematic time variations. However, we note that the W1 and W2 flux of Per emb 28 and HOPS 32 did increase by $\sim$ 1 mag within 6 years.
In addition, we note that among the 6 photospheric spectra we detect, Aqu MM4 and SerS MM16 seem to have a reported bolometric luminosity much higher than what it expected from the typical luminosity of a protostellar embryo itself (\ie 8.6 and 33.7 $L_\odot$, respectively). While, the beam of the sub-millimeter observations that computed $L_{\textrm{bol}}$ may encompassed several sources, this can suggest that at the time when the IR and submillimeter observations that allowed the computation of the bolometric luminosity where taken for these two Serpens sources (\ie in the 2000s for the Herschel and Spitzer surveys, and the IRAM 30m MAMBO survey by \citealt{Maury2011}), the accretion luminosity was significantly higher.

While variability of the \Brgamma line is commonly seen in Class Is and T-Tauri stars, having so many CO emitting sources in the Class 0 phase may point toward a different variability mechanisms. If ever observed in the future, CO overtone variability of absence thereof may be related to the embedded nature of such systems, where envelope infall would directly bring material to the inner region, explaining the presence of such (redshifted) dense and hot molecular gas.

It remains difficult to constrain the accreting scenario in which Class 0s would fit, \ie whether the inner disk is magnetized and turbulent enough to drive the accretion or if the accretion consists of direct envelope streams transiting through the highly magnetized disk upper layers. Disk winds can be expected to be active in the latter case as molecular outflows and jets are commonly detected in protostars. However, the CO overtone emission is consistent with redshifted emission, or emission consistent with no velocity shift. In the first case, we may exhibit this streamer scenario, while in the second case an inner disk origin centered on the source reference frame is more likely. However, whether the accretion is funneled by the protostellar embryo magnetic field remains an unsettled question in the Class 0 phase. The \Brgamma line in Class 0s is not shifted like it is in the Class Is, and the spectrally resolved \Brgamma line profiles do not show any hints of ubiquitous redshifted absorption in the Class 0 spectra, which is generally expected if \Brgamma were to trace an accretion column of a magnetospherically-accreting source. It is also interesting that the intense mass accretion rate expected in the Class 0 phase may easily crush the protostellar embryo magnetosphere. 
Using Equation 1. of \citet{Koenigl1991}, this would happen if the mass accretion rate exceeds $10^{-6}\beta_{0.5}^{7/2}B_{1.5}^{2}R_{1.5}^{5/2}\left(GM_{0.5}\right)^{-1/2}\,\textrm{M}_\odot \textrm{yr}^{-1}$, where the fiducial values are $\beta_{0.5}=0.5$, $B_{1.5} = 1.5$ kG (stellar magnetic field), $R_{1.5} = 1.5 \textrm{R}_\odot$ (stellar radius), $M_{0.5} = 0.5 \textrm{M}_\odot$ (stellar mass).
The mean mass accretion rate of low-mass Class Is is $\sim$ 10$^{-6} - 10^{-7}\,\textrm{M}_\odot \textrm{yr}^{-1}$ (using the results from  \citealt{Antoniucci2008,Watson2016,Fiorellino2021,Fiorellino2023}), and Class I protostellar embryo magnetic fields strength $\geq 2$ kG have been observed (\eg \citealt{JohnsKrull2009}). It is still unclear how early Class 0 protostellar embryos develop  dynamo-generated magnetic field, and if the primordial/fossil magnetic field could dominate the magnetic field budget. In their radiation hydrodynamic simulations, \citet{Bhandare2020} observed convection in the outer layers of the second hydrostatic cores,
arguing that early evolution of these objects may enable the generation of convective dynamo \citep{Chabrier2006}. Given the higher mass accretion rate encountered in Class 0s, and the nascent protostellar embryo characteristics (mass, magnetic field, radius), it is possible that the mass inflow would overcome the nascent magnetosphere (see also the discussion of \citealt{RodriguezAC2022,Cleaver2023}), which could greatly modify our HI recombination and CO overtone emission lines interpretation. 

\begin{figure}[!tbh]
\centering
\includegraphics[scale=0.4,clip,trim= 0cm 0cm 0cm 0cm]{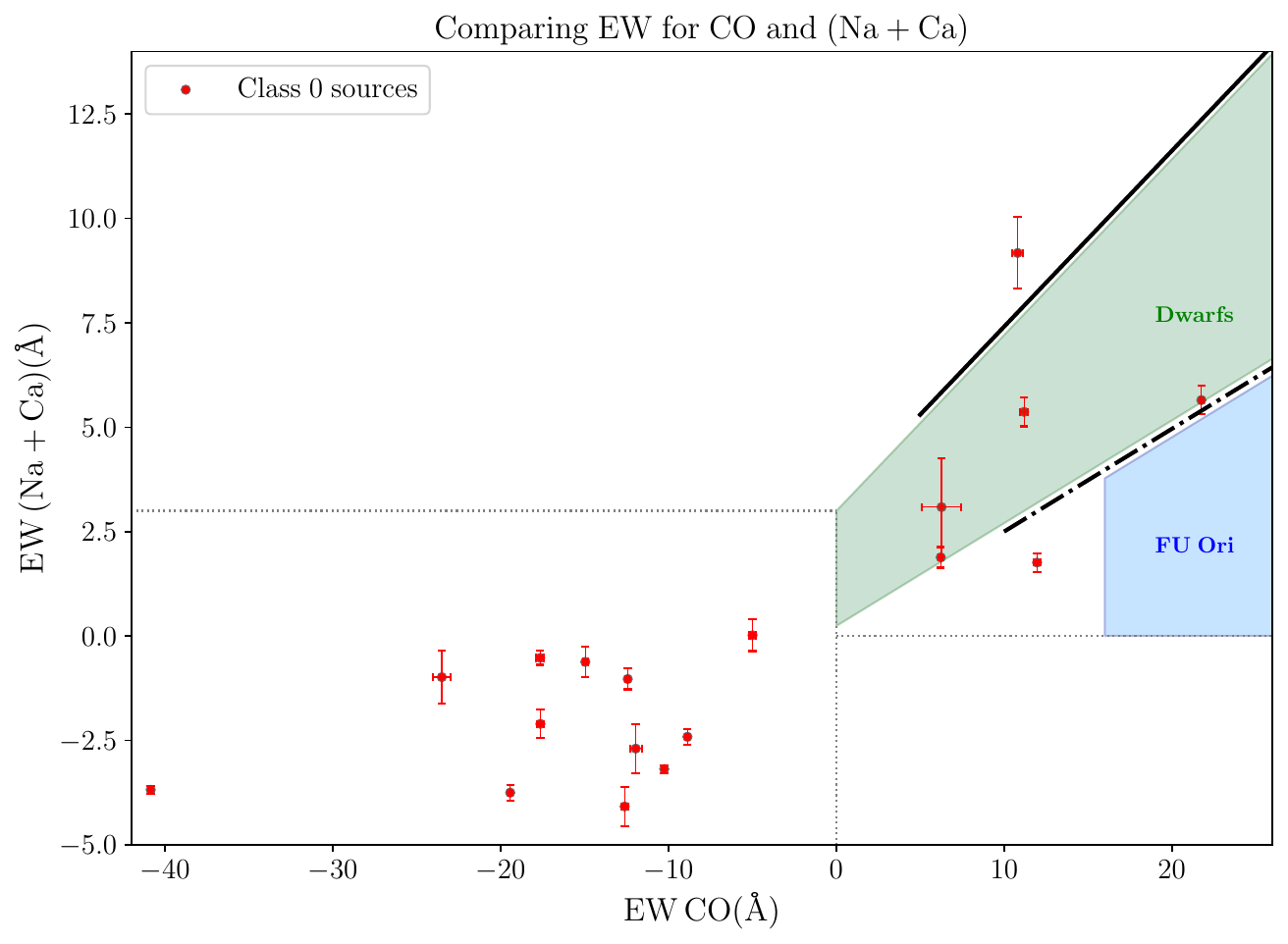}
\caption{\small EW of lines of Na (2.206 $\mu$m and 2.209 $\mu$m ) and Ca (2.263 $\mu$m and
2.266 $\mu$m) vs. the \COfirstband band. 
The Figure is adapted from the Figure 4 of \citet{Connelley2010} (and Figure 3 of \citealt{Greene1995}).
The solid dark line, and dot-dashed dark line, correspond to dwarf (luminosity Class V) stars and giant (luminosity Class III) stars, respectively (using the SpeX spectral library).
The green region thus corresponds to dwarf stars, while the blue region is consistent with FU Orionis type objects (see \citealt{Connelley2018}). The Class 0 photospheric spectra are most consistent with dwarf
stars. One source, HOPS 164, lies below the dwarf region of the diagram. However, its CO absorption is smaller than the typical CO absorption seen in FU Orionis-like stars with typical EW(CO) $\geq$ 20 \AA.}
\label{fig:_CO_Ca_Na}
\end{figure}

Interestingly, the six photospheric spectra do not have detected \Brgamma emission. These sources are either not undergoing strong accretion activity, or they might be consistent with the scenario mentioned above, \ie where a recent accretion outburst crushed the magnetosphere. In this later case, the \Brgamma emission-related processes are turned off, the veiling would be high, and deep CO absorptions would be seen, which characterizes FUor-type objects \citep{Connelley2010,Connelley2018}. Figure \ref{fig:_CO_Ca_Na} presents a 2D diagram of the EW of CO overtone vs. the EW of Na+Ca for our sample of Class 0 protostars. This figure shows that the equivalent widths of the photospheric spectra are consistent with dwarf spectra. We note that HOPS 164 has a strong CO absorption compared to Na I and Ca I lines, but its CO absorption is not deep enough (EW(CO) < 20 \AA) to be consistent FUor-type objects \citep{Connelley2018}.
Archival WISE+NEOWISE photometry do not see any evidence of increase or decrease of
flux with time for these 6 sources, while FUor-type objects usually show fading trends, but this is not exclusive. These 6 objects are also sub-luminous compared to typical FUor-type objects. Based on this, our results thus suggest that these photospheric spectra correspond to sources that are not undergoing episodes of strong accretion or ejection activity.

\subsection{The impact of protostellar environment}
\label{sec:disc_enviroment}

Several surveys of low-mass Class I objects \citep{Carr1989,Doppmann2005,Connelley2010,Fiorellino2021} intermediate-mass YSOs \citep{Ishii2001} and massive YSOs \citep{Martins2010,Cooper2013,Pomohaci2017,HsiehTH2021} reported 15-25\% of CO overtone emission detection rate. While the dissipation of the envelope reveals the heated inner disk and thus the CO overtone detection rate, the dissipation of the disk decreases the CO overtone emission probability. The proportion of CO emitters is thus thought to correlate with the evolution of YSOs \citep{Martins2010,RamirezTannus2017}. However, the external environment of protostellar envelopes and disks can also play a role in the accretion properties of these objects. In our sample of Class 0 objects, the Orion protostars (all located in the Orion A molecular cloud), which are not of systematically higher mass or higher luminosity compared to the other objects of the sample, show CO overtone emission in 66\% of cases, against 50\% in Perseus and 33 \% in Serpens-Aquila, which two regions forming lower mass stars compared to Orion. The feedback from the surrounding massive in the making may affect the accretion of YSOs in the cloud. \citet{Kryukova2012} showed that within the Orion molecular cloud, the luminosities of the protostars seems to depend on the local surface density of the YSOs, suggesting higher accretion in such environment. Future dedicated NIR observations of protostars in irradiated (or affected by a mechanical feedback induced by massive stars) regions shall further quantify the impact of the external environment on the accretion properties.

\section{Conclusions and summary}
\label{sec:ccl}

We present new observations of medium resolution (R $\sim$ 3300) NIR $K$-band spectroscopy for a sample of 26 Class 0 protostars in the Perseus, Serpens, and Orion molecular clouds. The H$_2$, \Brgamma, and CO overtone emission bands features are detected toward $\sim$ 90, 62, and 50 \% of the sample, respectively. We also detect several [Fe II] lines, a common jet tracer, as well as Ca I and Na I emission lines. The photosphere is detected in 6 sources, exhibiting CO, Na and Ca absorption features consistent with dwarf spectra, and indicative of low-veiling. To quantify these results further, we performed statistical comparisons with a sample of Class I $K$-band spectra taken from the literature. The main results and conclusions of this comparison are as follows:

\begin{enumerate}
    \item Analysis of the \Brgamma and CO overtone emission equivalent width shows that the accretion luminosity is on average higher in Class 0 than in Class I systems, suggesting that Class 0s accrete more vigorously. We reach the same conclusions using the line luminosity values, but difficulties in determining extinction make the accretion luminosities uncertain. Using Monte Carlo simulations, we find that the distribution of \Brgamma and CO overtone emission equivalent width and luminosity values are significantly higher in Class 0 than Class I protostars.
    \item The much higher detection rate of CO overtone in Class 0s (50 \% against $\leq$15 \% in Class Is) indicates that episodes of high accretion activity is more frequent in Class 0 systems.
    \item The H$_2$ excitation is similar between Class 0s and Is, pointing toward outflow/jet shock origin collisionally exciting H$_2$. However, the H$_2$ emission lines are much more luminous in Class 0s, suggesting stronger shocks, \ie larger shock velocity and/or pre-shock density.
    \item The Class 0 CO overtone emission is either consistent with redshifted emission or no shift. While the latter would suggest an inner disk origin, clear redshifted emission suggests that the dense and hot gas associated with the CO emitting regions can correspond to material infalling at $\sim$15-40 km s$^{-1}$ directly on the central region of the protostar.
    \item \Brgamma emission lines usually attributed to the magnetospheric accretion column in T-Tauri systems exhibit clearly different velocity profiles between the two protostellar phases: while Class Is systems have blueshifted centroids and in a few sources a clear redshifted absorption, Class 0s exhibit symmetric \Brgamma profiles with no velocity shifts. This could point toward an accretion mechanism of different nature in Class 0 systems.
\end{enumerate}

This study reveals the NIR accretion properties of a statistically relevant sample of Class 0 protostellar systems. These objects exhibit systematic difference with the more evolved Class I objects: Class 0 accretion appears more vigorous and episodes of high accretion activity more frequent. Further modeling of the emission could reveal relevant further insights on the kinematics of the CO and \Brgamma emission line regions. JWST diagnostics of the spatial location of these emissions and the extinction of these embedded regions could enable more quantitative analysis of the actual mass accretion rate of these deeply embedded systems.

\begin{acknowledgments}

The authors thank Curtis Dewitt, Marion Villenave, Kyle Kaplan, and Sylvie Cabrit for helpfull discussions about the analysis of NIR lines and the statistical analysis. The authors also thank Eleonora Fiorellino for sharing the VLT KMOS of their NGC1333 Class I data with us. We also thank Stefan Laos for providing the Keck MOSFIRE data of protostars observed in 2019. 
Some of the data presented herein were obtained at the W. M. Keck Observatory, which is operated as a scientific partnership among the California Institute of Technology, the University of California and the National Aeronautics and Space Administration. The Observatory was made possible by the generous financial support of the W. M. Keck Foundation.
The authors wish to recognize and acknowledge the very significant cultural role and reverence that the summit of Maunakea has always had within the indigenous Hawaiian community.  We are most fortunate to have the opportunity to conduct observations from this mountain.
\end{acknowledgments}

\textit{Facilities:} Keck:I (MOSFIRE), VLT.


\textit{Software:}  Astropy \citep{Astropy2018}, Specutils, Lifeline, matplotlib \citep{Hunter2007}, numpy \citep{vanderWalt2011,Harris2020}, panda, scipy \citep{Jones2001}, emcee \citep{ForemanMackey2013}, Spectres \citep{Carnall2017} python packages. PypeIt \citet{Prochaska2020}, DS9 \citep{Joye2003}, CDS, Vizier, Simbdad softwares.

\bibliography{ms}
\bibliographystyle{apj}

\newpage
\clearpage
\newpage

\appendix
\addcontentsline{toc}{section}{Appendix}
\renewcommand{\thesection}{\Alph{section}}

\section{\normalfont{ Imperfect telluric line correction}}
\label{app:tell_line}

\begin{figure*}[!tbh]
\centering
\includegraphics[scale=0.4,clip,trim= 4cm 0.5cm 4cm 0cm]{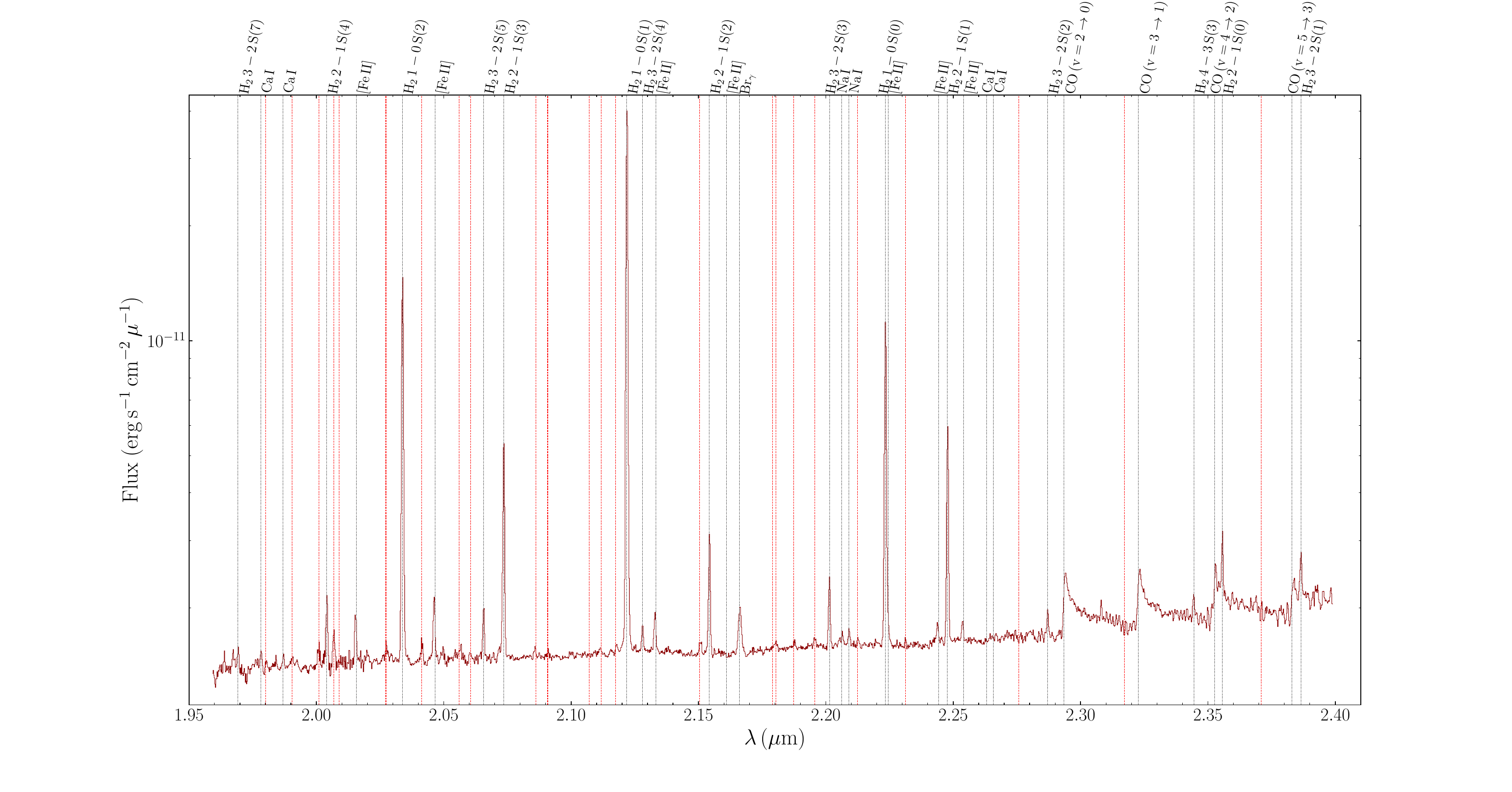}
\caption{\small Same as Figure \ref{fig:ex_spec_hops250} top panel, but the red vertical dashed lines mark telluric lines that are not perfectly subtracted by the PypeIt pipeline, resulting in small spikes in the 1D extracted spectra.
}
\label{fig:tell_lines_hops250}
\end{figure*}

Figure \ref{fig:tell_lines_hops250} shows the same spectra presented in Figure \ref{fig:ex_spec_hops250} with the locations of the main telluric lines overlaid. The strongest lines that we identified in the raw 2D detector files are not perfectly subtracted by the pipeline, such that residual spikes can remain in the final co-added 1D spectra. We thus carefully check for these residual telluric lines when performing the line identification in the Class 0 $K$-band spectra.

\section{\normalfont{Class 0 Sample Spectral Line parameters }}
\label{app:line_param}

Tables \ref{t.C0_line_obs1}, \ref{t.C0_line_obs2}, \ref{t.C0_line_obs3} present the line parameters results derived from the analysis of the Class 0 $K$-band spectra (see Section \ref{sec:analysis_tech}) for \HtonetoOStwo,\HtonetoOSone, \Brgamma, \HtonetoOSO, \HttwotooneSone (Table \ref{t.C0_line_obs1}), \HttwotooneSfour, \HtthreetotwoSfive, \HttwotooneSthree, \HtthreetotwoSfour, \HttwotooneStwo (Table \ref{t.C0_line_obs2}), and \HtthreetotwoSthree, \HtthreetotwoStwo, \HtfourtothreeSthree, \HttwotooneSO, \HtthreetotwoSone (Table \ref{t.C0_line_obs3}).

\begin{sidewaystable}[h!]
\centering
\begin{minipage}{\textwidth}
\scriptsize
\caption[]{Line parameters}
\label{t.C0_line_obs1}
\setlength{\tabcolsep}{0.08em} 
\begin{tabular}{p{0.08\linewidth}cccccccccccccccc}
\hline \hline \noalign{\smallskip}
      & \multicolumn{3}{c}{$\rm  H_{2}\,1-0\,S(2) $} & \multicolumn{3}{c}{$\rm  H_{2}\,1-0\,S(1) $} & \multicolumn{3}{c}{$\rm Br_{\gamma}$} & \multicolumn{3}{c}{$\rm  H_{2}\,1-0\,S(0) $} & \multicolumn{3}{c}{$\rm  H_{2}\,2-1\,S(1) $}\\ 
  & EW & Flux & FWHM  & EW & Flux & FWHM  & EW & Flux & FWHM  & EW & Flux & FWHM  & EW & Flux & FWHM \\ 
  & \AA&$ 10^{-19}\,\textrm{W}\,\textrm{m}^{-2}$&$\textrm{km}\,\textrm{s}^{-1}$ & \AA&$ 10^{-19}\,\textrm{W}\,\textrm{m}^{-2}$&$\textrm{km}\,\textrm{s}^{-1}$ & \AA&$ 10^{-19}\,\textrm{W}\,\textrm{m}^{-2}$&$\textrm{km}\,\textrm{s}^{-1}$ & \AA&$ 10^{-19}\,\textrm{W}\,\textrm{m}^{-2}$&$\textrm{km}\,\textrm{s}^{-1}$ & \AA&$ 10^{-19}\,\textrm{W}\,\textrm{m}^{-2}$&$\textrm{km}\,\textrm{s}^{-1}$\\ 
\noalign{\smallskip}  \hline \noalign{\smallskip} 
HOPS 50&-6.0$\pm$0.2&7.2$\pm$0.2&66.2$\pm$4.0&-16.2$\pm$0.1&19.8$\pm$0.1&55.3$\pm$3.1&-6.0$\pm$0.1&6.9$\pm$0.1&218.6$\pm$7.0&-7.3$\pm$0.1&8.5$\pm$0.1&59.4$\pm$4.1&-1.7$\pm$0.1&2.0$\pm$0.1&---\\ 
\noalign{\smallskip} 
HOPS 60&-25.3$\pm$0.2&77.9$\pm$0.7&106.5$\pm$7.7&-62.7$\pm$0.2&242.4$\pm$0.6&78.5$\pm$2.6&-3.1$\pm$0.0&12.5$\pm$0.1&179.0$\pm$6.5&-12.7$\pm$0.1&57.6$\pm$0.3&87.4$\pm$3.9&-3.7$\pm$0.1&17.4$\pm$0.3&55.1$\pm$6.2\\ 
\noalign{\smallskip} 
HOPS 87&---&47.2$\pm$0.5&64.3$\pm$0.5&---&140.5$\pm$0.5&61.3$\pm$2.8&$\leq $-58.4&$\leq $0.1&---&---&40.7$\pm$0.3&9.8$\pm$2.6&---&10.1$\pm$0.1&23.1$\pm$3.4\\ 
\noalign{\smallskip} 
HOPS 164&-29.7$\pm$0.5&8.1$\pm$0.1&71.9$\pm$2.1&-66.2$\pm$0.2&21.2$\pm$0.1&67.3$\pm$3.5&$\leq $0.5&$\leq $0.2&---&-15.0$\pm$0.2&5.2$\pm$0.1&60.9$\pm$4.7&-5.2$\pm$0.1&1.9$\pm$0.0&45.8$\pm$3.1\\ 
\noalign{\smallskip} 
HOPS 171&-1.6$\pm$0.4&0.8$\pm$0.2&83.9$\pm$22.0&-2.0$\pm$0.1&1.2$\pm$0.1&43.3$\pm$12.5&-7.7$\pm$0.1&4.7$\pm$0.1&181.5$\pm$4.6&$\leq $0.8&$\leq $0.5&---&$\leq $0.2&$\leq $0.2&---\\ 
\noalign{\smallskip} 
HOPS 203&-58.7$\pm$0.8&11.2$\pm$0.2&68.6$\pm$3.5&-113.1$\pm$0.6&25.9$\pm$0.1&40.0$\pm$6.1&-2.8$\pm$0.3&0.5$\pm$0.1&206.9$\pm$31.7&-61.9$\pm$0.4&13.6$\pm$0.1&82.7$\pm$10.0&-9.4$\pm$0.2&2.1$\pm$0.0&27.1$\pm$5.3\\ 
\noalign{\smallskip} 
HOPS 250&-73.0$\pm$0.2&106.5$\pm$0.3&63.2$\pm$1.8&-193.6$\pm$0.2&307.0$\pm$0.3&46.7$\pm$3.0&-4.3$\pm$0.1&6.6$\pm$0.1&153.6$\pm$4.9&-46.7$\pm$0.1&75.2$\pm$0.2&37.7$\pm$2.8&-19.7$\pm$0.1&32.4$\pm$0.1&24.3$\pm$2.7\\ 
\noalign{\smallskip} 
Per emb 24&-15.3$\pm$0.4&4.6$\pm$0.1&68.9$\pm$3.0&-36.6$\pm$0.2&12.2$\pm$0.1&56.2$\pm$4.3&-3.4$\pm$0.1&1.1$\pm$0.0&186.6$\pm$12.0&-9.6$\pm$0.1&3.3$\pm$0.0&49.3$\pm$6.4&-3.7$\pm$0.1&1.4$\pm$0.0&30.0$\pm$3.9\\ 
\noalign{\smallskip} 
Ser SMM3&-17.9$\pm$0.8&5.3$\pm$0.2&82.5$\pm$4.1&-42.1$\pm$0.4&14.1$\pm$0.1&80.9$\pm$3.5&$\leq $0.7&$\leq $0.3&---&-8.9$\pm$0.3&3.2$\pm$0.1&79.2$\pm$6.8&-4.6$\pm$0.3&1.7$\pm$0.1&73.6$\pm$9.1\\ 
\noalign{\smallskip} 
Aqu MM4&$\leq $2.9&$\leq $0.4&---&$\leq $0.8&$\leq $0.2&---&$\leq $0.4&$\leq $0.1&---&$\leq $0.4&$\leq $0.2&---&$\leq $0.3&$\leq $0.1&---\\ 
\noalign{\smallskip} 
S68N & ---  & ---  & --- &-9.0$\pm$0.6&0.0$\pm$0.0&---&$\leq $1.9&$\leq $0.0&---&$\leq $1.6&$\leq $0.0&---&$\leq $2.1&$\leq $0.0&---\\ 
\noalign{\smallskip} 
HOPS 32&-52.0$\pm$0.1&44.2$\pm$0.1&---&-144.4$\pm$0.1&140.0$\pm$0.1&13.5$\pm$7.5&-4.0$\pm$0.2&4.1$\pm$0.2&230.5$\pm$15.9&-24.9$\pm$0.1&28.7$\pm$0.2&---&-11.4$\pm$0.1&13.7$\pm$0.1&---\\ 
\noalign{\smallskip} 
HOPS 44&-12.6$\pm$0.7&1.5$\pm$0.1&80.9$\pm$7.0&-7.3$\pm$0.9&1.2$\pm$0.1&242.1$\pm$33.5&$\leq $1.8&$\leq $0.4&---&$\leq $1.6&$\leq $0.4&---&$\leq $0.9&$\leq $0.2&---\\ 
\noalign{\smallskip} 
Per emb 8&-62.7$\pm$0.2&20.4$\pm$0.1&110.7$\pm$10.8&-103.2$\pm$0.3&39.9$\pm$0.1&118.3$\pm$12.2&-6.6$\pm$0.2&2.6$\pm$0.1&187.6$\pm$8.2&-31.8$\pm$0.2&13.5$\pm$0.1&100.9$\pm$11.1&-9.2$\pm$0.1&4.1$\pm$0.1&91.1$\pm$11.8\\ 
\noalign{\smallskip} 
Per emb 25&-2.0$\pm$0.1&0.8$\pm$0.0&141.7$\pm$10.5&-0.9$\pm$0.1&0.4$\pm$0.0&39.9$\pm$11.4&-3.2$\pm$0.1&1.7$\pm$0.1&232.1$\pm$9.6&-2.2$\pm$0.1&1.2$\pm$0.0&111.1$\pm$9.4&$\leq $0.2&$\leq $0.1&---\\ 
\noalign{\smallskip} 
Per emb 26&-133.1$\pm$0.4&16.2$\pm$0.1&90.5$\pm$7.5&-296.2$\pm$0.4&55.8$\pm$0.1&88.7$\pm$8.6&-1.6$\pm$0.5&0.3$\pm$0.1&244.9$\pm$87.4&-76.8$\pm$0.3&17.9$\pm$0.1&68.3$\pm$10.4&-25.7$\pm$0.4&5.8$\pm$0.1&59.6$\pm$8.9\\ 
\noalign{\smallskip} 
Per emb 28&-4.0$\pm$0.1&8.3$\pm$0.2&100.5$\pm$6.1&-6.2$\pm$0.1&16.3$\pm$0.3&67.8$\pm$12.0&-2.9$\pm$0.2&8.4$\pm$0.4&232.5$\pm$24.2&-1.2$\pm$0.1&3.8$\pm$0.2&24.3$\pm$5.0&-0.3$\pm$0.1&1.0$\pm$0.2&---\\ 
\noalign{\smallskip} 
Per emb 21&---&9.0$\pm$0.0&70.0$\pm$1.6&---&49.1$\pm$0.0&51.0$\pm$1.6&$\leq $35.6&$\leq $0.1&---&---&13.3$\pm$0.0&23.1$\pm$4.4&---&5.0$\pm$0.0&31.1$\pm$1.5\\ 
\noalign{\smallskip} 
Aqu MM11&-4.9$\pm$0.2&1.6$\pm$0.1&49.2$\pm$4.0&-12.5$\pm$0.1&4.9$\pm$0.0&---&-0.8$\pm$0.1&0.3$\pm$0.0&161.8$\pm$23.8&-3.2$\pm$0.1&1.4$\pm$0.0&---&-1.0$\pm$0.1&0.4$\pm$0.0&---\\ 
\noalign{\smallskip} 
Ceph mm&-2.2$\pm$0.1&3.6$\pm$0.2&34.4$\pm$5.5&-5.0$\pm$0.0&9.5$\pm$0.1&---&$\leq $0.2&$\leq $0.3&---&-3.2$\pm$0.1&6.3$\pm$0.1&68.7$\pm$9.4&-0.3$\pm$0.0&0.6$\pm$0.1&---\\ 
\noalign{\smallskip} 
Ser emb 2&-3.6$\pm$0.3&1.1$\pm$0.1&---&-5.4$\pm$0.1&1.8$\pm$0.0&---&-1.3$\pm$0.1&0.4$\pm$0.0&100.0$\pm$20.8&$\leq $0.4&$\leq $0.1&---&-0.6$\pm$0.1&0.2$\pm$0.0&---\\ 
\noalign{\smallskip} 
Ser emb 15&-22.1$\pm$0.4&5.6$\pm$0.1&86.4$\pm$4.6&-51.2$\pm$0.2&16.1$\pm$0.1&80.3$\pm$4.9&-3.8$\pm$0.2&1.2$\pm$0.1&147.7$\pm$10.7&-11.5$\pm$0.2&4.2$\pm$0.1&72.8$\pm$5.9&-4.5$\pm$0.2&1.7$\pm$0.1&45.1$\pm$5.6\\ 
\noalign{\smallskip} 
Aqu MM5&-3.2$\pm$0.1&5.5$\pm$0.2&67.5$\pm$6.7&-6.8$\pm$0.0&14.0$\pm$0.1&10.5$\pm$2.6&-3.3$\pm$0.0&7.0$\pm$0.1&72.7$\pm$4.7&-1.5$\pm$0.0&3.5$\pm$0.1&---&-0.4$\pm$0.0&1.1$\pm$0.1&20.7$\pm$22.4\\ 
\noalign{\smallskip} 
Aqu MM8&$\leq $0.8&$\leq $0.2&---&$\leq $0.3&$\leq $0.1&---&$\leq $0.5&$\leq $0.1&---&-1.2$\pm$0.1&0.3$\pm$0.0&---&$\leq $0.6&$\leq $0.2&---\\ 
\noalign{\smallskip} 
Ser emb 22&$\leq $0.7&$\leq $0.6&---&$\leq $0.2&$\leq $0.2&---&-5.9$\pm$0.1&10.6$\pm$0.1&213.3$\pm$5.0&$\leq $0.2&$\leq $0.4&---&-0.5$\pm$0.0&1.3$\pm$0.1&---\\ 
\noalign{\smallskip} 
SerS MM16&-3.4$\pm$0.6&0.5$\pm$0.1&60.0$\pm$11.5&-5.1$\pm$0.1&0.9$\pm$0.0&---&$\leq $0.7&$\leq $0.1&---&-2.3$\pm$0.2&0.4$\pm$0.0&49.9$\pm$15.6&$\leq $0.4&$\leq $0.1&---\\ 
\noalign{\smallskip} 
\hline
\smallskip
\end{tabular}
\tablecomments{\small Line parameters derived from the spectral analysis of the Class 0 \textit{Keck} NIR spectroscopic observations. Equivalent width (EW), line flux, and line FWHM (where the resolution of the instruments has been subtracted in quadrature) are shown for \HtonetoOStwo,\HtonetoOSone, \Brgamma, \HtonetoOSO, \HttwotooneSone. Upper limits are shown in case of non-detections. Dashes indicate the cases where a line is not covered by the observations, or if it was not possible to retrieve the line parameter (\ie the equivalent width for non-continuum detected sources).}
\end{minipage}
\end{sidewaystable}


\begin{sidewaystable}[h!]
\centering
\begin{minipage}{\textwidth}
\scriptsize
\caption[]{Line parameters}
\label{t.C0_line_obs2}
\setlength{\tabcolsep}{0.08em} 
\begin{tabular}{p{0.08\linewidth}cccccccccccccccc}
\hline \hline \noalign{\smallskip}
  & \multicolumn{3}{c}{$\rm  H_{2}\,2-1\,S(4) $} & \multicolumn{3}{c}{$\rm  H_{2}\,3-2\,S(5) $} & \multicolumn{3}{c}{$\rm  H_{2}\,2-1\,S(3) $} & \multicolumn{3}{c}{$\rm  H_{2}\,3-2\,S(4) $} & \multicolumn{3}{c}{$\rm  H_{2}\,2-1\,S(2) $}\\ 
  & EW & Flux & FWHM  & EW & Flux & FWHM  & EW & Flux & FWHM  & EW & Flux & FWHM  & EW & Flux & FWHM \\ 
  & \AA&$ 10^{-19}\,\textrm{W}\,\textrm{m}^{-2}$&$\textrm{km}\,\textrm{s}^{-1}$ & \AA&$ 10^{-19}\,\textrm{W}\,\textrm{m}^{-2}$&$\textrm{km}\,\textrm{s}^{-1}$ & \AA&$ 10^{-19}\,\textrm{W}\,\textrm{m}^{-2}$&$\textrm{km}\,\textrm{s}^{-1}$ & \AA&$ 10^{-19}\,\textrm{W}\,\textrm{m}^{-2}$&$\textrm{km}\,\textrm{s}^{-1}$ & \AA&$ 10^{-19}\,\textrm{W}\,\textrm{m}^{-2}$&$\textrm{km}\,\textrm{s}^{-1}$\\ 
\noalign{\smallskip}  \hline \noalign{\smallskip} 
HOPS 50&-1.2$\pm$0.3&1.4$\pm$0.4&---&-0.5$\pm$0.1&0.6$\pm$0.1&106.4$\pm$48.3&-1.9$\pm$0.2&2.3$\pm$0.3&67.6$\pm$9.7&-0.2$\pm$0.0&0.2$\pm$0.1&---&-0.9$\pm$0.2&1.1$\pm$0.2&87.9$\pm$28.9\\ 
\noalign{\smallskip} 
HOPS 60&-2.8$\pm$0.3&8.2$\pm$0.8&284.7$\pm$80.5&-0.7$\pm$0.0&2.3$\pm$0.1&77.4$\pm$41.5&-7.5$\pm$0.1&25.6$\pm$0.4&92.4$\pm$9.4&-0.9$\pm$0.0&3.3$\pm$0.2&211.3$\pm$13.3&-2.8$\pm$0.1&10.6$\pm$0.3&113.4$\pm$14.3\\ 
\noalign{\smallskip} 
HOPS 87&---&1.7$\pm$0.3&213.2$\pm$74.9&---&0.5$\pm$0.0&59.2$\pm$8.8&---&6.5$\pm$0.1&72.8$\pm$4.0&---&0.2$\pm$0.0&59.3$\pm$29.4&---&3.0$\pm$0.1&54.5$\pm$7.3\\ 
\noalign{\smallskip} 
HOPS 164&$\leq $3.9&$\leq $1.0&---&$\leq $0.5&$\leq $0.1&---&-6.9$\pm$0.6&2.0$\pm$0.2&87.1$\pm$13.6&$\leq $0.4&$\leq $0.1&---&-3.4$\pm$0.4&1.1$\pm$0.1&90.6$\pm$12.3\\ 
\noalign{\smallskip} 
HOPS 171&$\leq $1.2&$\leq $0.6&---&$\leq $0.4&$\leq $0.2&---&-1.0$\pm$0.3&0.6$\pm$0.2&---&$\leq $0.3&$\leq $0.2&---&$\leq $0.5&$\leq $0.3&---\\ 
\noalign{\smallskip} 
HOPS 203&$\leq $3.0&$\leq $0.6&---&$\leq $1.2&$\leq $0.2&---&-16.2$\pm$0.9&3.2$\pm$0.2&83.9$\pm$8.1&-1.4$\pm$0.3&0.3$\pm$0.1&241.9$\pm$100.7&-4.9$\pm$0.6&1.0$\pm$0.1&66.9$\pm$13.4\\ 
\noalign{\smallskip} 
HOPS 250&-4.0$\pm$0.3&5.7$\pm$0.4&66.7$\pm$9.6&-3.0$\pm$0.1&4.3$\pm$0.1&54.2$\pm$6.8&-20.5$\pm$0.1&30.8$\pm$0.2&46.0$\pm$2.5&-1.3$\pm$0.1&2.0$\pm$0.1&45.5$\pm$4.1&-8.3$\pm$0.1&12.6$\pm$0.1&47.5$\pm$3.6\\ 
\noalign{\smallskip} 
Per emb 24&-1.9$\pm$0.8&0.6$\pm$0.3&210.3$\pm$79.2&-0.5$\pm$0.1&0.2$\pm$0.0&---&-4.9$\pm$0.4&1.6$\pm$0.1&112.6$\pm$11.8&$\leq $0.3&$\leq $0.1&---&-1.9$\pm$0.3&0.6$\pm$0.1&64.7$\pm$13.1\\ 
\noalign{\smallskip} 
Ser SMM3&$\leq $3.2&$\leq $1.0&---&$\leq $0.8&$\leq $0.3&---&-4.8$\pm$0.9&1.5$\pm$0.3&---&$\leq $1.0&$\leq $0.3&---&-2.4$\pm$0.6&0.8$\pm$0.2&85.7$\pm$14.2\\ 
\noalign{\smallskip} 
Aqu MM4&$\leq $4.0&$\leq $0.4&---&$\leq $0.5&$\leq $0.1&---&$\leq $3.1&$\leq $0.5&---&$\leq $0.4&$\leq $0.1&---&-0.7$\pm$0.3&0.2$\pm$0.1&---\\ 
\noalign{\smallskip} 
S68N & ---  & ---  & ---  & ---  & ---  & --- &$\leq $1.2&$\leq $0.0&---&$\leq $1.2&$\leq $0.0&---&$\leq $1.6&$\leq $0.0&---\\ 
\noalign{\smallskip} 
HOPS 32&$\leq $1.1&$\leq $0.9&---&-2.6$\pm$0.2&2.3$\pm$0.2&30.8$\pm$14.5&-12.7$\pm$0.2&11.4$\pm$0.2&---&-1.0$\pm$0.1&0.9$\pm$0.1&---&-6.1$\pm$0.1&6.0$\pm$0.1&---\\ 
\noalign{\smallskip} 
HOPS 44&-13.4$\pm$2.0&1.3$\pm$0.2&296.7$\pm$66.9&$\leq $1.9&$\leq $0.2&---&-4.1$\pm$0.5&0.5$\pm$0.1&---&$\leq $0.7&$\leq $0.1&---&-4.4$\pm$0.5&0.8$\pm$0.1&166.3$\pm$58.1\\ 
\noalign{\smallskip} 
Per emb 8&$\leq $1.4&$\leq $0.5&---&-2.2$\pm$0.4&0.7$\pm$0.1&135.3$\pm$37.4&-9.2$\pm$0.3&3.2$\pm$0.1&106.5$\pm$11.7&$\leq $0.9&$\leq $0.3&---&-4.6$\pm$0.2&1.8$\pm$0.1&107.9$\pm$16.5\\ 
\noalign{\smallskip} 
Per emb 25&$\leq $0.5&$\leq $0.2&---&$\leq $0.6&$\leq $0.3&---&-1.7$\pm$0.1&0.8$\pm$0.0&87.0$\pm$18.0&$\leq $0.4&$\leq $0.2&---&-0.9$\pm$0.1&0.5$\pm$0.0&79.7$\pm$15.4\\ 
\noalign{\smallskip} 
Per emb 26&-16.1$\pm$1.1&1.7$\pm$0.1&163.7$\pm$35.9&-3.3$\pm$0.6&0.5$\pm$0.1&146.2$\pm$47.9&-31.6$\pm$0.5&4.5$\pm$0.1&89.6$\pm$8.4&-1.5$\pm$0.3&0.2$\pm$0.1&64.1$\pm$41.9&-10.1$\pm$0.3&1.9$\pm$0.1&66.6$\pm$12.3\\ 
\noalign{\smallskip} 
Per emb 28&$\leq $0.4&$\leq $0.9&---&$\leq $1.1&$\leq $2.5&---&-1.8$\pm$0.3&4.2$\pm$0.7&177.2$\pm$43.8&$\leq $0.2&$\leq $0.5&---&-0.8$\pm$0.1&2.3$\pm$0.4&138.1$\pm$21.9\\ 
\noalign{\smallskip} 
Per emb 21&---&0.4$\pm$0.1&72.7$\pm$37.3&---&0.3$\pm$0.0&97.0$\pm$13.6&---&3.4$\pm$0.0&59.5$\pm$2.5&---&0.2$\pm$0.0&---&---&2.1$\pm$0.0&37.5$\pm$2.7\\ 
\noalign{\smallskip} 
Aqu MM11&-0.8$\pm$0.4&0.2$\pm$0.1&---&$\leq $0.2&$\leq $0.1&---&-1.4$\pm$0.3&0.5$\pm$0.1&69.1$\pm$11.6&$\leq $0.2&$\leq $0.1&---&-0.5$\pm$0.2&0.2$\pm$0.1&105.6$\pm$37.3\\ 
\noalign{\smallskip} 
Ceph mm&$\leq $0.7&$\leq $1.1&---&$\leq $0.3&$\leq $0.5&---&-1.5$\pm$0.1&2.6$\pm$0.2&162.4$\pm$41.5&$\leq $0.2&$\leq $0.4&---&-1.1$\pm$0.1&1.9$\pm$0.2&151.5$\pm$13.2\\ 
\noalign{\smallskip} 
Ser emb 2&$\leq $2.5&$\leq $0.7&---&$\leq $0.3&$\leq $0.1&---&-1.9$\pm$0.4&0.6$\pm$0.1&---&$\leq $0.3&$\leq $0.1&---&-1.2$\pm$0.3&0.4$\pm$0.1&137.0$\pm$34.3\\ 
\noalign{\smallskip} 
Ser emb 15&-3.4$\pm$1.2&0.7$\pm$0.3&84.3$\pm$36.4&$\leq $0.5&$\leq $0.1&---&-8.1$\pm$0.5&2.3$\pm$0.1&90.8$\pm$5.4&$\leq $0.3&$\leq $0.1&---&-2.8$\pm$0.3&0.9$\pm$0.1&63.4$\pm$9.8\\ 
\noalign{\smallskip} 
Aqu MM5&$\leq $0.7&$\leq $1.2&---&$\leq $0.1&$\leq $0.2&---&-1.8$\pm$0.1&3.3$\pm$0.2&113.8$\pm$15.8&$\leq $0.1&$\leq $0.3&---&-0.6$\pm$0.1&1.3$\pm$0.2&84.8$\pm$14.9\\ 
\noalign{\smallskip} 
Aqu MM8&$\leq $1.3&$\leq $0.4&---&$\leq $0.3&$\leq $0.1&---&-0.4$\pm$0.4&0.1$\pm$0.1&---&$\leq $0.3&$\leq $0.1&---&$\leq $0.5&$\leq $0.1&---\\ 
\noalign{\smallskip} 
Ser emb 22&-0.9$\pm$0.6&0.7$\pm$0.5&---&$\leq $0.2&$\leq $0.2&---&-0.5$\pm$0.2&0.5$\pm$0.2&---&$\leq $0.1&$\leq $0.2&---&-0.6$\pm$0.1&1.1$\pm$0.2&150.7$\pm$46.2\\ 
\noalign{\smallskip} 
SerS MM16&-2.5$\pm$1.5&0.4$\pm$0.2&46.5$\pm$14.9&$\leq $0.6&$\leq $0.1&---&-2.3$\pm$0.8&0.4$\pm$0.1&76.2$\pm$18.7&$\leq $0.5&$\leq $0.1&---&-1.8$\pm$0.7&0.3$\pm$0.1&274.8$\pm$89.5\\ 
\noalign{\smallskip} 
\hline
\smallskip
\end{tabular}
\tablecomments{\small Same as Table \ref{t.C0_line_obs1} for \HttwotooneSfour, \HtthreetotwoSfive, \HttwotooneSthree, \HtthreetotwoSfour, \HttwotooneStwo.}
\end{minipage}
\end{sidewaystable}


\begin{sidewaystable}[h!]
\centering
\begin{minipage}{\textwidth}
\scriptsize
\caption[]{Line parameters}
\label{t.C0_line_obs3}
\setlength{\tabcolsep}{0.08em} 
\begin{tabular}{p{0.08\linewidth}cccccccccccccccc}
\hline \hline \noalign{\smallskip}
  & \multicolumn{3}{c}{$\rm  H_{2}\,3-2\,S(3) $} & \multicolumn{3}{c}{$\rm  H_{2}\,3-2\,S(2) $} & \multicolumn{3}{c}{$\rm  H_{2}\,4-3\,S(3) $} & \multicolumn{3}{c}{$\rm  H_{2}\,2-1\,S(0) $} & \multicolumn{3}{c}{$\rm  H_{2}\,3-2\,S(1) $}\\ 
  & EW & Flux & FWHM  & EW & Flux & FWHM  & EW & Flux & FWHM  & EW & Flux & FWHM  & EW & Flux & FWHM \\ 
  & \AA&$ 10^{-19}\,\textrm{W}\,\textrm{m}^{-2}$&$\textrm{km}\,\textrm{s}^{-1}$ & \AA&$ 10^{-19}\,\textrm{W}\,\textrm{m}^{-2}$&$\textrm{km}\,\textrm{s}^{-1}$ & \AA&$ 10^{-19}\,\textrm{W}\,\textrm{m}^{-2}$&$\textrm{km}\,\textrm{s}^{-1}$ & \AA&$ 10^{-19}\,\textrm{W}\,\textrm{m}^{-2}$&$\textrm{km}\,\textrm{s}^{-1}$ & \AA&$ 10^{-19}\,\textrm{W}\,\textrm{m}^{-2}$&$\textrm{km}\,\textrm{s}^{-1}$\\ 
\noalign{\smallskip}  \hline \noalign{\smallskip} 
HOPS 50&-0.5$\pm$0.1&0.6$\pm$0.1&43.6$\pm$25.1&$\leq $0.3&$\leq $0.3&---&-0.5$\pm$0.1&0.6$\pm$0.1&---&$\leq $0.4&$\leq $0.5&---&$\leq $0.8&$\leq $1.0&---\\ 
\noalign{\smallskip} 
HOPS 60&-0.9$\pm$0.0&3.8$\pm$0.1&58.0$\pm$9.6&$\leq $0.1&$\leq $0.4&---&$\leq $0.1&$\leq $0.8&---&$\leq $0.1&$\leq $0.8&---&-0.5$\pm$0.0&3.5$\pm$0.3&12.7$\pm$25.2\\ 
\noalign{\smallskip} 
HOPS 87&---&1.1$\pm$0.0&58.4$\pm$9.3&---&0.3$\pm$0.1&---&$\leq $-62.4&$\leq $0.3&---&---&2.1$\pm$0.1&13.1$\pm$4.3&---&1.1$\pm$0.2&---\\ 
\noalign{\smallskip} 
HOPS 164&-1.2$\pm$0.1&0.4$\pm$0.0&67.9$\pm$19.4&$\leq $0.8&$\leq $0.3&---&$\leq $0.7&$\leq $0.2&---&-2.3$\pm$0.3&0.8$\pm$0.1&92.4$\pm$30.0&$\leq $1.3&$\leq $0.4&---\\ 
\noalign{\smallskip} 
HOPS 171&$\leq $0.2&$\leq $0.2&---&$\leq $0.6&$\leq $0.4&---&$\leq $0.4&$\leq $0.4&---&$\leq $0.4&$\leq $0.4&---&$\leq $0.3&$\leq $0.3&---\\ 
\noalign{\smallskip} 
HOPS 203&-3.7$\pm$0.3&0.7$\pm$0.1&96.5$\pm$8.1&$\leq $0.4&$\leq $0.1&---&$\leq $0.9&$\leq $0.2&---&-5.6$\pm$0.5&1.3$\pm$0.1&153.4$\pm$29.5&$\leq $1.9&$\leq $0.4&---\\ 
\noalign{\smallskip} 
HOPS 250&-4.5$\pm$0.1&7.0$\pm$0.1&56.2$\pm$4.1&-1.3$\pm$0.1&2.1$\pm$0.1&---&-1.4$\pm$0.1&2.6$\pm$0.2&32.8$\pm$12.2&-2.9$\pm$0.1&6.5$\pm$0.2&---&-2.2$\pm$0.1&4.8$\pm$0.2&29.7$\pm$6.2\\ 
\noalign{\smallskip} 
Per emb 24&-0.7$\pm$0.1&0.3$\pm$0.0&---&$\leq $0.5&$\leq $0.2&---&$\leq $0.5&$\leq $0.2&---&-0.7$\pm$0.2&0.3$\pm$0.1&---&-1.0$\pm$0.2&0.4$\pm$0.1&37.8$\pm$14.9\\ 
\noalign{\smallskip} 
Ser SMM3&-3.4$\pm$0.4&1.1$\pm$0.1&159.3$\pm$32.6&$\leq $0.9&$\leq $0.3&---&$\leq $1.2&$\leq $0.5&---&$\leq $1.2&$\leq $0.5&---&$\leq $1.6&$\leq $0.7&---\\ 
\noalign{\smallskip} 
Aqu MM4&$\leq $0.3&$\leq $0.1&---&$\leq $0.3&$\leq $0.2&---&$\leq $0.2&$\leq $0.1&---&$\leq $0.5&$\leq $0.2&---&$\leq $0.7&$\leq $0.4&---\\ 
\noalign{\smallskip} 
S68N&$\leq $1.9&$\leq $0.0&---&$\leq $1.8&$\leq $0.0&---&$\leq $3.6&$\leq $0.0&---&$\leq $3.7&$\leq $0.0&---&$\leq $3.0&$\leq $0.0&---\\ 
\noalign{\smallskip} 
HOPS 32&-3.6$\pm$0.1&3.9$\pm$0.1&13.3$\pm$7.7&-0.5$\pm$0.1&0.7$\pm$0.2&---&$\leq $0.6&$\leq $0.9&---&-2.9$\pm$0.4&4.8$\pm$0.7&96.5$\pm$27.3&$\leq $0.5&$\leq $0.9&---\\ 
\noalign{\smallskip} 
HOPS 44&-3.5$\pm$0.6&0.8$\pm$0.1&169.1$\pm$19.9&-4.2$\pm$0.5&1.3$\pm$0.2&221.6$\pm$27.6&$\leq $1.4&$\leq $0.5&---&$\leq $2.6&$\leq $0.8&---&$\leq $1.6&$\leq $0.5&---\\ 
\noalign{\smallskip} 
Per emb 8&-2.5$\pm$0.2&1.0$\pm$0.1&148.1$\pm$16.8&$\leq $0.9&$\leq $0.4&---&$\leq $0.5&$\leq $0.3&---&-2.1$\pm$0.3&1.1$\pm$0.2&---&-1.3$\pm$0.5&0.7$\pm$0.3&---\\ 
\noalign{\smallskip} 
Per emb 25&$\leq $0.2&$\leq $0.1&---&$\leq $0.7&$\leq $0.5&---&$\leq $0.4&$\leq $0.3&---&$\leq $0.5&$\leq $0.4&---&$\leq $0.5&$\leq $0.4&---\\ 
\noalign{\smallskip} 
Per emb 26&-3.5$\pm$0.3&0.8$\pm$0.1&62.4$\pm$15.2&$\leq $1.2&$\leq $0.3&---&$\leq $2.4&$\leq $0.7&---&-6.2$\pm$0.8&2.1$\pm$0.3&96.9$\pm$25.8&-3.4$\pm$0.7&1.2$\pm$0.2&---\\ 
\noalign{\smallskip} 
Per emb 28&$\leq $0.3&$\leq $0.8&---&$\leq $0.3&$\leq $1.0&---&$\leq $0.4&$\leq $1.5&---&$\leq $0.4&$\leq $1.7&---&$\leq $0.7&$\leq $3.3&---\\ 
\noalign{\smallskip} 
Per emb 21&---&1.1$\pm$0.0&74.1$\pm$4.0&---&0.2$\pm$0.0&---&$\leq $28.1&$\leq $0.2&---&---&0.2$\pm$0.1&---&$\leq $74.2&$\leq $0.4&---\\ 
\noalign{\smallskip} 
Aqu MM11&$\leq $0.2&$\leq $0.1&---&$\leq $0.3&$\leq $0.2&---&$\leq $0.4&$\leq $0.2&---&$\leq $0.2&$\leq $0.1&---&$\leq $0.6&$\leq $0.3&---\\ 
\noalign{\smallskip} 
Ceph mm&-0.7$\pm$0.1&1.3$\pm$0.1&43.3$\pm$9.5&$\leq $0.2&$\leq $0.4&---&$\leq $0.3&$\leq $0.6&---&$\leq $0.4&$\leq $0.8&---&$\leq $0.3&$\leq $0.6&---\\ 
\noalign{\smallskip} 
Ser emb 2&$\leq $0.4&$\leq $0.1&---&$\leq $0.3&$\leq $0.1&---&$\leq $0.6&$\leq $0.2&---&$\leq $0.7&$\leq $0.3&---&$\leq $0.8&$\leq $0.3&---\\ 
\noalign{\smallskip} 
Ser emb 15&-1.2$\pm$0.1&0.4$\pm$0.1&53.9$\pm$15.1&$\leq $0.4&$\leq $0.2&---&$\leq $1.1&$\leq $0.5&---&$\leq $0.7&$\leq $0.4&---&$\leq $0.8&$\leq $0.4&---\\ 
\noalign{\smallskip} 
Aqu MM5&-0.3$\pm$0.0&0.7$\pm$0.1&---&$\leq $0.1&$\leq $0.4&---&$\leq $0.3&$\leq $1.0&---&$\leq $0.1&$\leq $0.4&---&$\leq $0.1&$\leq $0.5&---\\ 
\noalign{\smallskip} 
Aqu MM8&$\leq $0.4&$\leq $0.1&---&$\leq $0.5&$\leq $0.1&---&$\leq $1.0&$\leq $0.2&---&$\leq $0.6&$\leq $0.1&---&$\leq $1.6&$\leq $0.3&---\\ 
\noalign{\smallskip} 
Ser emb 22&$\leq $0.1&$\leq $0.3&---&$\leq $0.1&$\leq $0.4&---&$\leq $0.1&$\leq $0.5&---&$\leq $0.2&$\leq $0.7&---&$\leq $0.2&$\leq $0.8&---\\ 
\noalign{\smallskip} 
SerS MM16&$\leq $0.6&$\leq $0.1&---&$\leq $0.4&$\leq $0.1&---&$\leq $1.2&$\leq $0.2&---&$\leq $1.3&$\leq $0.2&---&$\leq $2.0&$\leq $0.3&---\\ 
\noalign{\smallskip} 
\hline
\smallskip
\end{tabular}
\tablecomments{\small Same as Table \ref{t.C0_line_obs1} for \HtthreetotwoSthree, \HtthreetotwoStwo, \HtfourtothreeSthree, \HttwotooneSO, \HtthreetotwoSone.}
\end{minipage}
\end{sidewaystable}

\section{\normalfont{On the Class 0 nature of the 6 photospheric spectra}}
\label{app:photospheres_disc}

Six sources of our sample exhibit photospheric absorption features (CO overtone, Ca, and Na absorption lines) but no \Brgamma emission and only faint or no H$_2$ emission is detected, suggesting a somewhat low accretion activity. In Class I objects the photosphere is detected most of the time ($\sim$ 70\% of objects), yet \ion{H}{1} lines are usually detected and used to measure the accretion. We thus list in this Appendix additional details that of those presented in Table \ref{t.C0_source_charac} about these Class 0 protostars, in order to discuss their nature. The inclination of outflow cavities with respect to the line of sight is known to impact the classification of the objects, but at this early stage it is relatively difficult to quantify it. Indeed, more evolved protostars close to being edge-on have a red SED due to the disk blocking stellar photons, which make them appear as embedded objects. However these sources have a clear extended envelope which would easily extinct the NIR light of edge-on systems. Detecting NIR light in these protostars thus suggest they are not close to being edge-on systems.

Aqu MM4: The SED modeling of this source has been performed in \citet{Pokhrel2023} combining 2MASS, Spitzer, Herschel, and JCMT SCUBA-2 archival data. The authors confirmed the embedded Class 0 nature of the object ($T_{\textrm{bol}} < 70$ K, $\alpha_{4.5, 24}$>0.3). \citet{Mottram2017} JCMT HARP observations of CO and its isotopologues revealed a clear molecular outflow morphology, whose characteristics was consistent with the rest of the WILL Class 0 sample. Its outflow has also been characterized with \textit{Herschel}-PACS in \citet{Karska2018}, targeting CO $J_{\textrm{up}} = 14-26$. However, we highlight that the submillimeter dust emission peak from the high angular resolution ALMA archival observations (program ID 2018.1.00197.S) is shifted by $\sim 8.5^{\prime\prime}$ from the NIR source we targeted. We originally selected the source from its coordinates provided by the MAMBO and Herschel data from \citet{Maury2011,Mottram2017,Karska2018} that exhibits a $\sim 3^{\prime\prime}$ shift with the NIR source, consistent with the orientation of the blueshifted cavity. Therefore, the NIR source may not correspond to the Class 0 source seen in the FIR and submillimeter. The ALMA data has a short integration time, and we cannot determine whether the NIR source is associated with the envelope of the submillimeter source. The nature of this NIR source is thus unclear and the subsequent analysis must be taken with caution.

Aqu MM8: The SED modeling of \citet{Pokhrel2023} confirmed the Class 0 nature of this object. The submillimeter dust emission peak from the high angular resolution ALMA archival observations (program ID 2018.1.00197.S) is shifted by $\sim 2.5^{\prime\prime}$ from the NIR source. No clear CO (1$\rightarrow 0$) outflow is detected in the ALMA data, but a diffuse \HtonetoOSone emission is seen in the maps of \citet{ZhangM2015}.

Aqu MM11: In their SED modeling, \citet{Pokhrel2023} found a IR spectral index $\alpha_{4.5, 24}$ of 1.74, and a $T_{\textrm{bol}}$ of 87.6 K against the value of 54 K found by \citet{Maury2011} using MAMBO data. \citet{Pokhrel2023} classified Aqu MM11 as a Class I for its $T_{\textrm{bol}} > 70$ K.
The submillimeter dust emission peak from the high angular resolution ALMA archival observations (program ID 2019.1.01792.S) matches well the NIR source coordinate ($\sim 0.4^{\prime\prime}$ shift).

SerS MM16: The SED modeling of \citet{Pokhrel2023} confirmed the Class 0 nature of this object.
The NIR source detected in UKIDSS is $\sim 2.5^{\prime\prime}$ away from the peak of emission in the ALMA observations presented in \citet{Plunkett2018} (the associated source in their paper is called serps16/18), but within the extended protostellar envelope.

S68N: This source is a bona fide Class 0 object, analyzed in different studies \citep{WolfChase1998,Enoch2009,Enoch2011}.
The SED modeling of \citet{Pokhrel2023} confirmed the Class 0 nature of this object.
It was recently characterized in the NIR \citep{Greene2018}, and submillimeter studies \citep{LeGouellec2019a,Tychoniec2019,Galametz2019,Maury2019}. Several fragments of the protostellar envelope have been detected with ALMA 870 $\mu$m observations \citep{LeGouellec2019a}, whose one of them is bright in the NIR.

HOPS 164: In their SED modeling, \citep{Furlan2016} found a $T_\textrm{bol}$ of 50 K, and a low alpha of $\alpha_{4.5, 24}$ = -0.27, suggesting a hot inner disk. ALMA and VLA high angular resolution data of the sources are presented in \citet{Tobin2020}. The ALMA disk is $\sim 0.9^{\prime\prime}$ from the NIR source in the 2MASS image.

\section{\normalfont{ Additional Class 0 to Class I comparisons}}
\label{app:add_comp_plots}

\begin{figure*}[!tbh]
\centering
\subfigure{\includegraphics[scale=0.5,clip,trim= 1cm 0cm 2cm 0.8cm]{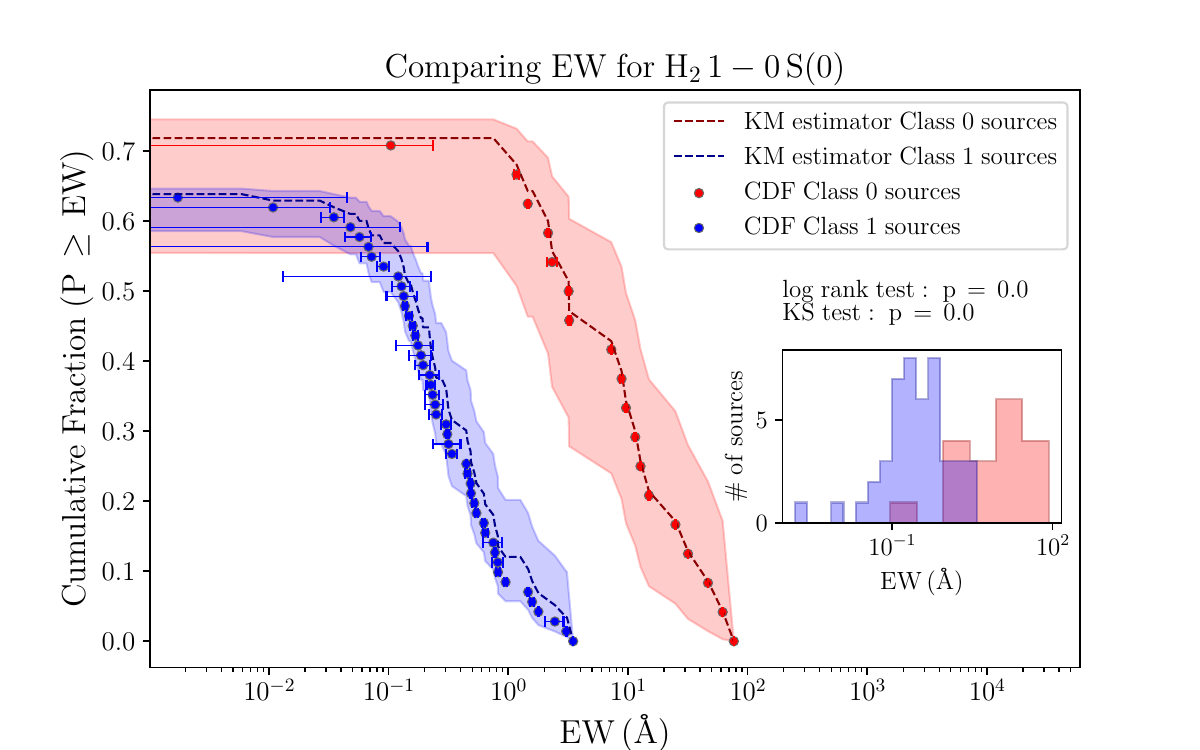}}
\subfigure{\includegraphics[scale=0.5,clip,trim= 1cm 0cm 2cm 0.8cm]{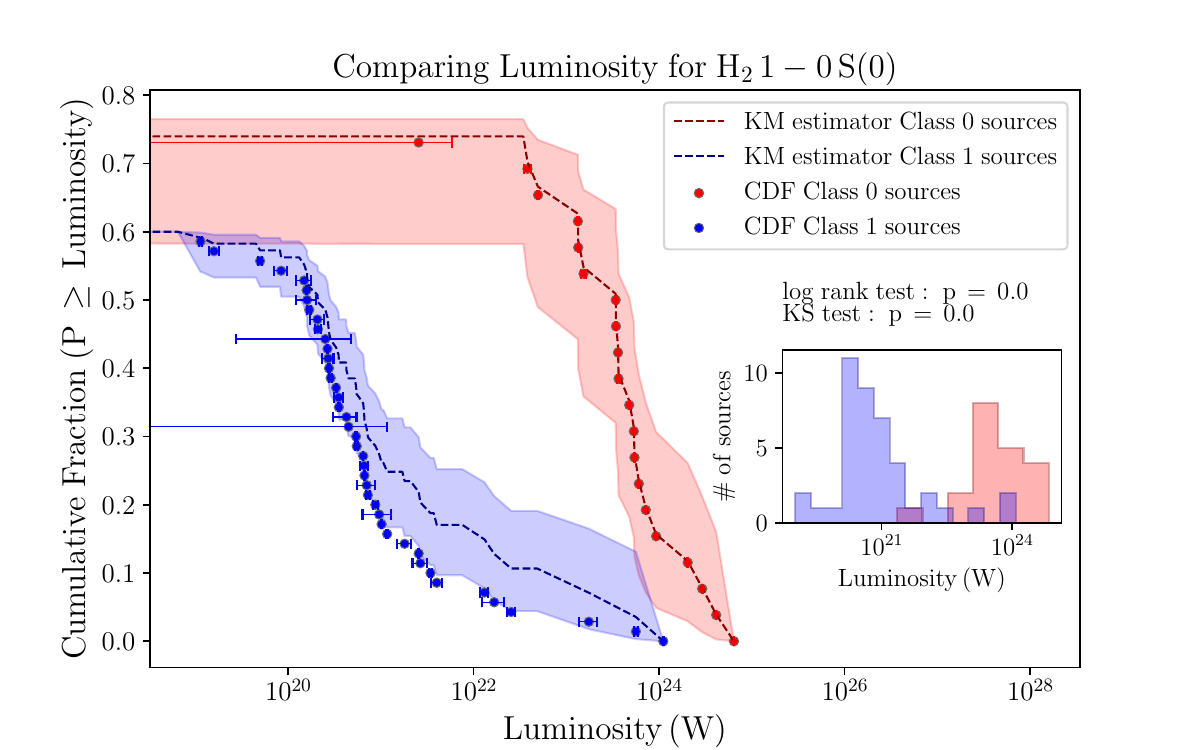}}
\subfigure{\includegraphics[scale=0.5,clip,trim= 1cm 0cm 2cm 0.8cm]{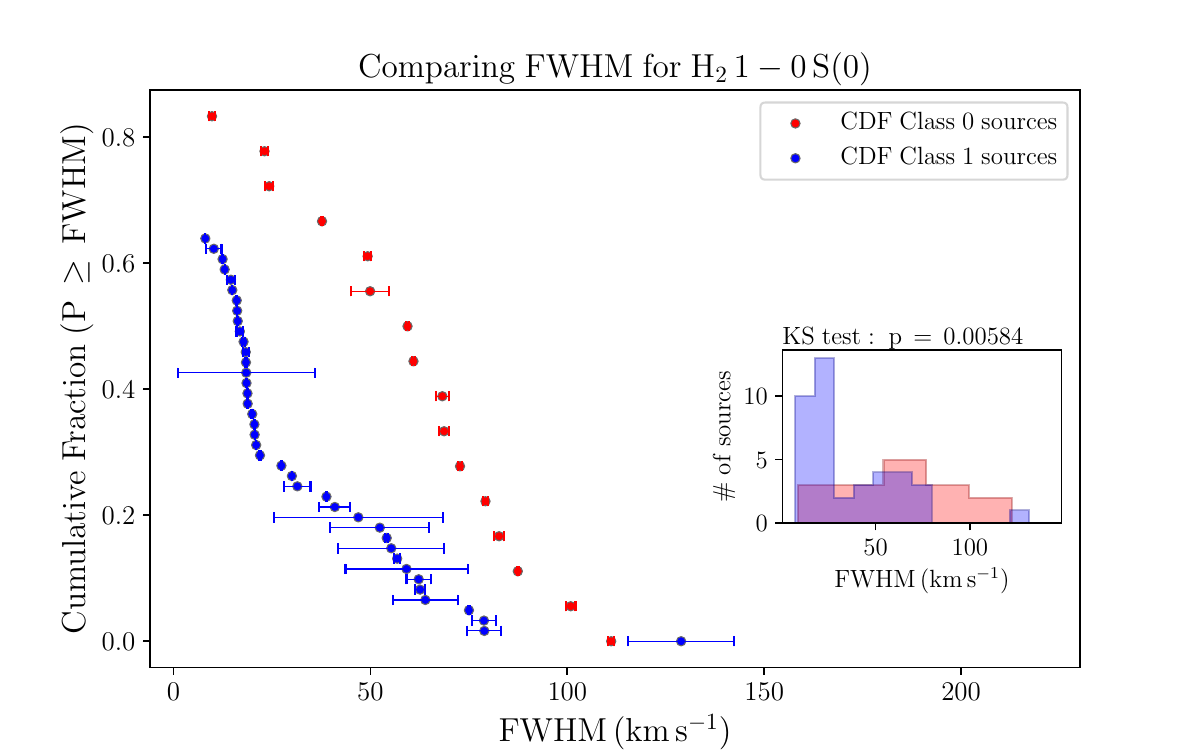}}
\subfigure{\includegraphics[scale=0.5,clip,trim= 1cm 0cm 2cm 0.8cm]{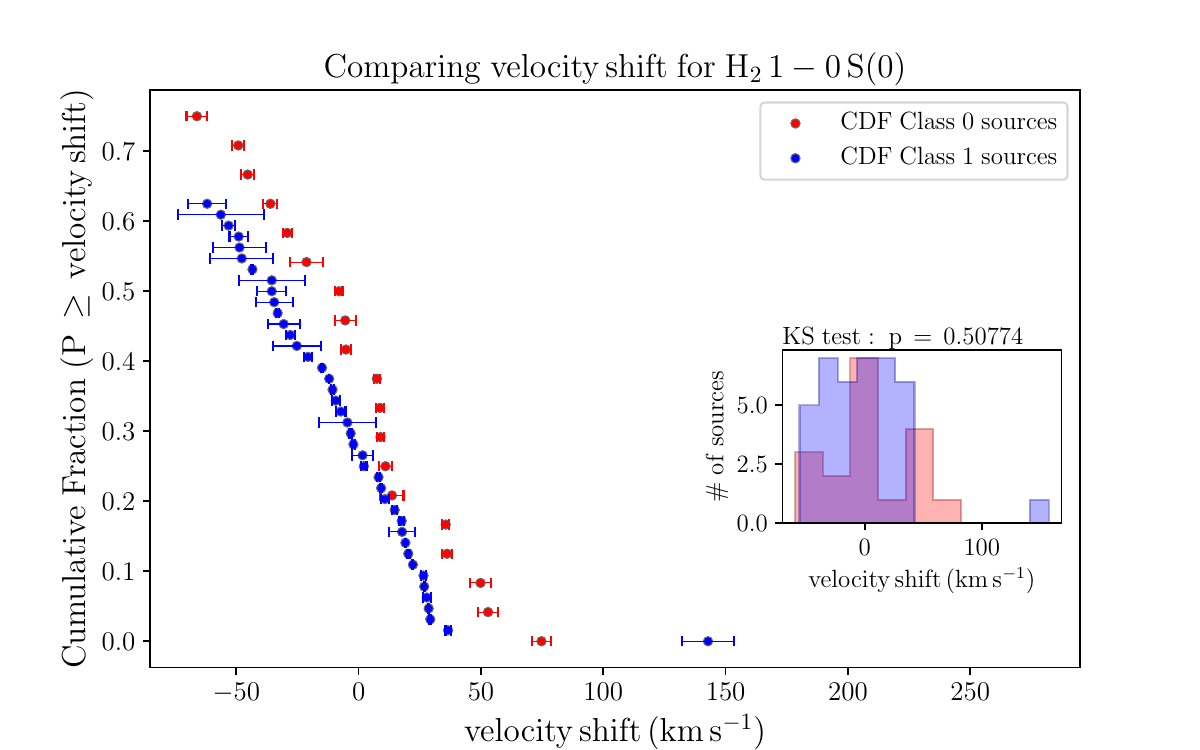}}
\caption{\small Same as Figure \ref{fig:C0_C1_comp_H2_1-0_S1} for the \HtonetoOSO emission lines.}
\label{fig:C0_C1_comp_H2_1-0_S0}
\vspace{0.2cm}
\end{figure*}

\begin{figure*}[!tbh]
\centering
\subfigure{\includegraphics[scale=0.5,clip,trim= 1cm 0cm 2cm 0.8cm]{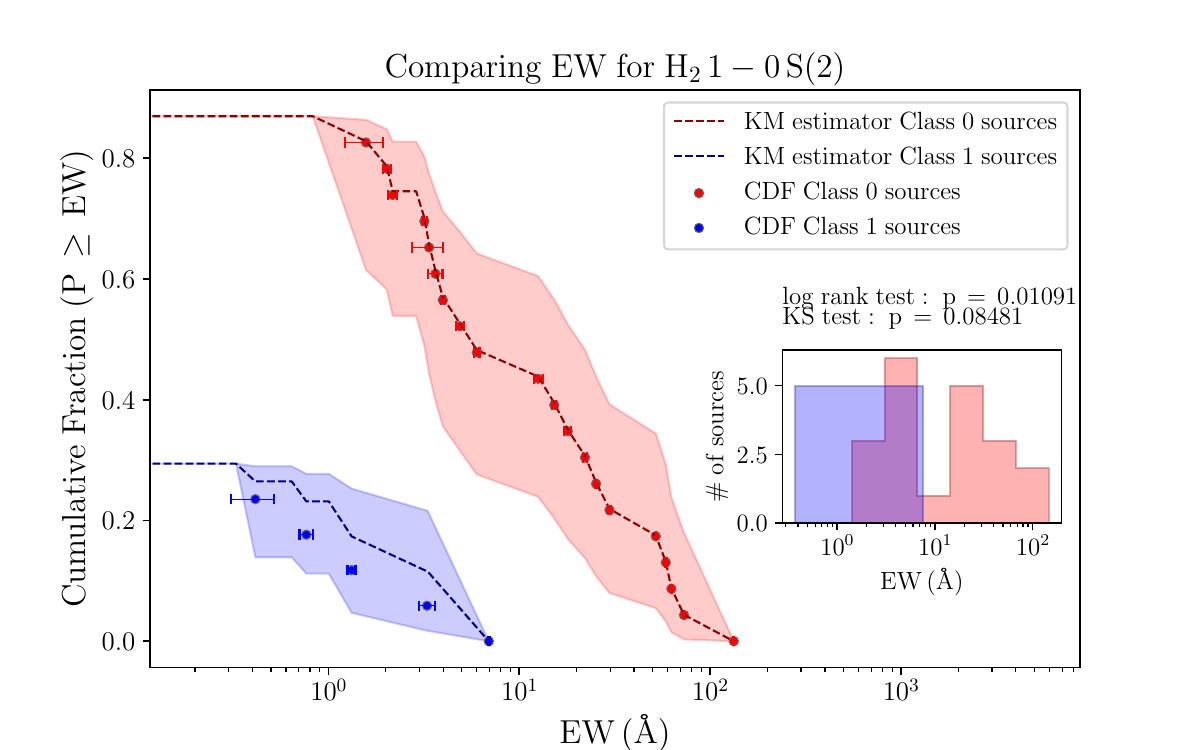}}
\subfigure{\includegraphics[scale=0.5,clip,trim= 1cm 0cm 2cm 0.8cm]{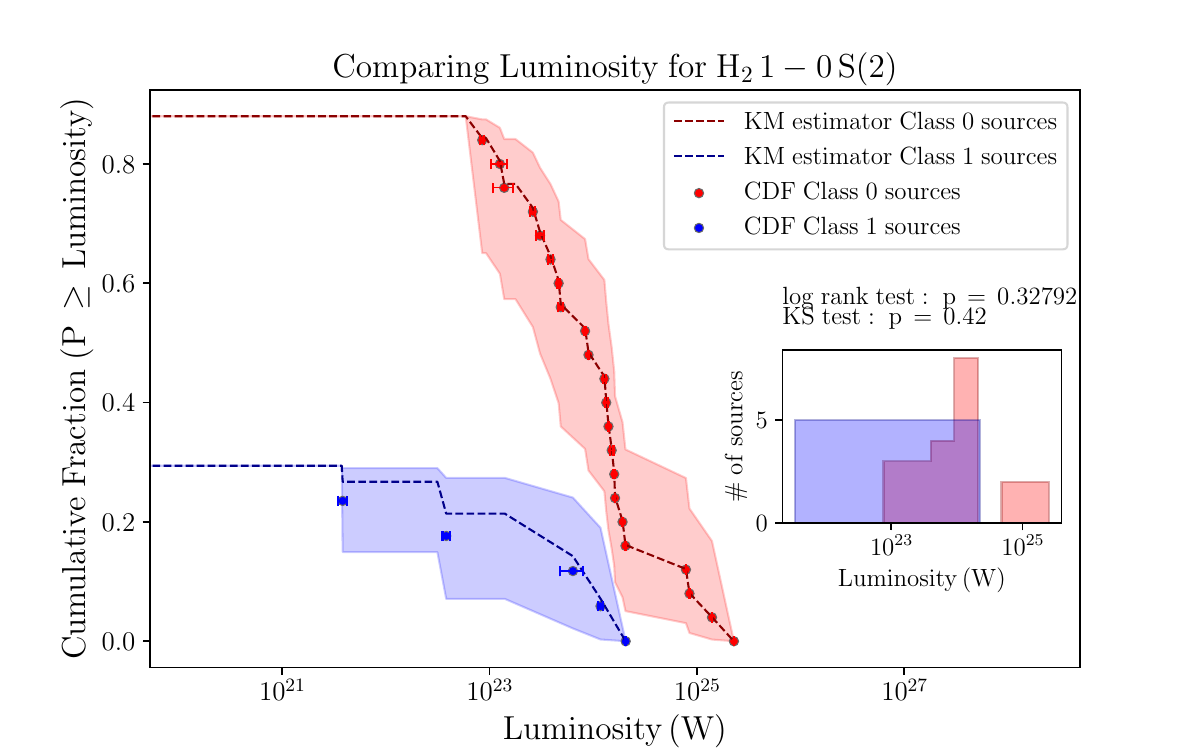}}
\subfigure{\includegraphics[scale=0.5,clip,trim= 1cm 0cm 2cm 0.8cm]{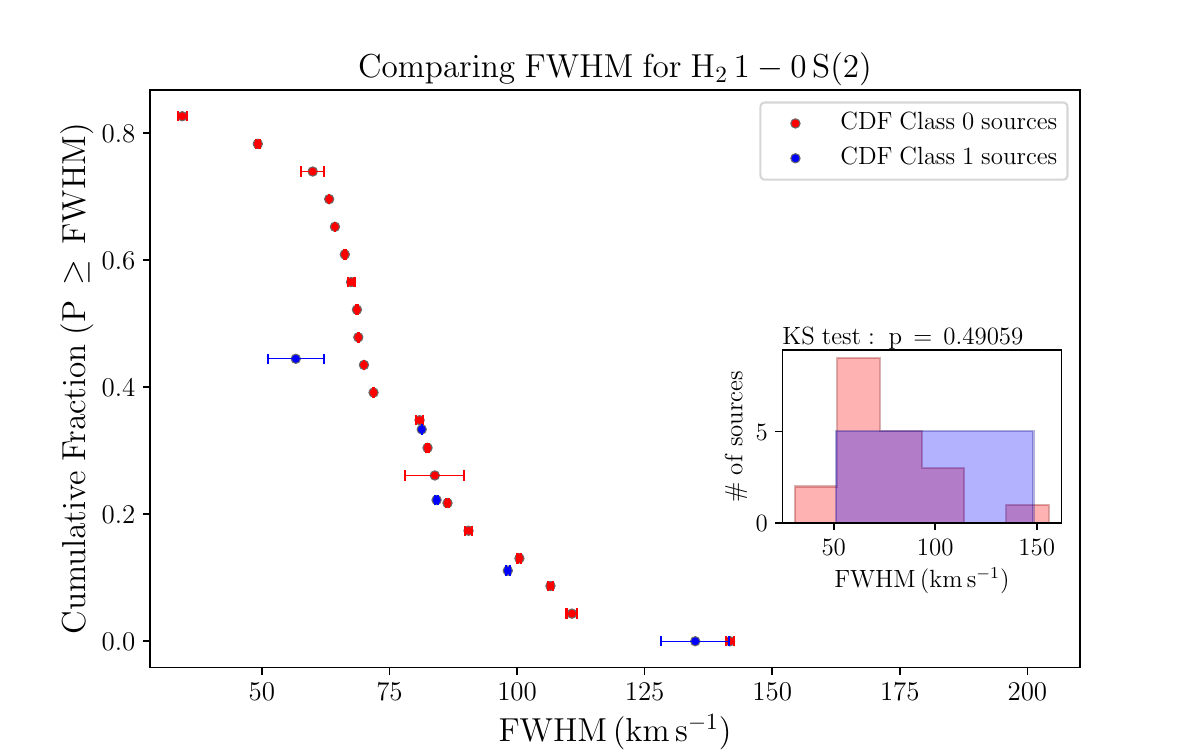}}
\subfigure{\includegraphics[scale=0.5,clip,trim= 1cm 0cm 2cm 0.8cm]{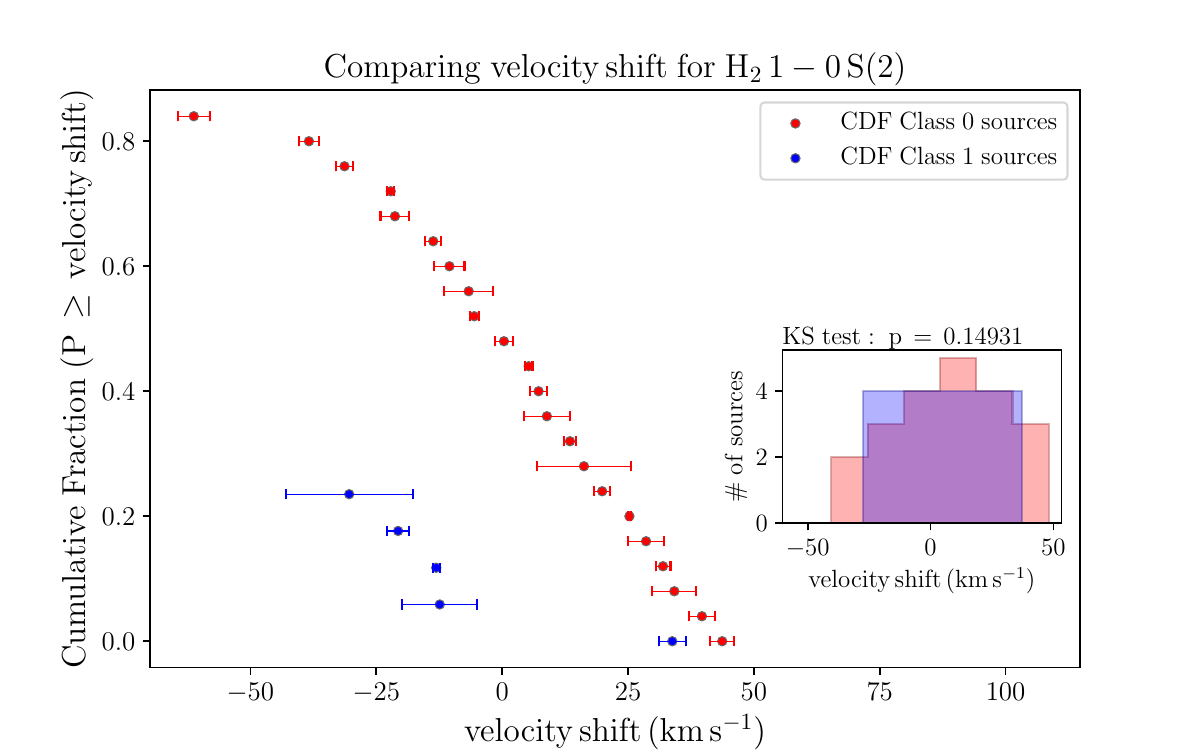}}
\caption{\small Same as Figure \ref{fig:C0_C1_comp_H2_1-0_S1} for the \HtonetoOStwo emission lines.}
\label{fig:C0_C1_comp_H2_1-0_S2}
\vspace{0.2cm}
\end{figure*}

\begin{figure*}[!tbh]
\centering
\subfigure{\includegraphics[scale=0.5,clip,trim= 1cm 0cm 2cm 0.8cm]{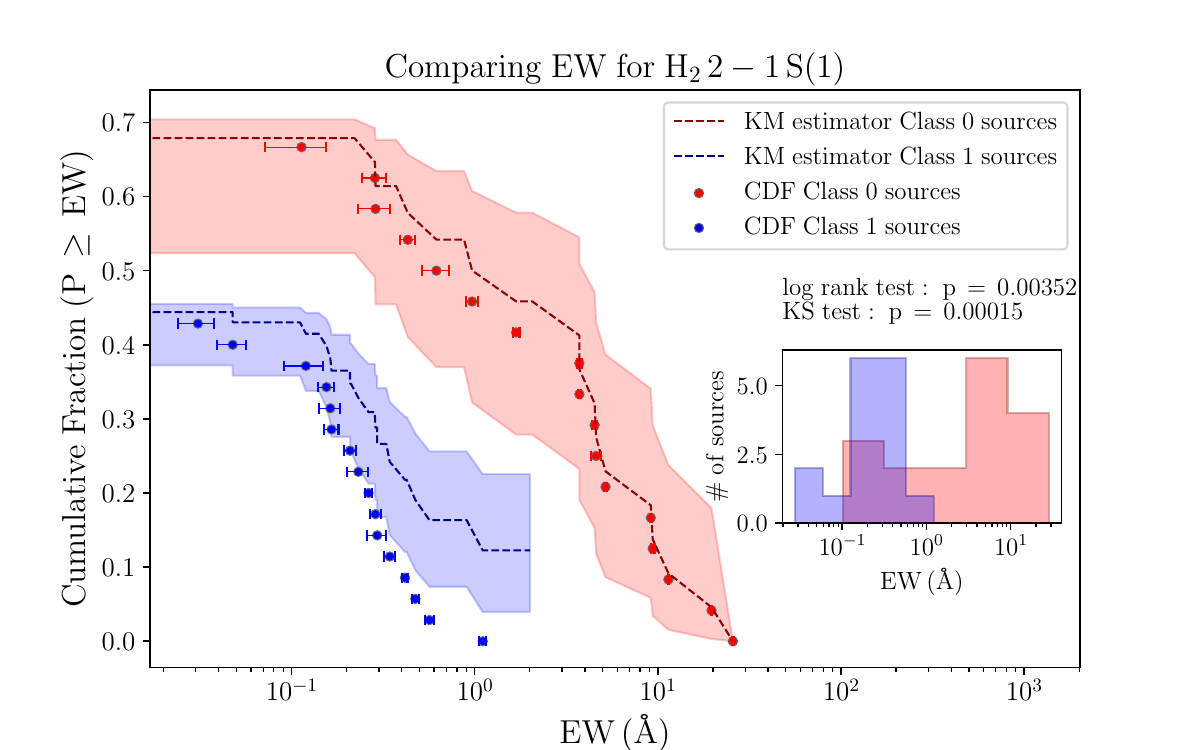}}
\subfigure{\includegraphics[scale=0.5,clip,trim= 1cm 0cm 2cm 0.8cm]{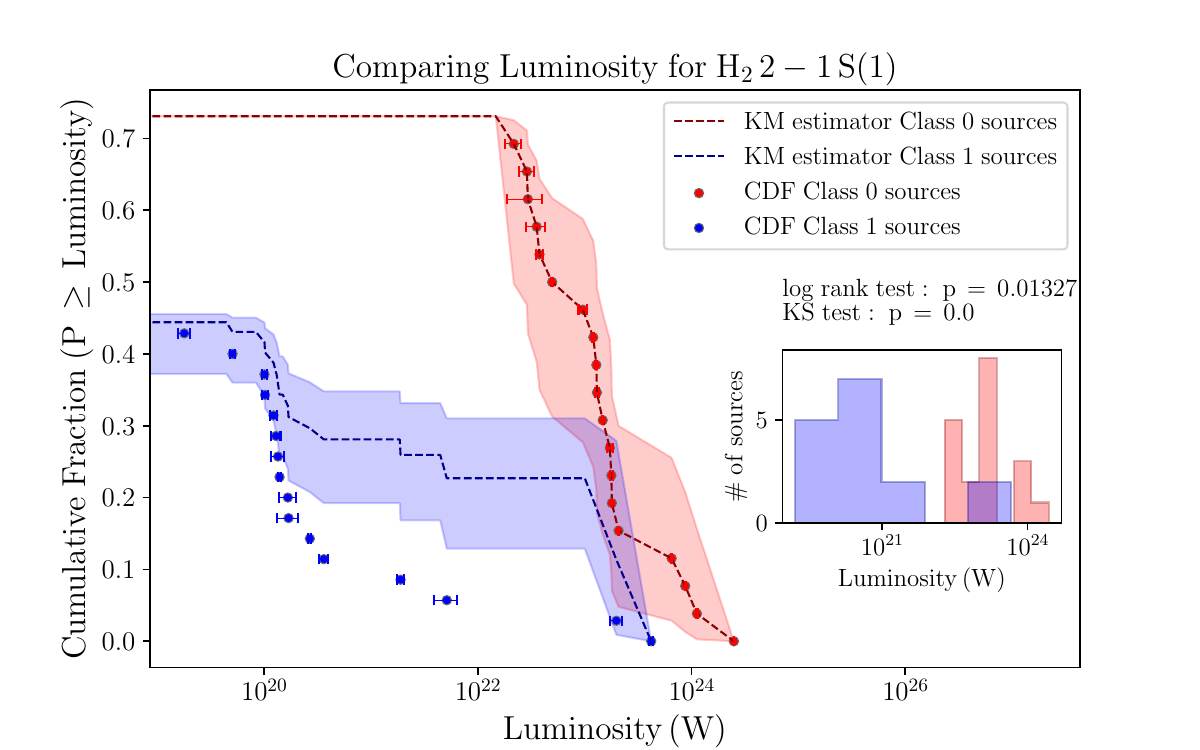}}
\subfigure{\includegraphics[scale=0.5,clip,trim= 1cm 0cm 2cm 0.8cm]{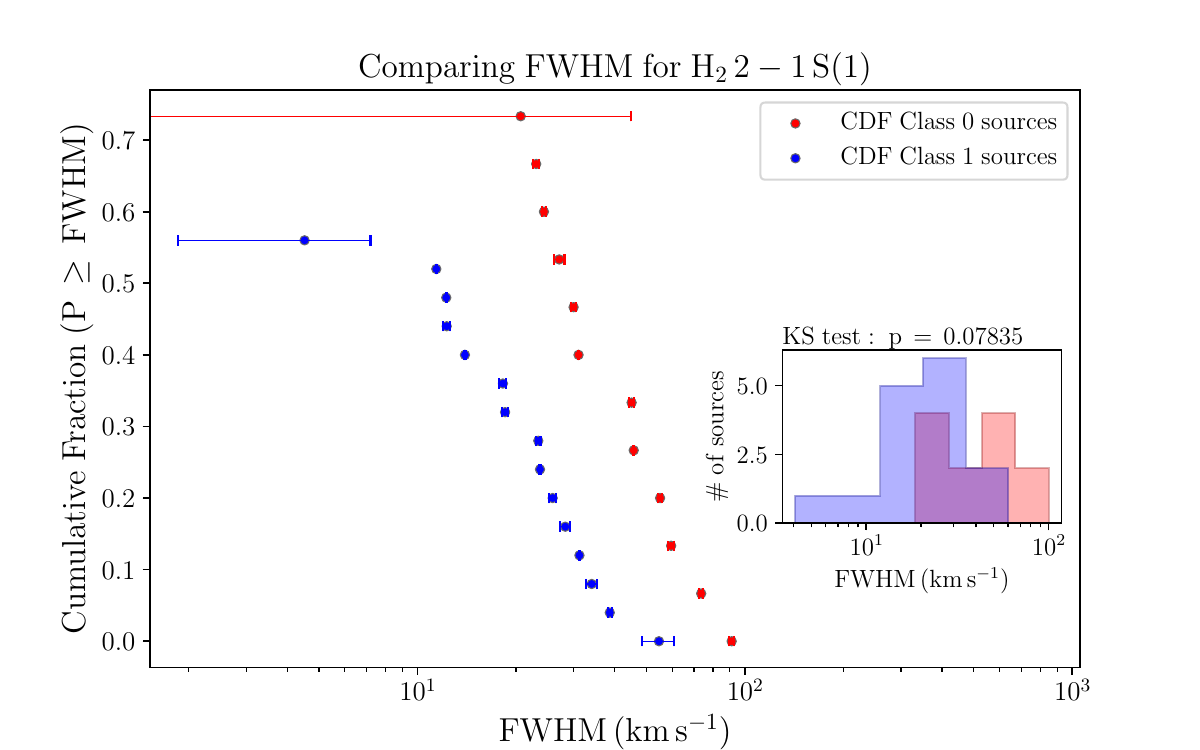}}
\subfigure{\includegraphics[scale=0.5,clip,trim= 1cm 0cm 2cm 0.8cm]{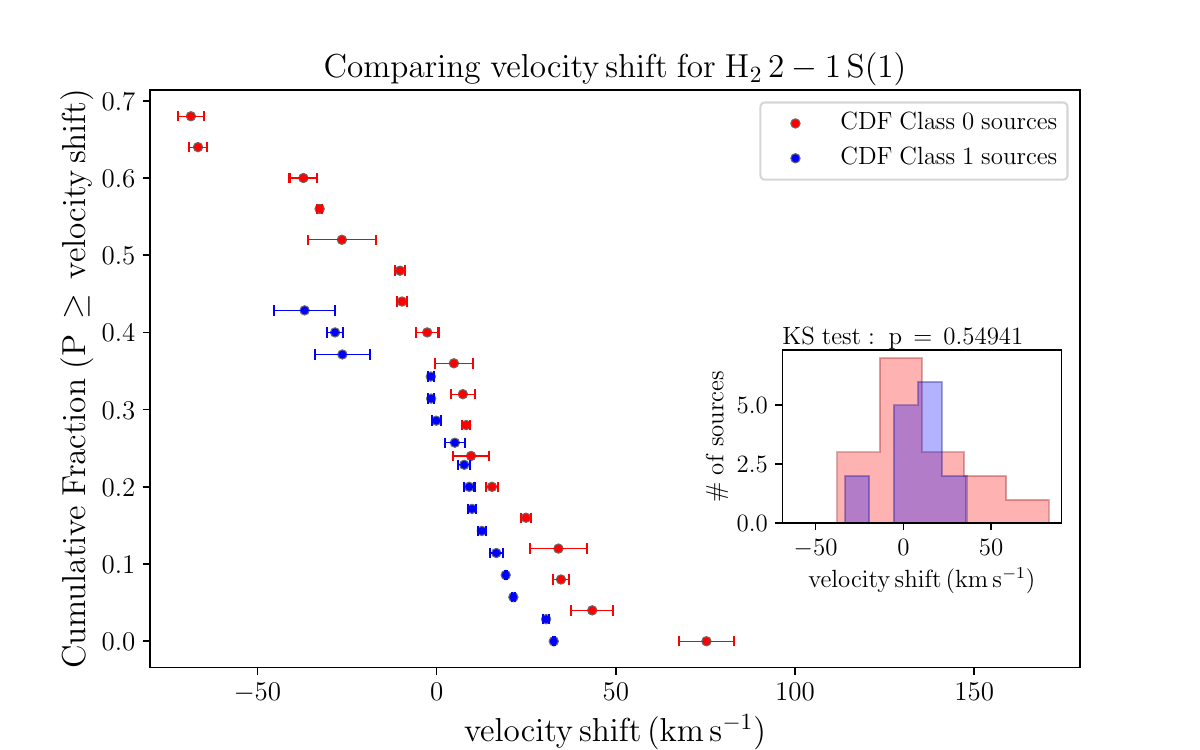}}
\caption{\small Same as Figure \ref{fig:C0_C1_comp_H2_1-0_S1} for the \HttwotooneSone emission lines.}
\label{fig:C0_C1_comp_H2_2-1_S1}
\vspace{0.2cm}
\end{figure*}

\begin{figure*}[!tbh]
\centering
\subfigure{\includegraphics[scale=0.5,clip,trim= 1cm 0.3cm 1.5cm 1cm]{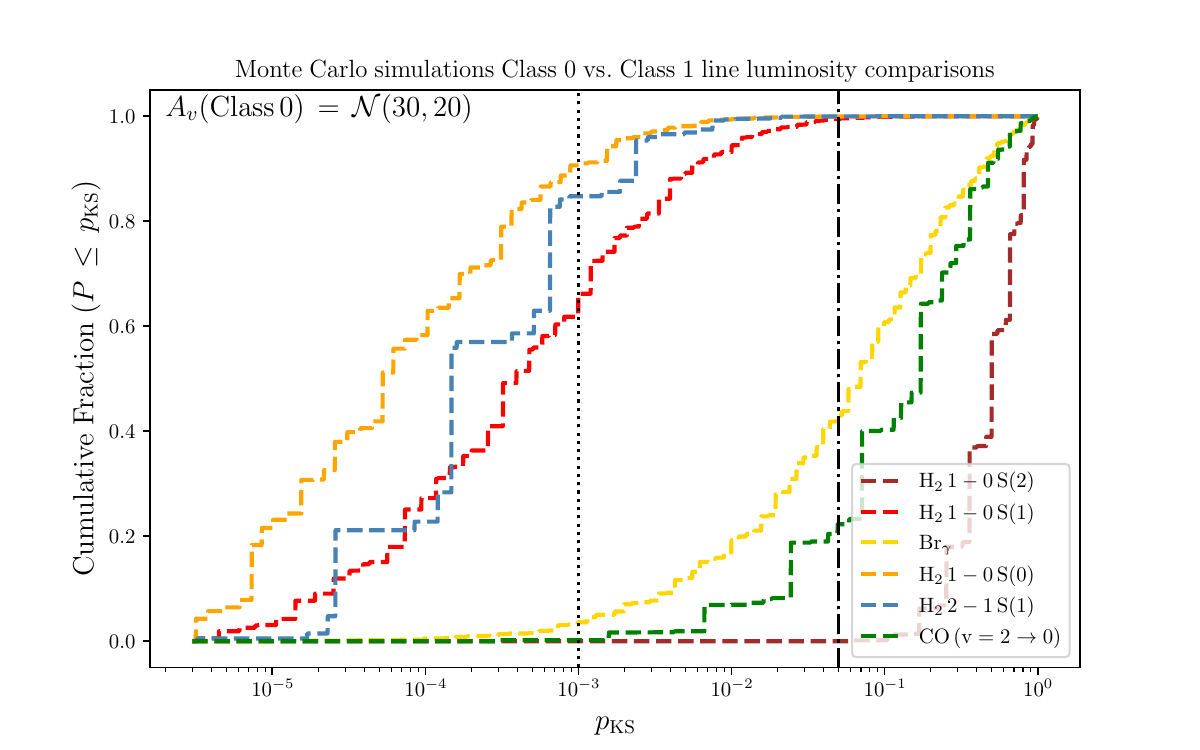}}
\subfigure{\includegraphics[scale=0.5,clip,trim= 1cm 0.3cm 1.5cm 1cm]{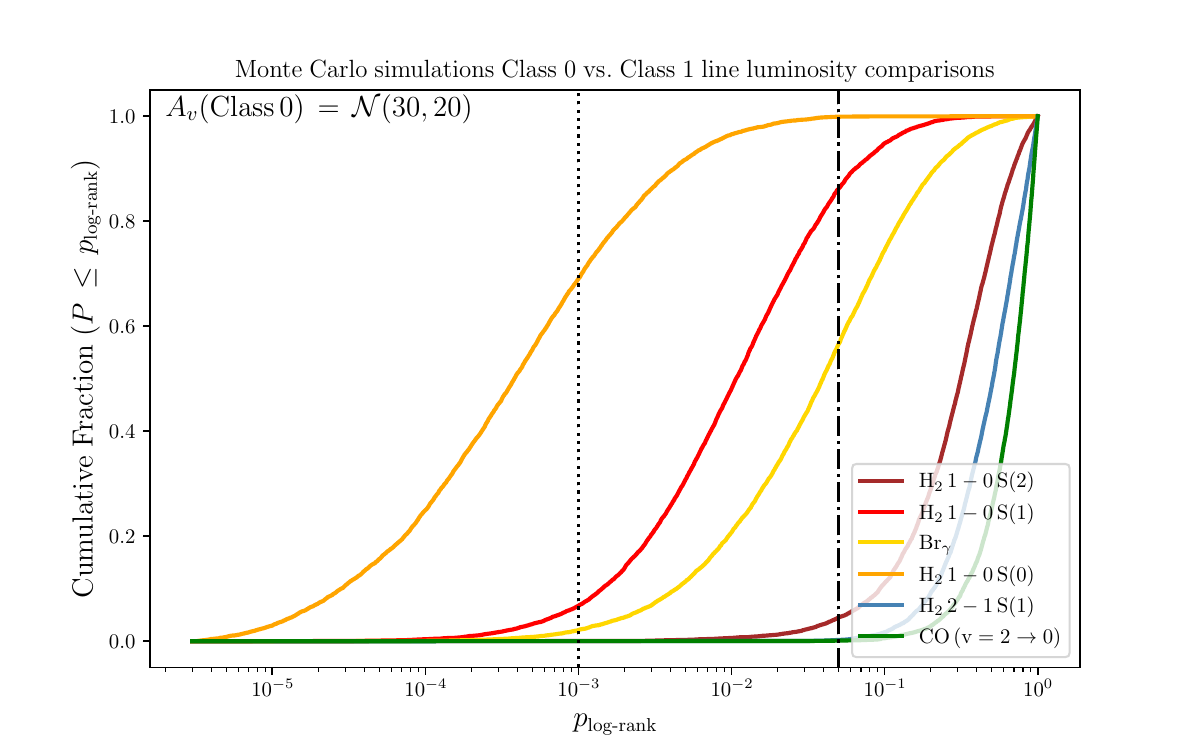}}
\caption{\small Same as Figure \ref{fig:C0_C1_comp_MC_simu_20_50} for normal distribution of $A_v$ centered on 30 mag, with a dispersion of 20 mag for the synthetic population of Class 0 line luminosities.}
\label{fig:C0_C1_comp_MC_simu_20_30}
\end{figure*}

\begin{figure*}[!tbh]
\centering
\subfigure{\includegraphics[scale=0.5,clip,trim= 1cm 0.3cm 1.5cm 1cm]{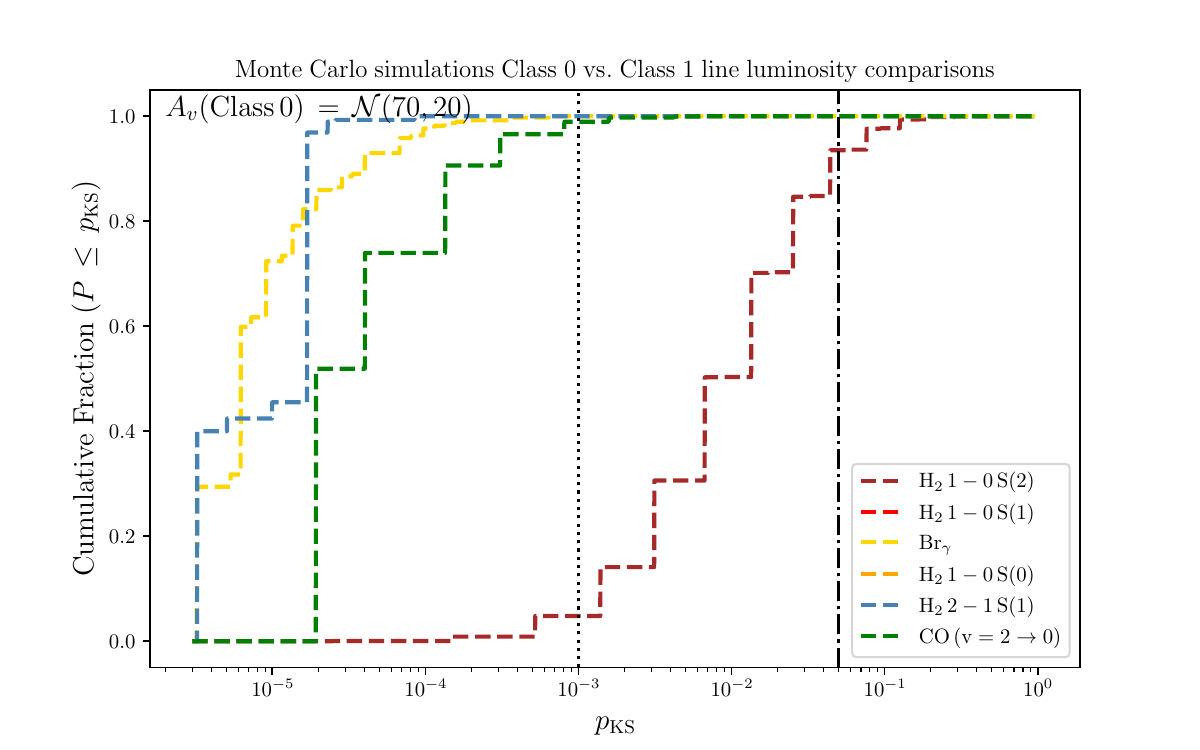}}
\subfigure{\includegraphics[scale=0.5,clip,trim= 1cm 0.3cm 1.5cm 1cm]{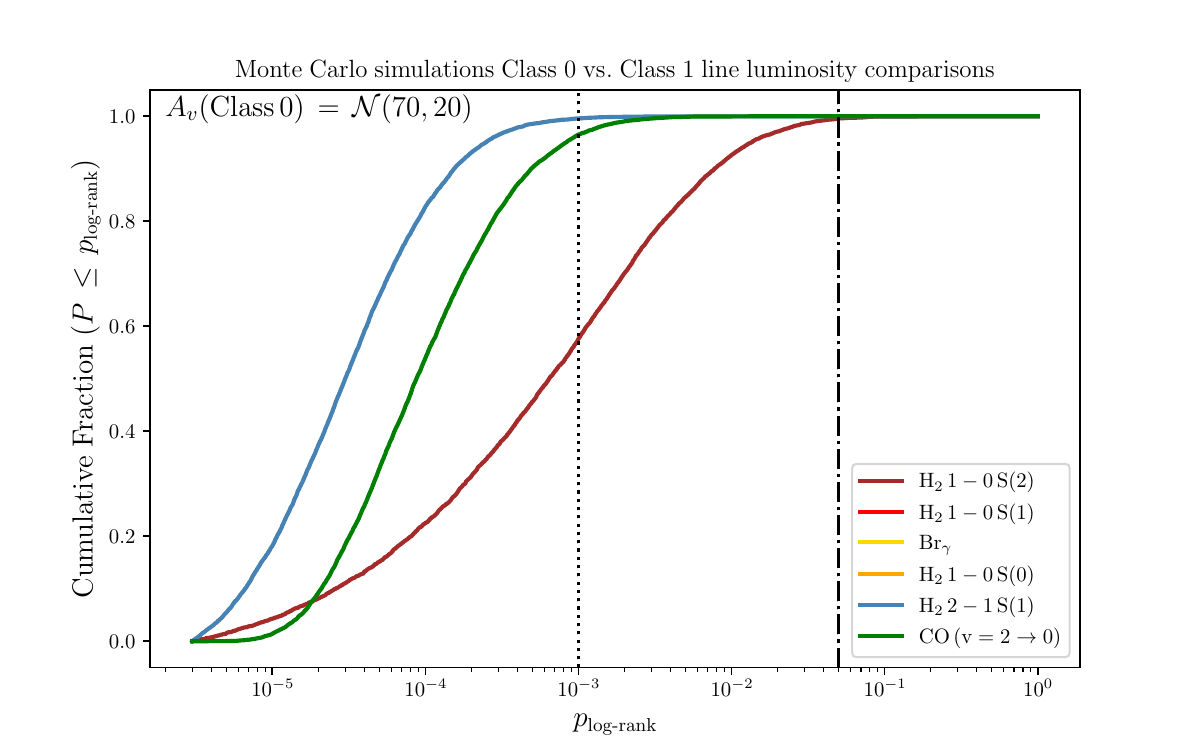}}
\caption{\small Same as Figure \ref{fig:C0_C1_comp_MC_simu_20_50} for normal distribution of $A_v$ centered on 30 mag, with a dispersion of 20 mag for the synthetic population of Class 0 line luminosities. Distributions are not shown if all the corresponding $p$-values are $\leq\,10^{-6}$.}
\label{fig:C0_C1_comp_MC_simu_20_70}
\end{figure*}

We present in Figures \ref{fig:C0_C1_comp_H2_1-0_S0}, \ref{fig:C0_C1_comp_H2_1-0_S2}, and \ref{fig:C0_C1_comp_H2_2-1_S1} additional Class 0 and Class I statistical comparisons using the \HtonetoOSO, \HtonetoOStwo, and \HttwotooneSone lines (see Section \ref{sec:comp_ClassI}). Figures \ref{fig:C0_C1_comp_MC_simu_20_30} and \ref{fig:C0_C1_comp_MC_simu_20_70} present the results of the Monte Carlo simulation performed in Section \ref{sec:stats_comp}, where the normal distributions of the Class 0 $A_v$ extinction values are centered on 30 mag and 70 mag, respectively. While the latter case exhibits clear statistical differences between the Class 0 and I emission line luminosities, the case with the $A_v$ distribution centered on 30 mag is not fully consistent with the distribution of line luminosities being drawn from statistically different parent distributions. However, \Brgamma, \HtonetoOSone, and \HtonetoOSO can be considered marginally different. As Class 0 protostars are expected to be deeply embedded in their dense circumstellar envelope, we find it reasonable to consider that Class 0s have $A_v \gtrsim$ 50 mag \citep{Greene2018,Laos2021}, which is the mean value we use for our Class 0 visual extinction distribution in Section \ref{sec:stats_comp}.

\end{document}